\documentclass[fleqn,usenatbib]{mnras}

\usepackage{newtxtext,newtxmath}

\usepackage[T1]{fontenc}
\usepackage{threeparttable}
\usepackage{float}
\usepackage{ulem}
\usepackage{longtable}
\usepackage{xcolor}
\usepackage{fix-cm}
\DeclareRobustCommand{\VAN}[3]{#2}
\let\VANthebibliography\thebibliography
\def\thebibliography{\DeclareRobustCommand{\VAN}[3]{##3}\VANthebibliography}

\usepackage{graphicx}	
\usepackage{amsmath}	
\usepackage{upgreek}

\title[ATOMS -]{ATOMS: ALMA Three-millimeter Observations of Massive Star-forming regions – \uppercase\expandafter{\romannumeral19}. The origin of SiO emission}

\author[Rong Liu et al.]
{Rong Liu,$^{1,2}$\thanks{E-mail: liu\underline{~~}rong@bao.ac.cn}
Tie Liu,$^{3}$
Izaskun Jiménez-Serra,$^{4}$
Jin-Zeng Li,$^{1}$
Jesús Martín-Pintado,$^{4}$
Xunchuan Liu,$^{3}$
\newauthor
Chang Won Lee,$^{5,6}$
Patricio Sanhueza,$^{7,8}$
James O. Chibueze,$^{9,10}$
Víctor M. Rivilla,$^{4}$
Mika Juvela,$^{11}$
\newauthor
Laura Colzi,$^{4}$
Leonardo Bronfman,$^{12}$
Hong-Li Liu,$^{13}$
Miguel Sanz-Novo,$^{4}$
Álvaro López-Gallifa,$^{4}$
\newauthor
Shanghuo Li,$^{15}$
Andrés Megías,$^{4}$
David San Andrés,$^{4}$
Guido Garay,$^{12,20}$
Jihye Hwang,$^{5}$
Jianwen Zhou,$^{16}$
\newauthor
Fengwei Xu,$^{17,18}$
Antonio Martínez-Henares,$^{4}$
Anindya Saha,$^{19}$
Hafiz Nazeer$^{19}$
\\
$^{1}$National Astronomical Observatories of China, Chinese Academy of Sciences, Beijing, 100012, China\\
$^{2}$University of Chinese Academy of Sciences, Beijing 100049, Peoples Republic of China  \\
$^{3}$Shanghai Astronomical Observatory, Chinese Academy of Sciences, 80 Nandan Road, Shanghai 200030, Peoples Republic of China \\
$^{4}$Centro de Astrobiología (CAB), CSIC-INTA, Carretera de Ajalvir km 4, 28850 Torrejón de Ardoz, Spain\\
$^{5}$Korea Astronomy and Space Science Institute, 776 Daedeokdae-ro, Yuseong-gu, Daejeon 34055, Republic of Korea\\
$^{6}$University of Science and Technology, Korea (UST), 217 Gajeong-ro, Yuseong-gu, Daejeon 34113, Republic of Korea\\
$^{7}$National Astronomical Observatory of Japan, National Institutes of Natural Sciences, 2-21-1 Osawa, Mitaka, Tokyo 181-8588, Japan\\
$^{8}$Astronomical Science Program, The Graduate University for Advanced Studies, SOKENDAI, 2-21-1 Osawa, Mitaka, Tokyo 181-8588, Japan\\
$^{9}$Department of Mathematical Sciences, University of South Africa, Cnr Christian de Wet Rd and Pioneer Avenue, Florida Park, 1709, Roodepoort, South Africa\\
$^{10}$Department of Physics and Astronomy, Faculty of Physical Sciences, University of Nigeria, Carver Building, 1 University Road, Nsukka 410001, Nigeria\\
$^{11}$Department of Physics, P.O. Box 64, FI-00014, University of Helsinki, Finland\\
$^{12}$Departamento de Astronomía, Universidad de Chile, Las Condes, Santiago, Chile\\
$^{13}$School of Physics and Astronomy, Yunnan University, Kunming, 650091, PR China\\
$^{15}$Max Planck Institute for Astronomy, Königstuhl 17, D-69117 Heidelberg, Germany\\
$^{16}$Max-Planck-Institut für Radioastronomie, Auf dem Hügel 69, 53125 Bonn, Germany\\
$^{17}$Kavli Institute for Astronomy and Astrophysics, Peking University, Haidian District, Beijing 100871, PR China\\
$^{18}$Department of Astronomy, School of Physics, Peking University, Beijing 100871, PR China\\
$^{19}$Indian Institute of Space Science and Technology, Thiruvananthapuram 695 547, Kerala, India\\
$^{20}$Chinese Academy of Sciences South America Center for Astronomy, National Astronomical Observatories, CAS, Beĳing 100101, China\\
}
\date{Accepted XXX. Received YYY; in original form ZZZ}

\pubyear{2023}

\begin{document}
\label{firstpage}
\pagerange{\pageref{firstpage}--\pageref{lastpage}}
\maketitle
\begin{abstract}
The production of silicon monoxide (SiO) can be considered as a fingerprint of shock interaction. 
In this work, we use high-sensitivity observations of the SiO (2-1) and H$^{13}$CO$^{+}$ (1-0) emission to investigate the broad and narrow SiO emission toward 146 massive star-forming regions in the ATOMS survey. We detected SiO emission in 136 regions and distinguished broad and narrow components across the extension of 118 sources (including 58 UC H\textsc{ii} regions) with an average angular resolution of 2.5$^{\prime}$$^{\prime}$. 
The derived SiO luminosity ($L_\textup{SiO}$) across the whole sample shows that the majority of $L_\textup{SiO}$ (above 66$\%$) can be attributed to broad SiO, indicating its association with strong outflows.
The comparison of the ALMA SiO images with the filamentary skeletons identified from H$^{13}$CO$^{+}$ and in the infrared data (at 4.5, 8, and 24 $\upmu$m), further confirms that most SiO emission originates from outflows. However, note that for nine sources in our sample, the observed SiO emission may be generated by expanding UC H\textsc{ii} regions. There is a moderate positive correlation between the bolometric luminosity ($L_\textup{bol}$) and $L_\textup{SiO}$ for both components (narrow and broad). The UC H\textsc{ii} sources show a weaker positive correlation between $L_\textup{bol}$ and $L_\textup{SiO}$ and higher $L_\textup{SiO}$ compared to the sources without UC H\textsc{ii} regions. These results imply that the SiO emission from UC H\textsc{ii} sources might be affected by UV-photochemistry induced by UC H\textsc{ii} regions.

\end{abstract}


\section{Introduction}
The process of massive star formation provides a variety of feedback to their hosting clouds and the interstellar medium (ISM). The protostellar outflows and the ionizing radiation generated by ultra-compact (UC) H\textsc{ii} regions profoundly influence their natal cloud. These processes inject mass, momentum, and energy into their nearby environments \citep{bally2016protostellar, kuiper2016protostellar}. 
Indeed, the jets and outflows powered by massive stars interact with the surrounding medium forming shock waves \citep{bally2016protostellar, li2019formation}, and the ultraviolet radiation from the central massive stars ionizes and dissociates the surrounding ISM. The expansion of UC H\textsc{ii} regions is also expected to produce shocks that compress the ambient medium \citep{hosokawa2006dynamical,Zhu2023Hiiregion}. These processes change the density and gas turbulence of the clouds, promoting or inhibiting future star formation. Therefore, studying protostellar feedback is essential to deepen our understanding not only of the physical and chemical processes involved in massive star formation but also on how it affects the formation of the future generation of stars in the Milky Way and in external galaxies.

The shock activity in molecular clouds could serve as a fossil record of star formation feedback \citep{duarte2014sio, Cosentino2020}. Thus, it is crucial for us to unveil the origin of the shocks in star-forming regions (SFRs). SiO is the perfect tracer of interstellar shock waves because it experiences high depletion in quiescent regions compared to shocked regions \citep{martin1992sio}. In particular, the SiO abundance can be enhanced by up to six orders of magnitude in outflow regions \citep{martin1992sio}. 
Dust grains are known to be partially (or totally) destroyed through sputtering in shocks, 
producing the release of significant amounts of Si atoms and/or Si-bearing molecules into the gas phase \citep{schilke1997sio, 1997Caselli, 2008Jimnez}.

\begin{figure*}
\begin{minipage}[t]{0.25\linewidth}
    \vspace{0.5pt}
\centerline{\includegraphics[width=1.03\linewidth]{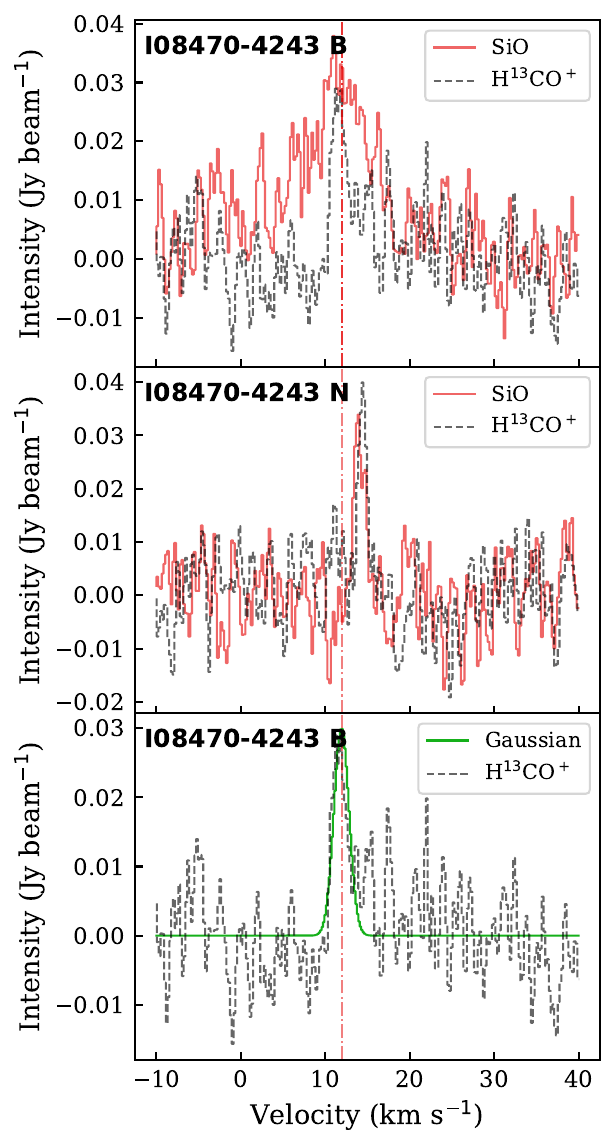}}
\end{minipage}
\begin{minipage}[t]{0.24\linewidth}
\vspace{0.5pt}  \centerline{\includegraphics[width=1.03\linewidth]{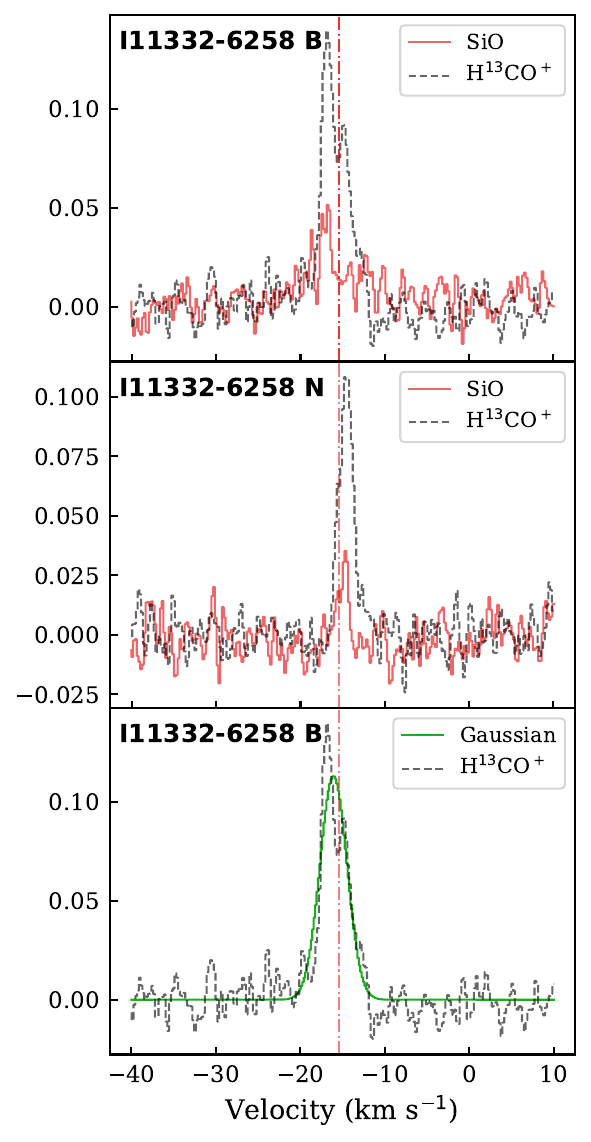}}
\end{minipage}
\begin{minipage}[t]{0.236\linewidth}
\vspace{0.5pt}  \centerline{\includegraphics[width=1.03\linewidth]{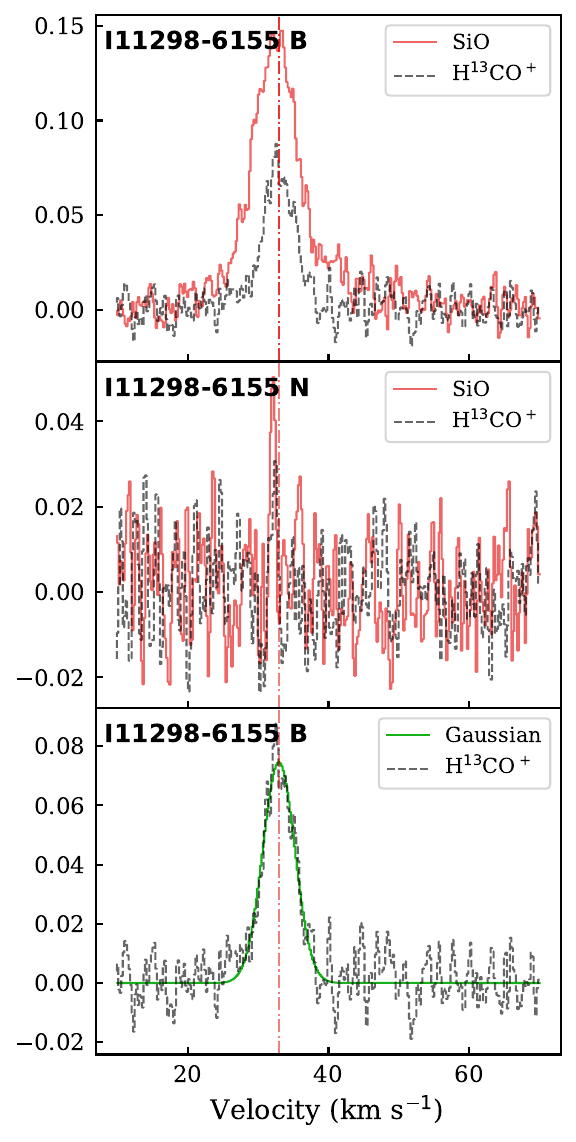}}   
\end{minipage}
\begin{minipage}[t]{0.24\linewidth}
\vspace{0.5pt}   \centerline{\includegraphics[width=1.03\linewidth]{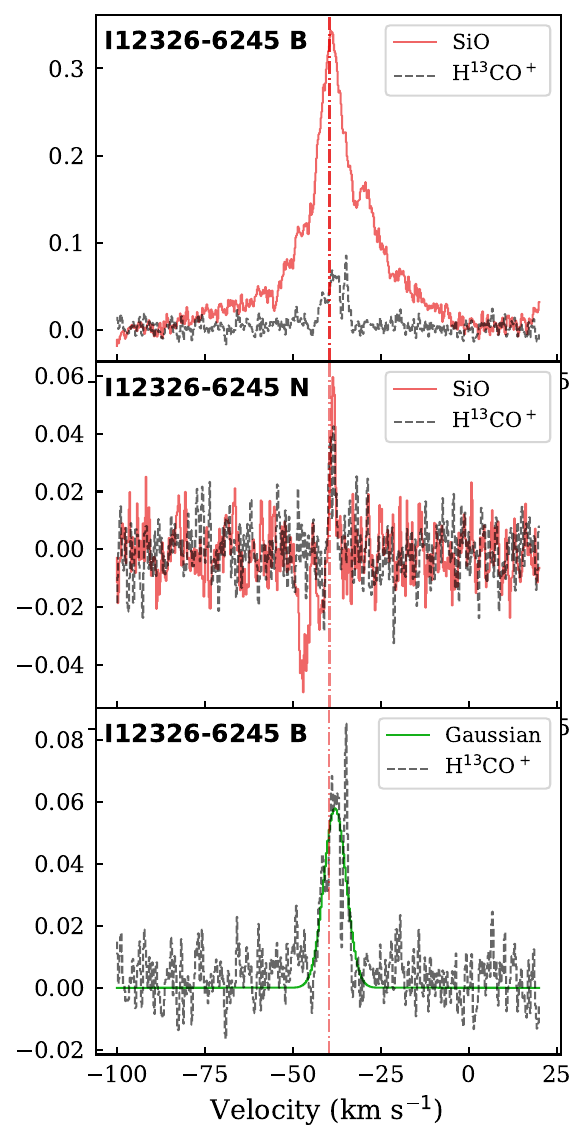}}
\end{minipage}
\caption{{\it Upper and middle panels:} Comparison between the line profiles of SiO and H$^{13}$CO$^+$ emission extracted from the positions associated with the broadest and the narrowest SiO line widths (top and middle panels). The labels ‘\textit{B}’ and ‘\textit{N}’ after the source name in the upper left part of each panel, correspond to the positions with the broadest and narrowest SiO emission marked by orange and blue rectangles in Figure~\ref{fig3}, respectively.
Red histograms represent the SiO spectra. The spectra of H$^{13}$CO$^+$ emission are shown in black histograms. The dashed-dotted vertical line indicates the systemic velocity of the source reported by \citet{liu2020atoms}. {\it Bottom panels:} Black histograms show the H$^{13}$CO$^+$ spectra extracted from position ‘\textit{B}’. Green curves report the Gaussian fits to the H$^{13}$CO$^+$ spectra. The spectra for all sources are available as supplementary material.}
\label{fig1}
\end{figure*}

Previous observations have reported that the line profiles of SiO emission frequently present two Gaussian velocity components: a bright broad component, and a weak narrow component, centered at velocities similar to the velocity of the ambient cloud \citep{martin1992sio,schilke1997sio,lefloch1998widespread,2004Jimnez,2005Jimnez,2006Jimnez,2008Jimnez,2009Jimnez,jimenez2010parsec,louvet2016tracing,2018Cosentino,2019Cosentino, Cosentino2020, 2022MNRASRong, 2023arXiv231013125T}. These two components are considered to have different origins. The broad components are attributed to the powerful outflows from protostellar objects \citep{qiu2007high,duarte2014sio}, and the narrow components are often related to the production of low-velocity shocks from either decelerated entrained material \citep{lefloch1998widespread}, young shocks through the interaction of the magnetic precursor \citep{2004Jimnez,2005Jimnez}, cloud-cloud collisions \citep{jimenez2010parsec,louvet2016tracing,2018Cosentino}, large-scale gas inflows, and expansion of UC H\textsc{ii} regions or supernova remnants \citep{2019Cosentino, Cosentino2020, Cosentino2022}. \citet{jimenez2009evolution} modeled the evolution of SiO line profiles along the evolution of magnetohydrodynamics (MHD) shock in young molecular outflows, showing that changes in the SiO line profiles are expected. At early times, the SiO emission is characterised by a narrow SiO component centered at ambient velocities (with line widths of 0.5 km s$^{-1}$, linked to the magnetic precursor) that evolves into a broad component line profile peaking at high velocities.

\citet{2022MNRASRong} conducted a study on SiO (2-1) emission using the observations from the Atacama Compact Array (ACA) 7 m Array to understand better the properties of the SiO shocked gas in different regions across the Galaxy and at different evolutionary stages. They found that a large amount of the observed SiO line profiles (60$\%$) within the sample presents two velocity components (one narrow and another one broad) in the large field of view (2$^{\prime}$) and the low angular resolution (13$^{\prime}$$^{\prime}$) of the ACA observations. In the present work, we now utilize the 12m and 7m combined data from the ATOMS survey to disentangle these different SiO components and investigate their origin. To do this, we image the SiO emission with a higher angular resolution (2.5$^{\prime}$$^{\prime}$) toward 146 massive SFRs using ALMA. 
In Sect.~\ref{sec2}, we provide details about the selected source sample and performed ALMA observations. The analysis method is presented in Sect.~\ref{sec3}. Sect.~\ref{sec4} describes the results of our analysis for the whole sample. The origin of the SiO emission is discussed in Sect.~\ref{sec5}. The summary is presented in Sect.~\ref{sec6}.

\section{Observations}
\label{sec2}
\subsection{The ATOMS Sample}
This paper investigates 146 massive SFRs from the ATOMS (‘ALMA Three-millimeter Observations of Massive Star-forming regions’) survey \citep{liu2020atoms, liu2020atomsII}. 
The sources were selected from UC H\textsc{ii} region candidates located in the Galactic plane and that contain bright CS $J$=2-1 emission ($T_\textup{b}$ > 2 K) \citep{bronfman1996cs}.
The physical parameters of 146 sources were taken from \citep{liu2020atoms, liu2020atomsII}.
Sources I08076-3556 and I11590-6452 are in reality low-mass SFRs, with mass values of 5.01 and 12.59 $M_{\odot}$, respectively. 
These two low-mass sources were excluded from the analysis. 
Source I08448-4343 is an intermediate-mass SFRs with an estimated mass of 39 $M_{\odot}$. The remaining sources (143 sources) are massive SFRs with estimated masses ranging from 126 to 2.5×10$^5$ $M_{\odot}$.
The distance of these sources covered in our sample ranges from 0.4 kpc to 13.0 kpc. The bolometric luminosity ranges from 16 to 8.1×10$^6$ $L_{\odot}$, while the dust temperature ranges from 18 to 46 K.
Outflow activity were found in 97 sources through HCO$^{+}$ and SiO emission (Baug et al. in prep). 76 sources were identified as UC H\textsc{ii} regions using H40$\alpha$ emission \citep{liu2021atomsIII,zhang2023atoms}. \citet{zhou2022atoms} searched for filamentary structures, as plotted in below, towards all sources making use of H$^{13}$CO$^{+}$ $J$=1-0 line data.
\begin{figure}
\begin{center}
\begin{minipage}[t]{0.45\linewidth}
\vspace{3pt}
\centerline{\includegraphics[width=2\linewidth]{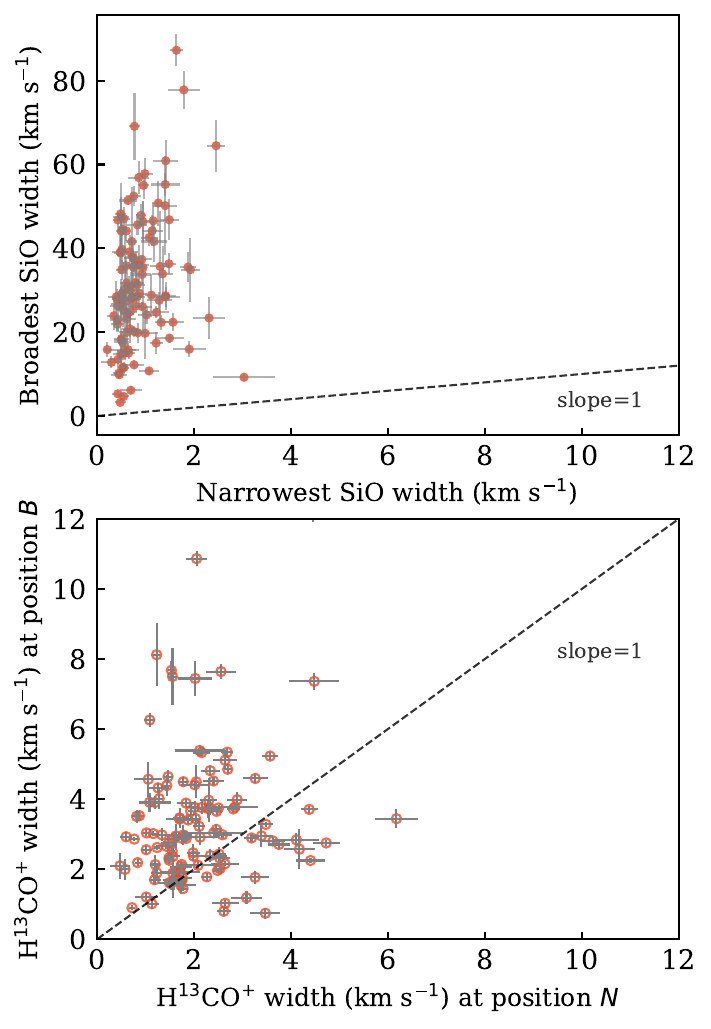}}
 \end{minipage}
\caption{{\it Top panel}: Comparison between the broadest SiO line width compared to the narrowest SiO line width measured for each source (see red filled circles). {\it Bottom panel}: Comparison between the H$^{13}$CO$^+$ line width extracted from position ‘\textit{B}’ versus the H$^{13}$CO$^+$ line width extracted from position ‘\textit{N}’. The red empty circles show each source. In both panels, the dashed line represents a 1:1 slope. The error bars in both plots represent the uncertainties associated with the measured line widths.}
\label{fig2}
\end{center}
\end{figure}

\begin{figure*}
\begin{minipage}[t]{0.48\linewidth}
    \vspace{3pt}
\centerline{\includegraphics[width=1.1\linewidth]{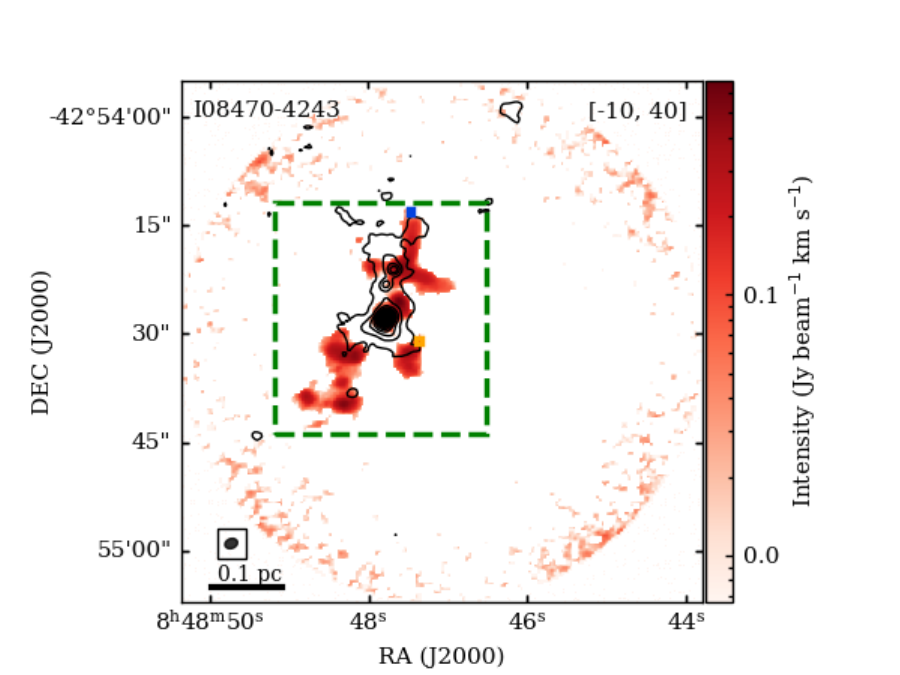}}
 \end{minipage}
\begin{minipage}[t]{0.48\linewidth}
    \vspace{3pt}   \centerline{\includegraphics[width=1.1\linewidth]{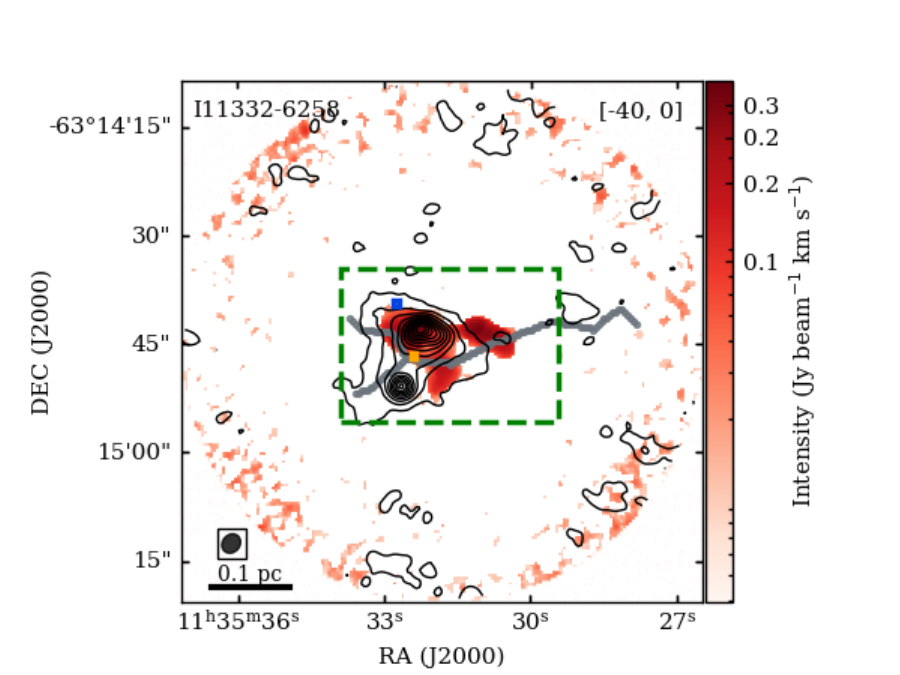}}
 \end{minipage}
\begin{minipage}[t]{0.48\linewidth}
    \vspace{3pt}    \centerline{\includegraphics[width=1.1\linewidth]{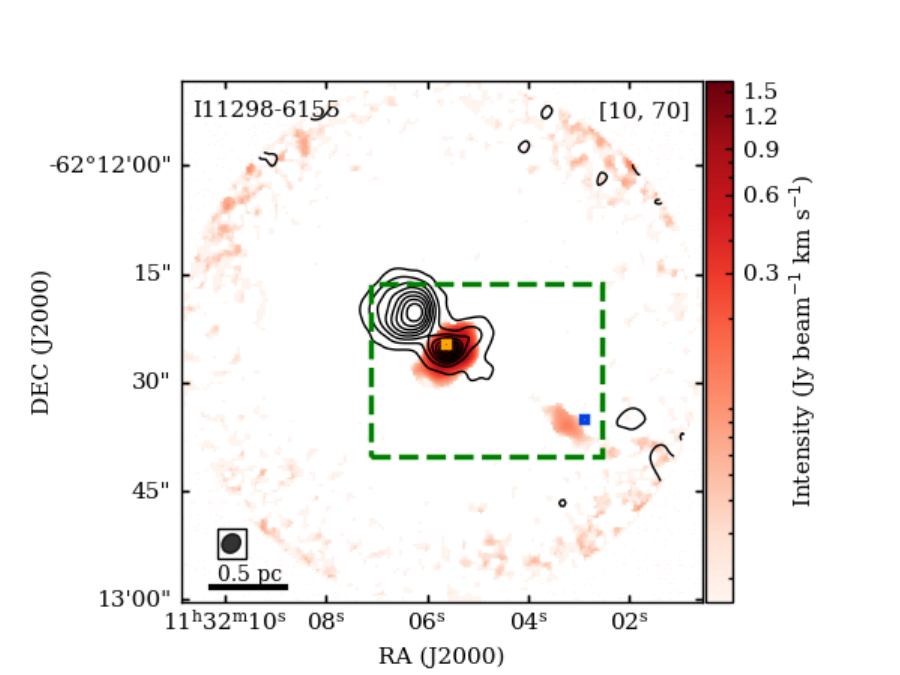}}
 \end{minipage}
\begin{minipage}[t]{0.48\linewidth}
    \vspace{3pt} \centerline{\includegraphics[width=1.1\linewidth]{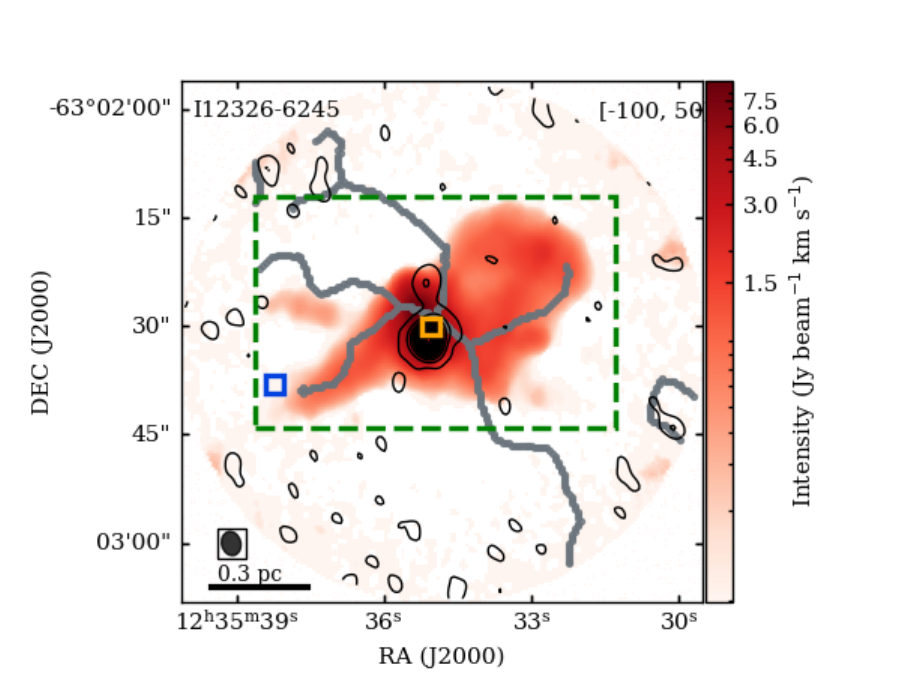}}
 \end{minipage} 
\caption{Four representative sources imaged with ALMA within the ATOMS program. The background corresponds to the SiO (2-1) integrated intensity maps. The black contours show the 3 mm continuum emission detected with ALMA, and contours are from 5$\sigma$ to the peak values in steps of 10$\sigma$. 
The bold gray lines represent the filament skeletons identified using H$^{13}$CO$^+$ as reported by \citet{zhou2022atoms}. The green dashed rectangle is the area of the SiO emission maps where the SiO line profiles have been decomposed into different velocity components. The orange and blue rectangles, with the broadest and narrowest SiO line widths, respectively, indicate the locations where SiO and H$^{13}$CO$^+$ have been extracted.
The field of view (FOV) is 72$^{\prime}$$^{\prime}$ corresponding with the FOV of the ALMA observations. All images have been primary-beam corrected. The source name and integrated velocity ranges (in km s$^{-1}$) are shown on the upper left and right corners, respectively. The beam size is reported in the lower left corner. The same images are provided for all sources within the supplementary material.}
\label{fig3}
\end{figure*}

\subsection{ALMA observations}
We use ALMA Band 3 data from the ATOMS survey (Project ID: 2019.1.00685.S; PI: Tie Liu). This survey has obtained both 12m-array and 7m-array data, and the details of the observations can be found in \citep{liu2020atoms, liu2020atomsII}. The spectral setup consisted of six narrow spectral windows (SPWs 1 to 6) at the lower sideband and two wide spectral windows (7 to 8) at the upper sideband. The narrow SPWs are in the range of 86.31–89.2 GHz, with a spectral resolution of ${\sim}$0.2 km s$^{-1}$, and the wide SPWs 7 and 8 are in the range of 97.52${\sim}$ 99.39 GHz and 99.46${\sim}$101.33 GHz, respectively, and employ a spectral resolution of ${\sim}$1.6 km s$^{-1}$. The wide SPWs 7 and 8 are used for continuum measurements. All data reductions were carried out using the CASA software package version 5.6 \citep{mcmullin2007casa}. The 7-m array data and 12-m array data are calibrated separately and then combined and cleaned. All images are primary-beam corrected. In this paper, the SiO \textit{J}=2-1 and H$^{13}$CO$^+$ $J$=1-0 lines are included in SPWs 1${\sim}$6 with spectral resolutions of  0.21 and 0.211 km s$^{-1}$, respectively. The maximum recoverable scale (MRS) is about 1$^{\prime}$. The typical beam size and channel rms noise level of these two lines are 2.5$^{\prime}$$^{\prime}$ and 8 mJy beam$^{-1}$, respectively \citep{liu2020atoms, liu2020atomsII}. The pixel angular size in each image is about 0.4$^{\prime}$$^{\prime}$. This implies that there are about 40 pixels per synthesized beam. Since our ALMA images do not include total power (TP) data, they may suffer from missing flux. However, note that as discussed by \citet{liu2020atoms}, the 12m array observations recover more than 92\% of the flux measured in the ACA images as inferred by averaging the H$^{13}$CO$^{+}$ emission over a 30 arcsecond-region in one example source.

\begin{figure*}
\begin{minipage}{0.7\linewidth}
    \vspace{3pt}
\centerline{\includegraphics[width=1.5\linewidth]{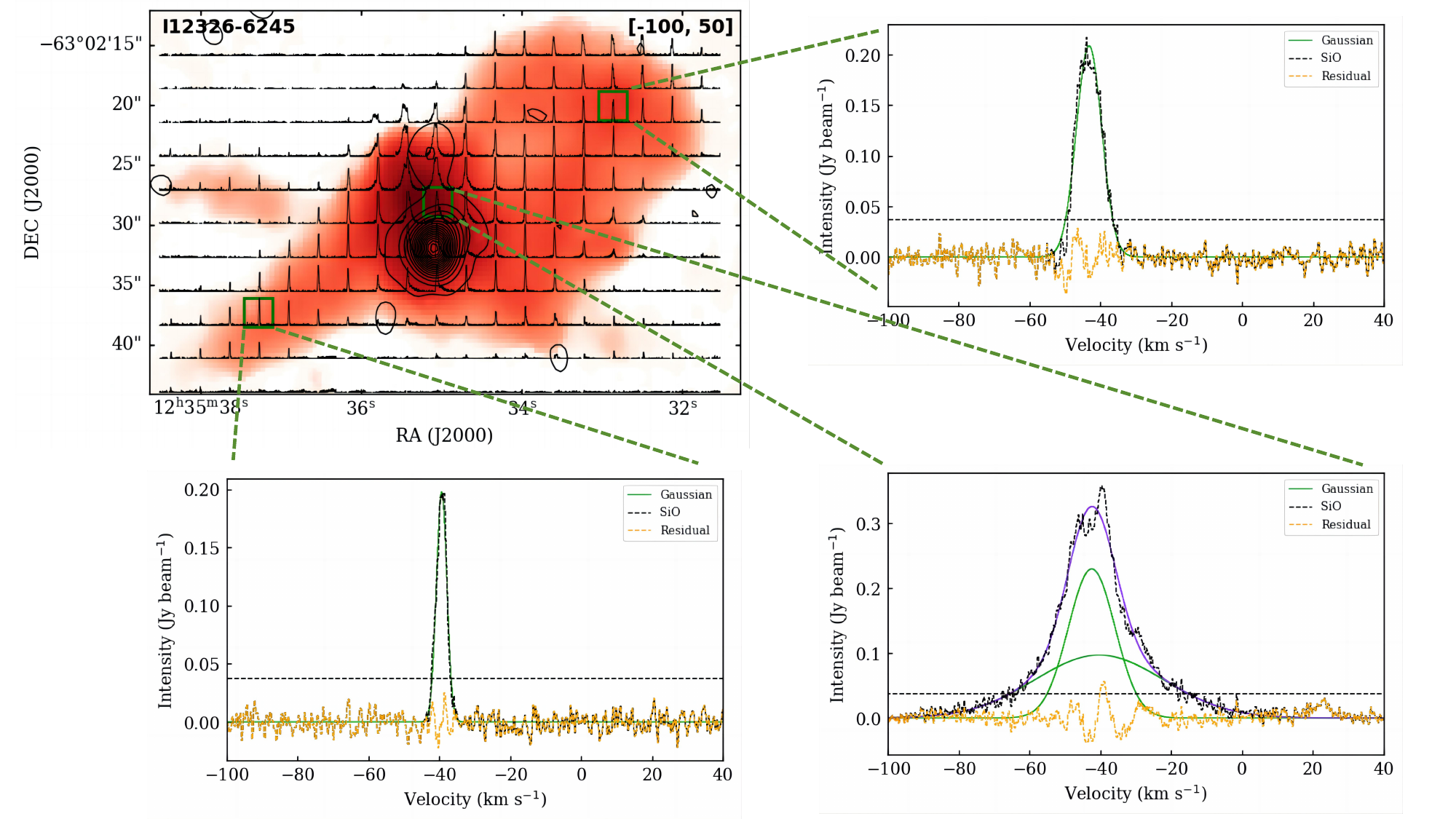}}
 \end{minipage}
\caption{Output of \texttt{scousepy} for source I12326-6245. In the upper left panel, we show the SiO (2-1) integrated intensity map where the SiO spectral grid map is overlaid. The grid size corresponds to the green rectangle size ($6\times6$ pixels$^2$) used in the fitted average spectra in \texttt{scousepy}.
The black contours report the 3 mm continuum emission, and contours are from 5$\sigma$ to the peak values in steps of 10$\sigma$. The displayed field of view corresponds to the green dotted rectangle in Figure~\ref{fig3}. Green dotted lines indicate the regions for which example spectra are also shown (see right upper panel and lower panels). In the latter, the black dashed and green solid lines represent the average SiO spectra and the Gaussian fit lines of the different components, respectively. The purple line is the total Gaussian fit obtained for these lines. The yellow solid line indicates the residuals and the black horizontal dashed lines show the 3$\sigma$ rms noise levels.}
\label{fig4}
\end{figure*}

\section{Methods}
\label{sec3}
The goal of this study is to investigate the different origins of the SiO emission toward the 146 massive SFRs in our sample. The basic parameters 
(distances from the sun, Distance; systemic velocities, $v_{\rm LSR}$; bolometric luminosity,  $L_\textup{bol}$; clump masses, $M$) of these sources are provided in Appendix~\ref{Appendix A}, taken from \citet{liu2020atoms}.
Firstly, we inspect the SiO emission in CASA to determine the velocity ranges in the observed spectra for each source. Then, we generate the moment 2 maps of the SiO emission and identify the locations where SiO presents the broadest and the narrowest line widths. Figure~\ref{fig1} shows the SiO and H$^{13}$CO$^+$ spectra extracted from these positions for four representative sources. The spectra were extracted over the aperture within a specific area, as described below.
Figure~\ref{fig2} presents the variations in the line widths of SiO and H$^{13}$CO$^+$ at these positions. The corresponding positions are marked by orange and blue rectangles in Figure~\ref{fig3} for the broadest and narrowest SiO line width, respectively. 

Due to the complex line profiles observed for the SiO emission, the broad and narrow velocity components were separated by carrying out a Gaussian spectral decomposition with the \texttt{Python} analysis tool \texttt{scousepy} \citep{henshaw2016molecular}. \texttt{scousepy} is a semi-automated multi-component spectral line decomposition tool that can fit the SiO spectra in every detected source. Initially, the fitted areas were defined in the generated moment 0 maps, constraining the SiO emission within the fitted areas above 3$\sigma$ rms noise. The fitted areas correspond with the green rectangle in Figure~\ref{fig3}. We calculated $\sigma$ rms noise values using the standard deviation from ten free channels without SiO emission and have listed these values in Table~\ref{tab:TableA1}. After this, we semi-automatically fitted the average spectra extracted from sub-areas of size $2\times2$, $4\times4$, $6\times6$, or $8\times8$ pixels$^2$. The sub-areas are used for the spectra extraction through the observed SiO emission. The output of these Gaussian fits was then used as input information for the Gaussian fitting of the line profiles extracted pixel-by-pixel \citep[see the details of the method in][]{henshaw2016molecular}. We have added the size of sub-areas and the number of sub-areas where the broad and narrow components were detected in Table~\ref{tab:TableA1}.

The criteria for a successful fitting of two Gaussian components included the narrow component having a peak flux above 3$\sigma$ and the velocity offset between these two components smaller than the line width of the broad component. If the observed spectrum did not satisfy this condition, we retained a single Gaussian fit and ensured its peak flux laid above 3$\sigma$. We obtain the Gaussian fitting parameters (peak flux, central velocity, and velocity dispersion) provided by \texttt{scousepy}. In Figure~\ref{fig4}, we show an example of the fitting procedure with \texttt{scousepy} used to decompose the broad and narrow components of the observed SiO emission.
To better disentangle the broad and narrow SiO components, we compare the measured SiO line widths with the derived H$^{13}$CO$^+$ line width, following a similar method to that described in \cite{2018Cosentino, Cosentino2020}. Its high critical density and low optical depth make H$^{13}$CO$^+$ a good tracer of dense gas. In addition, the abundance of H$^{13}$CO$^+$ is not expected to vary substantially with time. 
Indeed, by using a large sample of IRDC clumps at different evolutionary stages, \citet{2012ApJ...756...60S} found that the HCO$^{+}$ abundance varies by less than a factor of 3. In addition, chemical fractionation models show that the HCO$^+$/H$^{13}$CO$^+$ abundance ratio varies by less than a factor of 2 for time-scales between 10$^{5}$ and 10$^6$ yrs \citep{roueff2015}. The H$^{13}$CO$^+$ spectra were extracted from the positions with the broadest SiO emission detected toward each source, as representative of the emission arising from the ambient gas. In our analysis, we consider that SiO emission with line widths larger than that of the H$^{13}$CO$^+$ emission can be attributed to stronger shock activity. 
To better analyze the kinematic structure and the origin of the shocked gas, we show the spatial distribution of the broad and narrow components in the next section.

\begin{figure*}
\begin{minipage}[t]{0.7\linewidth}
    \vspace{3pt}
\centerline{\includegraphics[width=1.4\linewidth]{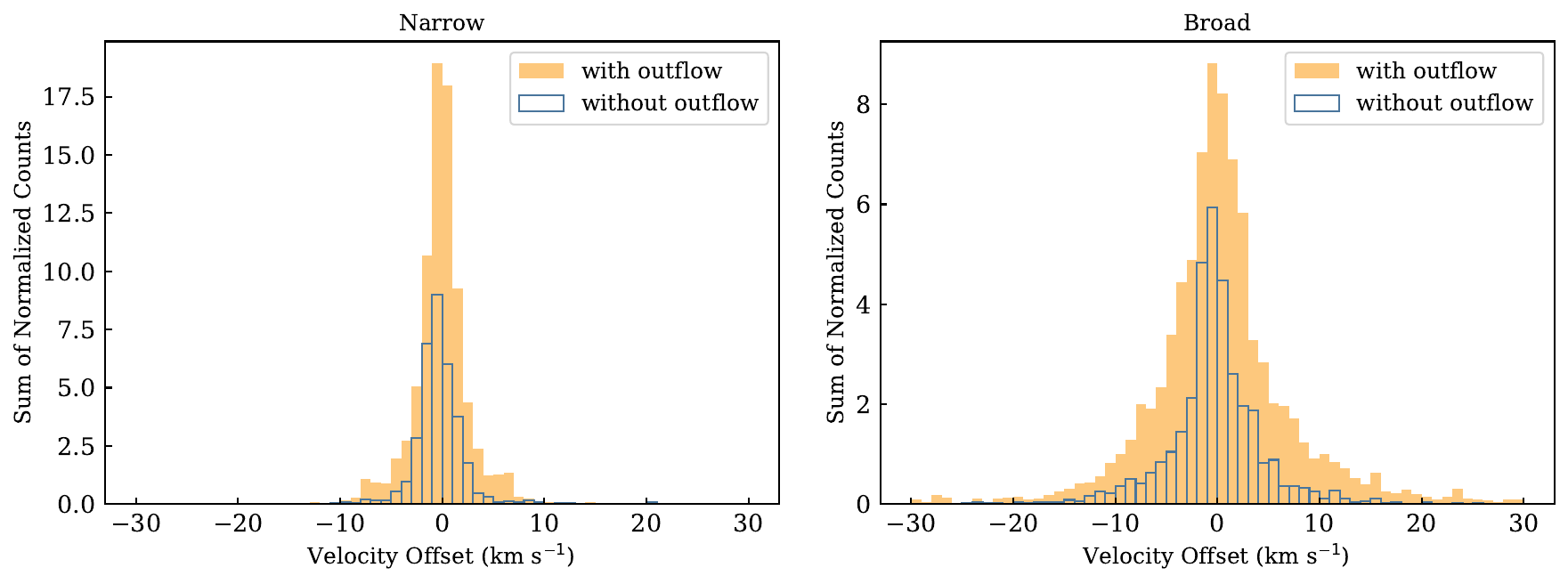}}
 \end{minipage}
\begin{minipage}[t]{0.7\linewidth}
    \vspace{3pt}
\centerline{\includegraphics[width=1.4\linewidth]{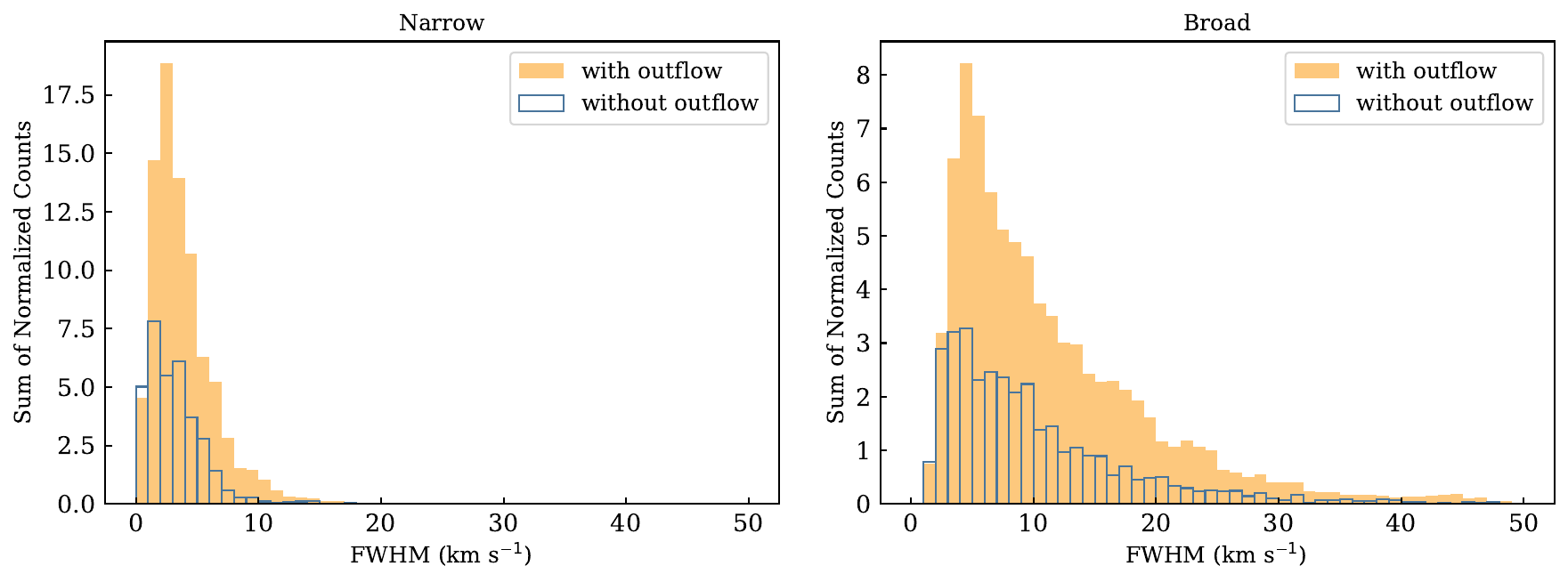}}
 \end{minipage}
\caption{The histogram shows the velocity offset compared to the systemic velocity and line width distribution of the broad and narrow components of the SiO emission for the entire sample. The open blue histograms correspond to the sources without outflow activity, and the orange-filled histograms represent the sources exhibiting outflow activity. The outflow sources can be found in Table~\ref{tab:TableA1}.}
\label{fig5}
\end{figure*}

\section{Results}
\label{sec4}
\subsection{Line profiles of SiO and \texorpdfstring{H$^{13}$CO$^+$} emission in active and quiescent regions}
\label{sec4.1}
The positions associated with the broadest and the narrowest SiO line widths were identified using the results from \texttt{scousepy} and SiO moment 2 maps. In the top and middle panels of Figure~\ref{fig1}, we present the line profiles of SiO and H$^{13}$CO$^+$ emission extracted from these positions toward four selected sources. 
The upper panels show the spectra extracted from the positions with the broadest SiO line widths, thought to mark the location with the strongest shock activity in each source. The middle panels present the spectra extracted from the position with the narrowest SiO line widths. These regions are likely undergoing a more gentle shock interaction. These regions are respectively labeled ‘\textit{B}’ and ‘\textit{N}’ after their source name in the upper left part of the top and middle panels of Figure~\ref{fig1}.
Toward positions ‘\textit{B}’, the SiO emission exhibits broader and more complex spectra than toward the ‘\textit{N}’ positions, while the H$^{13}$CO$^+$ emission shows small variations across both positions. 

In the bottom panels of Figure~\ref{fig1}, we show the H$^{13}$CO$^+$ spectra (black curves) with their Gaussian fits superimposed using green lines extracted from position ‘\textit{B}’. The Gaussian fits have been carried out with the \texttt{MADCUBA} analysis tool \citep{2019A&A...631A.159M}. Table~\ref{tab:TableA1} lists the H$^{13}$CO$^+$ line widths measured toward each source. The H$^{13}$CO$^+$ line widths varies from 0.7 to 12.2 km s$^{-1}$, with a mean value of 3.3 km s$^{-1}$.

In Figure~\ref{fig2}, we compare the range in the measured line widths of SiO and H$^{13}$CO$^+$ at positions ‘\textit{B}’ and ‘\textit{N}’ for each source. From this Figure, it is clear that the narrowest SiO component measured at position ‘\textit{N}’ is clearly narrower than the broadest SiO emission observed at position ‘\textit{B}’ toward each source. The difference in linewidths is apparent since narrow SiO shows linewidths narrower than 2 km s$^{-1}$ for the majority of sources, while the broadest SiO emission lies typically above 10 km s$^{-1}$. For SiO, the line width at position ‘\textit{B}’ ranges from 3.2 to 87.3 km s$^{-1}$, while at position ‘\textit{N}’, it varies from 0.2 to 3 km s$^{-1}$. 
For the H$^{13}$CO$^+$ emission, the line width displays only slight variations between the two positions, and its measured line widths are typically below 5 km s$^{-1}$ for most sources. The maximum line width ratio across both positions is approximately six.

To analyze the SiO line profiles, we employed the \texttt{scousepy} analysis tool. This tool performs the Gaussian fitting of the SiO line profiles pixel by pixel within the ALMA maps toward the observed sources. Multiple velocity components are fitted using multiple Gaussian profiles. After performing the Gaussian fitting, we compared the H$^{13}$CO$^+$ line width at position ‘\textit{B}’ with SiO line widths measured pixel by pixel to differentiate between the broad and narrow components of SiO for each spectrum. As mentioned above, if the linewidth of the SiO velocity component measured by \texttt{scousepy} is broader than the one of H$^{13}$CO$^+$, then that SiO component is classified as broad, while if the measured SiO linewidth is narrower than (or similar to) the line width of H$^{13}$CO$^+$, we consider that the SiO emission is narrow. We note that we use the H$^{13}$CO$^+$ linewidth measured toward position ‘\textit{B}’ because it represents an upper limit to the distribution of H$^{13}$CO$^+$ linewidths across each source, as this is the location with the strongest shock interaction found in each region. In any case, as shown in Figure$\,$\ref{fig2}, the H$^{13}$CO$^+$ linewidth upper limits typically cluster between 0.5 and 5-6 km s$^{-1}$, which indicates that the H$^{13}$CO$^+$ linewidths will be restricted to a narrow linewidth range.

\begin{figure*}
\begin{minipage}[t]{0.58\linewidth}
    \vspace{0.3pt}
\centerline{\includegraphics[width=1.35\linewidth]{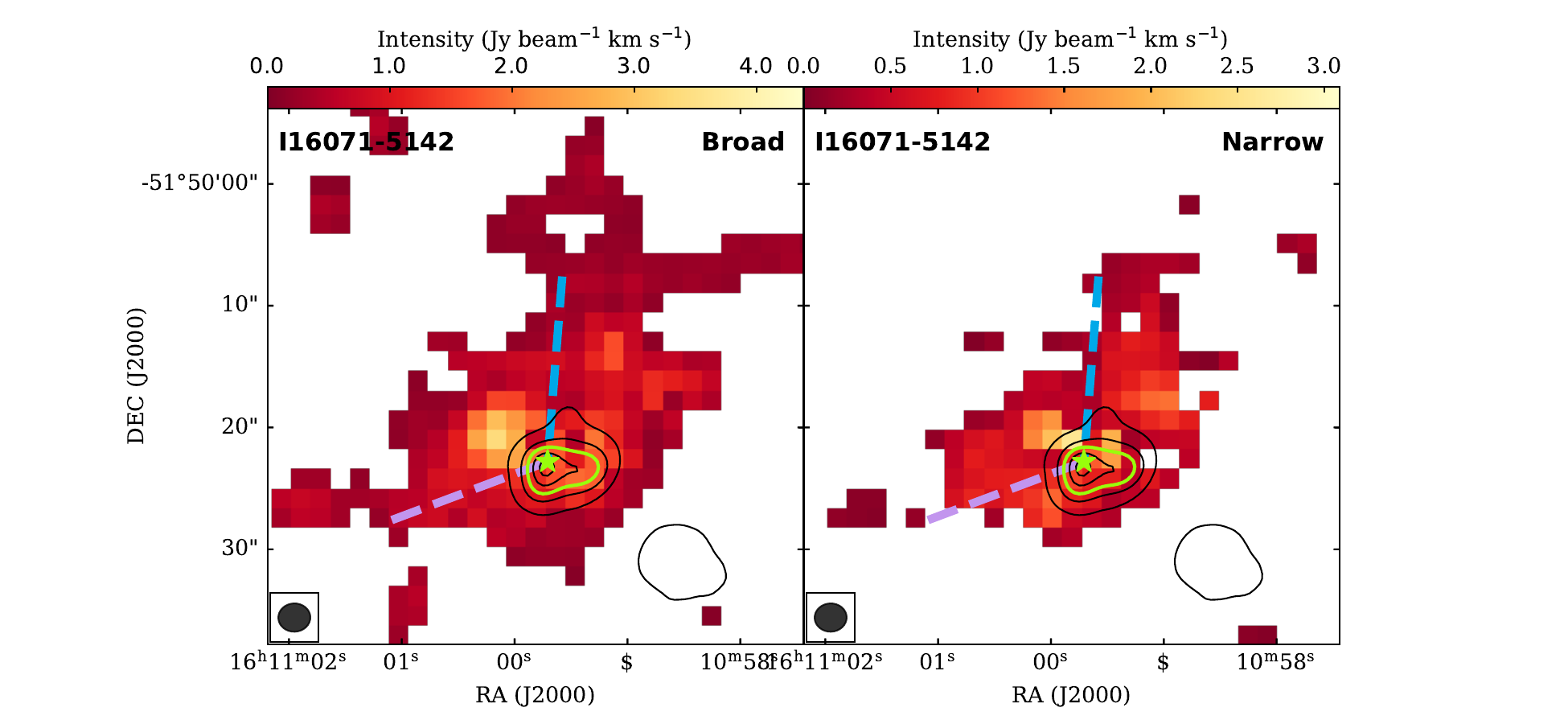}}
 \end{minipage}
\begin{minipage}[t]{0.58\linewidth}
\vspace{0.3pt}   \centerline{\includegraphics[width=1.35\linewidth]{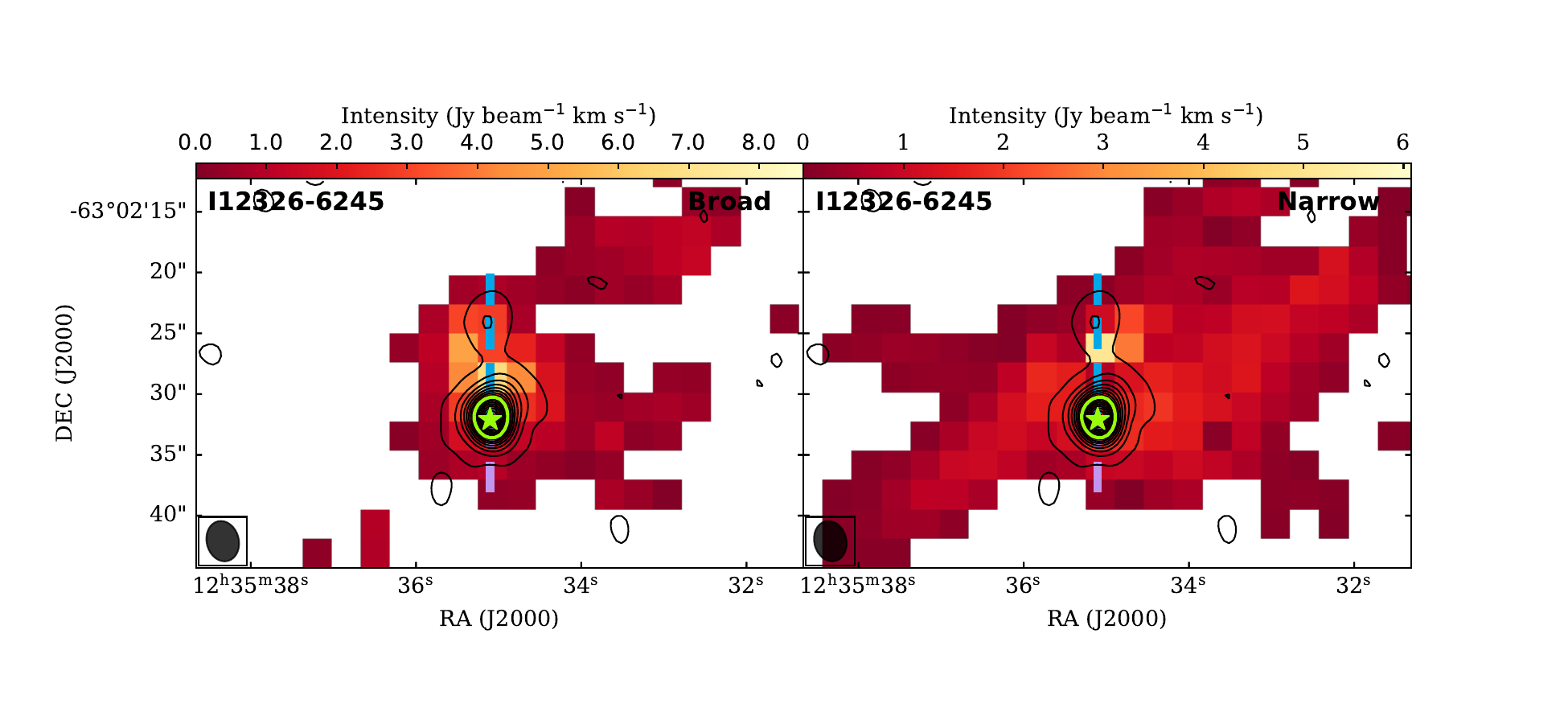}}
 \end{minipage} 
\caption{The two cases of decomposed sources are based on the different morphology of narrow and broad SiO components. The upper panels display the SiO integrated intensity maps of broad (left) and narrow (right) components for a source from \textit{Group A} and the lower panels present the SiO integrated intensity maps of broad (left) and narrow (right) components for a source from \textit{Group B}. The black contours are 3 mm continuum emission, and contours are from 5$\sigma$ to the peak values with a step of 10$\sigma$. The blue and purple dashed lines represent the blue and red lobes of identified outflows from Baug et al. (in prep), and the green star is the position of the protostar centrally positioned between the two lobes. The green contour is the half-peak value of the 3 mm continuum emission. The source name is shown on the upper left of the panel and the beam size is reported in the lower left corner. The shown field of view is the area of the green dashed rectangle presented in Figure~\ref{fig3}.}
\label{fig6}
\end{figure*}

\subsection{The general spatial distribution of SiO emission}
\label{general}

SiO (2-1) emission was detected towards 136 out of the 146 sources in our sample, whereas 10 sources had no SiO emission measured above the 3 $\sigma$ level. 
We classified the 136 detected sources into four different categories based on them hosting UC H\textsc{ii} regions and on the presence of SiO emission associated with filament skeletons seen in H$^{13}$CO$^+$ \citep{zhou2022atoms}: (1) Six sources without UC H\textsc{ii} regions and not associated with filament skeletons; (2) 60 sources without UC H\textsc{ii} regions but associated with filament skeletons; (3) Three sources hosting UC H\textsc{ii} regions but not associated with filament skeletons; (4) 66 sources hosting UC H\textsc{ii} regions and associated with filament skeletons. 

Figure~\ref{fig3} shows the morphology of the SiO emission as seen from the moment 0 maps toward four sources representative of the above categories. 
Source I08470-4243 is representative of category  (1), with no H40$\alpha$ emission being detected and no filament skeletons identified using H$^{13}$CO$^+$ emission. The SiO emission is associated with 3mm continuum emission (see green dashed rectangle). 
Category  (2) is exemplified by source I11332-6258, where the SiO emission is clearly associated with filament skeletons and 3mm continuum emission but no H40$\alpha$ emission is detected. We consider the lack of H40$\alpha$ emission in these two categories as a strong indicator that these sources are not associated with UC H\textsc{ii} regions. Therefore, in these sources the 3mm continuum emission measured with ALMA is dominated by thermal dust emission.
Sources I11298-6155 and I12326-6245 are illustrative of categories  (3) and (4), respectively. Both sources exhibit H40$\alpha$ emission, implying the presence of UC H\textsc{ii} regions. Toward source I11298-6155, the detected SiO emission coincides spatially with the 3mm continuum emission and it is very compact. In addition, it is not associated with any filament skeletons. 
Toward source I12326-6245, the SiO emission is associated with filament skeletons and 3mm continuum emission. SiO also displays a more extended distribution than in catergory (3). Similar images are available for all sources in the supplementary material.

\begin{figure*}
\begin{minipage}{0.5\linewidth}
    \vspace{3pt}
\centerline{\includegraphics[width=1.5\linewidth]{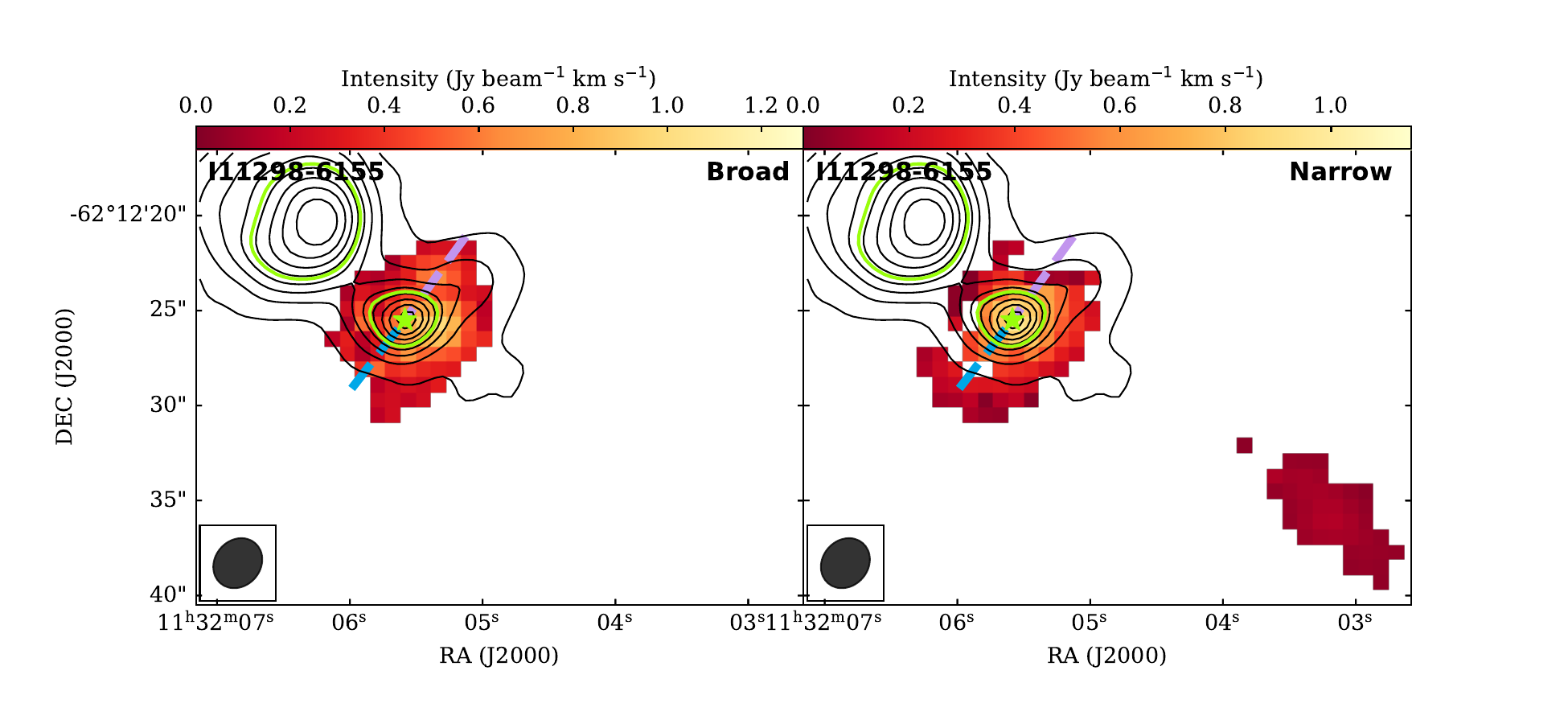}}
 \end{minipage}
 \begin{minipage}{0.5\linewidth}
    \vspace{3pt}
\centerline{\includegraphics[width=1.5\linewidth]{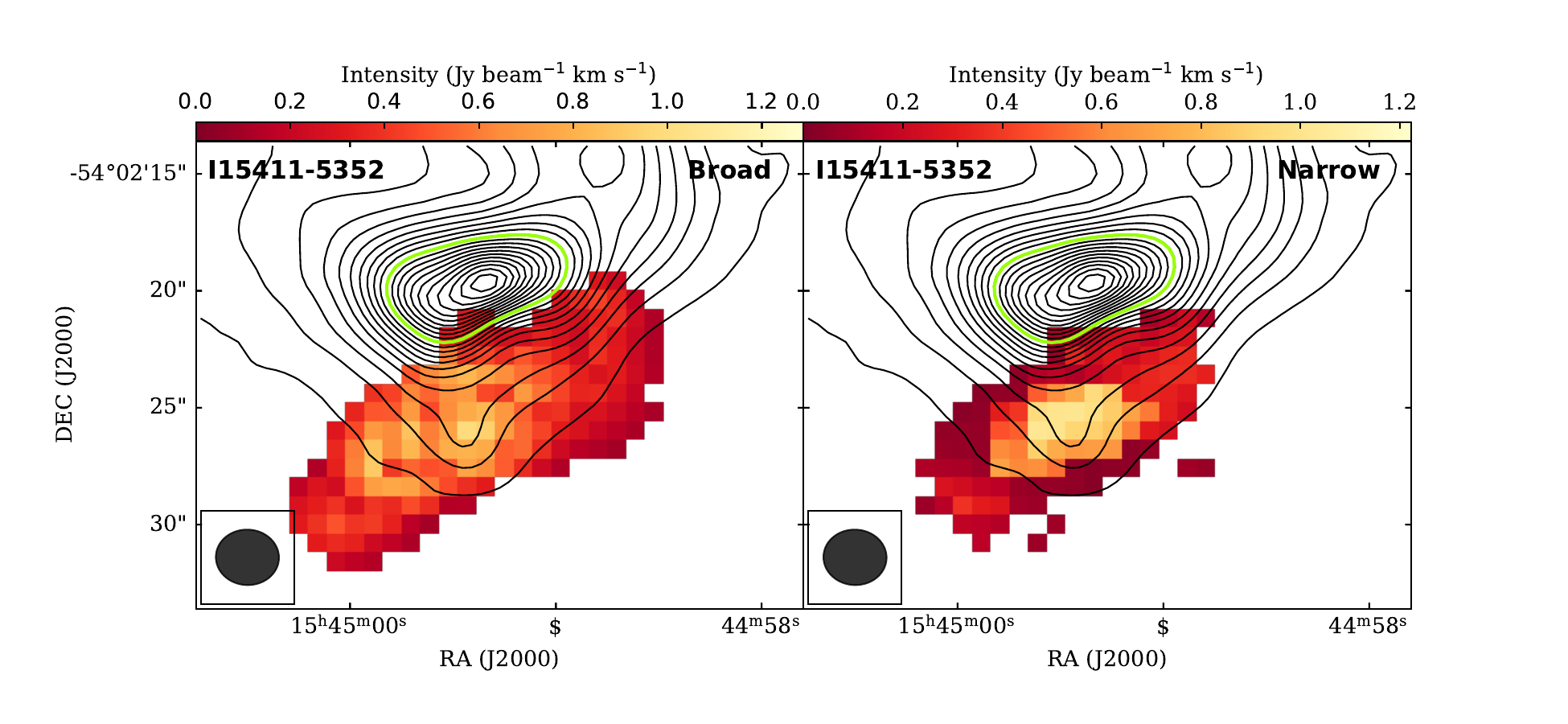}}
 \end{minipage}
 \begin{minipage}{0.5\linewidth}
    \vspace{3pt}
\centerline{\includegraphics[width=1.5\linewidth]{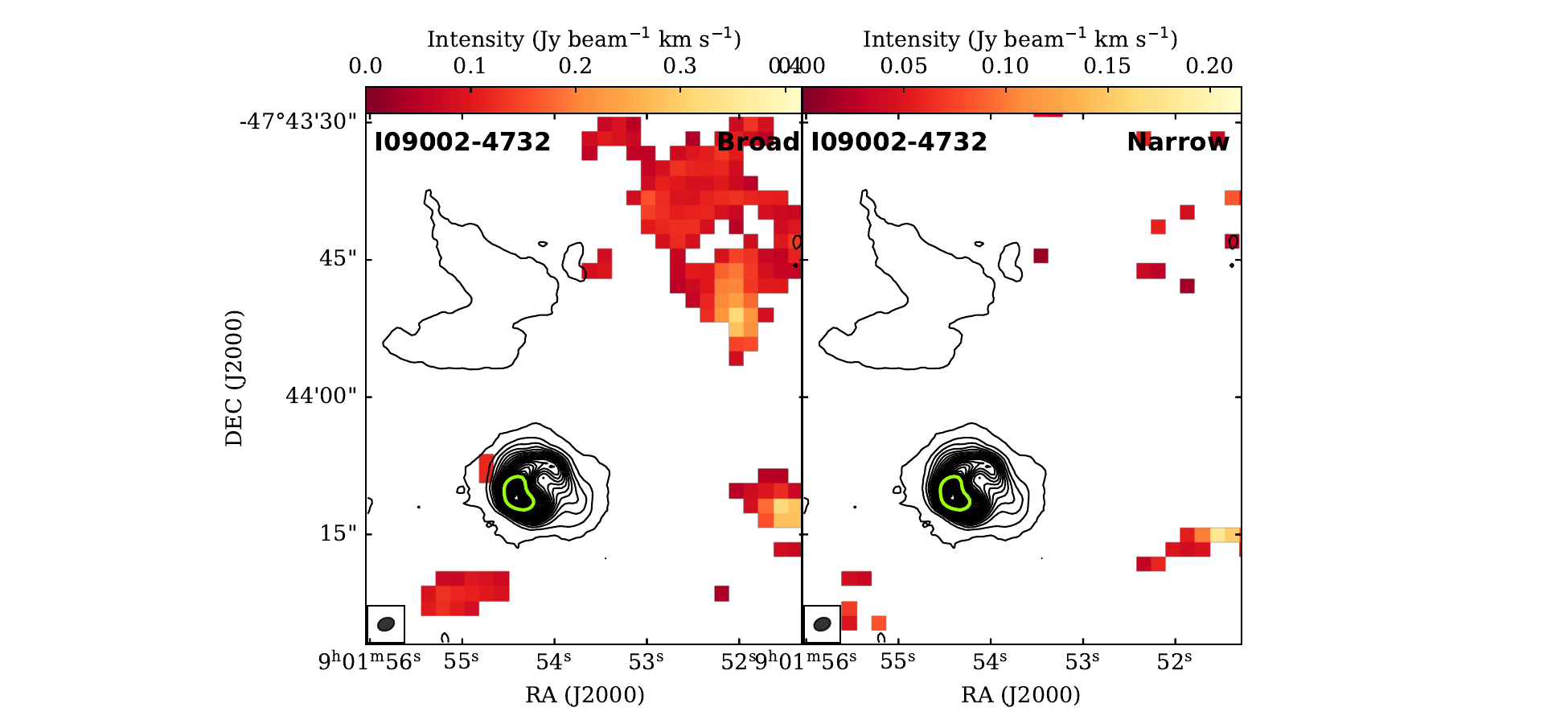}}
 \end{minipage}
\caption{The three classes of decomposed sources that host UC H\textsc{ii} regions based on their spatial distribution of the SiO and 3mm continuum emission: i) a {\it 'coincident'} source (upper panels); ii) an {\it 'offset'} source (middle panels); and iii) an {\it 'surrounded'} source (lower panels). The background image displays the SiO integrated intensity maps of the broad and narrow SiO components. The blue and purple dashed thick lines represent the blue and red lobes of identified outflows from Baug et al. (in prep.), and the green star indicates the position of the protostar centrally positioned between the two lobes (see also Figure \ref{fig6}).
The black contours show the 3 mm continuum emission detected with ALMA, and contours are from 5$\sigma$ to the peak values in steps of 10$\sigma$. The green contour is the half-peak value of the 3 mm continuum emission. The source name is shown on the upper left of the panel and the beam size is reported in the lower left corner.}
\label{fig7}
\end{figure*}

\subsection{Broad and narrow components of SiO emission}

Using the fitting results provided by \texttt{scousepy} and considering the H$^{13}$CO$^+$ line widths, we generated the histograms of the velocity and line width distributions for the narrow and broad SiO components to illustrate the velocity structure across the sample (see Section \ref{histograms}). We also produced the integrated intensity maps for both the narrow and broad components of SiO to investigate their spatial distribution and their association with the measured 3mm continuum emission (see Sections \ref{sec4.3.2} and \ref{sec4.3.3}). In this analysis, we have only used 118 out of the 136 sources with SiO emission. The exclusion of 18 sources from this analysis is due to absorption features in SiO line profiles caused by the presence of strong continuum sources \citep{2022ApJ...927...54O,2024MNRAS.528.7383C}. We show the SiO spectra of the excluded sources in Appendix~\ref{Appendix B}, and a list of these sources is available in Table~\ref{tab:TableA1}.

\subsubsection{General statistics across the whole sample}
\label{histograms}
We investigate the velocity and line width distribution of the broad and narrow components across the 118 sources for which we have performed the analysis with \texttt{scousepy}. Figure~\ref{fig5} shows the distribution of sources as a function of velocity offset with respect to the systemic velocity of the cloud \citep{liu2020atoms,2018MNRAS.473.1059U,2004A&A...426...97F} and as a function of line width. Because the count of fitted spectra is different from source to source, to avoid any statistical bias produced by sources with many fitted spectra we normalize the counts of fitted spectra per velocity bin by the total number of counts obtained for each source. The y-axes in Figure~\ref{fig5} represent the sum of normalized counts in each source falling into the same velocity bin.
The unfilled blue histograms correspond to the sources without known outflow activity, and the orange-filled histograms represent the sources exhibiting known outflow activity as inferred by Baug et al. (in prep).
To identify the outflows, Baug et al. (in prep) have employed the ALMA SiO and HCO$^{+}$ maps whose emission is brighter than the 5 $\sigma$ rms noise threshold. In our images, the extension of the lobes was marked by arrows. The location of the triggering star was identified by the peak of 3 mm continuum emission and the centrally positioned between the blue and red lobes. The sources with known outflow activity are indicated in Table~\ref{tab:TableA1}.

We used the Kolmogorov-Smirnov (KS) test to analyze the velocity offset and line width across sources with and without outflows considering both broad and narrow components together. The P-values generated by the KS test indicate the probability that the two samples originated from the same distribution. If the P-value is less than 0.05, this suggests that the samples likely come from different distributions. 
In our analysis, the velocity offset comparison between sources with and without outflows returned a P-value of 4.05$\times{10^{-31}}$ from the KS test. For the line width across these two categories, the P-value was 1.62$\times{10^{-10}}$. These results confirm that the sources with and without outflows come from distinct distributions.
For the narrow component (see Figure~\ref{fig5}, left panels), 
most of the emission with or without outflow present velocity offsets that lie within $\pm$10 km s$^{-1}$, indicating that this emission arises from molecular gas close to ambient cloud velocities. 
In contrast, the velocity offset of the broad SiO components is distributed across a wider range of velocities compared to the narrow SiO emission, even reaching values above 60 km s$^{-1}$. 
For the line width distribution of narrow SiO emission (left lower panel, Figure~\ref{fig5}), we find that the vast majority of fitted spectra have line widths $\leq$5-6 km $^{-1}$ suggesting that the gas may have been affected either by low-velocity shocks or by young shocks at the magnetic precursor stage \citep{2004Jimnez,2005Jimnez,duarte2014sio}. For the broad SiO emission, however, its line widths can be as large as 60 km s$^{-1}$, with a maximum of 87 km s$^{-1}$. Note that a few extreme values are omitted in Figure~\ref{fig5} to better display the details of the distributions. For a full view of the line widths, refer to Figure~\ref{fig2}.

\begin{figure*}
\begin{minipage}[t]{0.7\linewidth}
    \vspace{3pt}
\centerline{\includegraphics[width=1.6\linewidth]{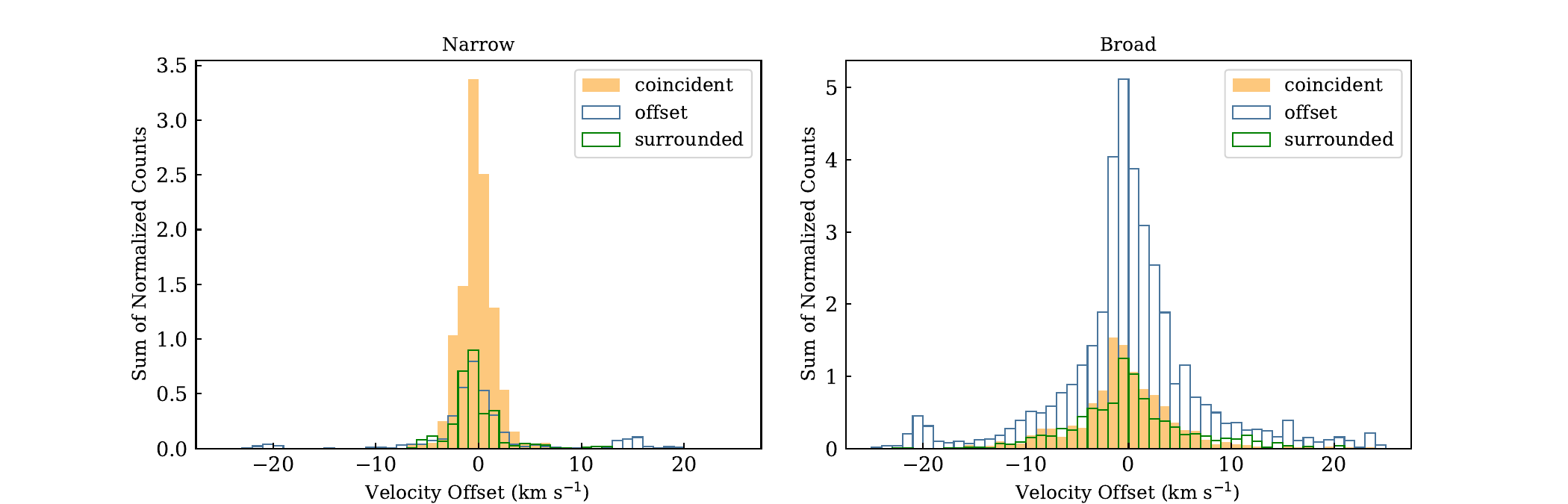}}
 \end{minipage}
\begin{minipage}[t]{0.7\linewidth}
    \vspace{3pt}
\centerline{\includegraphics[width=1.6\linewidth]{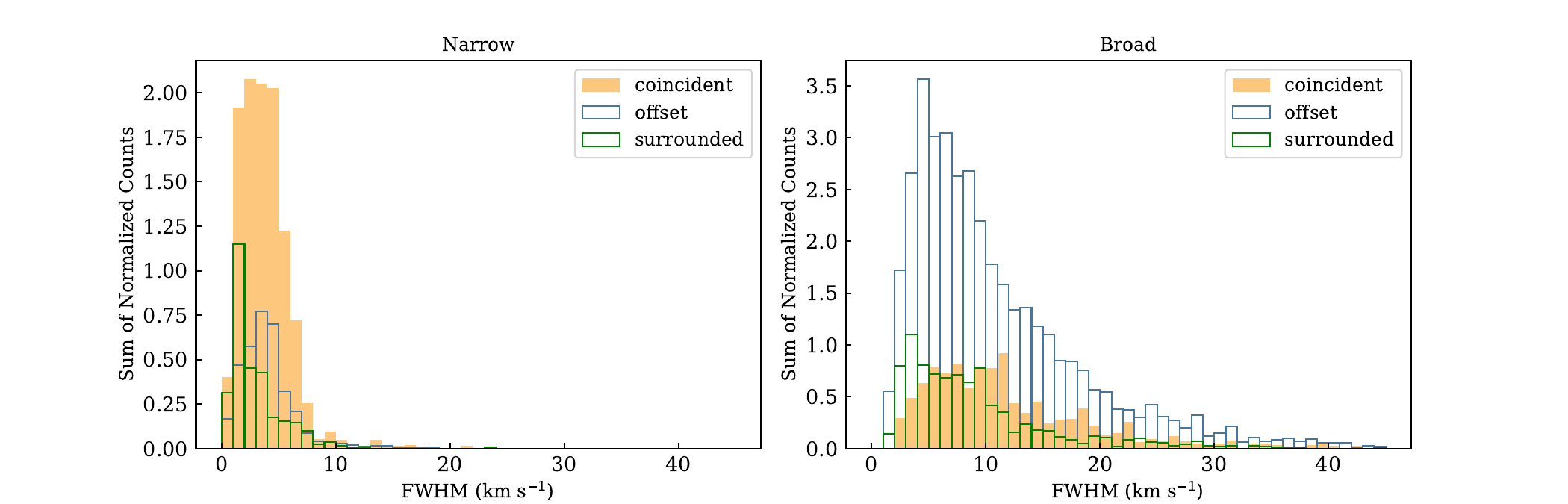}}
\end{minipage}
\caption{Histograms of the velocity offset compared with the systemic velocity and line width distribution of the broad and narrow components of SiO emission for the three UC H\textsc{ii} categories. The orange-filled histograms correspond to \textit{coincident} sources, and the open blue histograms represent the \textit{offset} sources. The open green histograms show the \textit{surrounded} sources. 
The three UC H\textsc{ii} categories can be found in Table~\ref{tab:TableA2}.}
\label{fig8}
\end{figure*}
\subsubsection{The spatial distribution of the different components of the SiO emission}
\label{sec4.3.2}

In this section we aim to elucidate the diverse spatial distribution between the narrow and broad SiO components across all observed categories. Figure \ref{fig6} presents the two distinct spatial distributions of the narrow and broad SiO components. The images for all the velocity-decomposed sources are available as supplementary material.

We classified the morphology between the narrow and broad components into two cases by comparing the spatial extension of the emission (in pixel counts) across the 118 decomposed sources: (\textit{A}) Broad SiO emission is more extended than the narrow emission (94 sources); (\textit{B}) Narrow SiO emission is more extended than the broad SiO emission (24 sources). The pixel counts for the broad and narrow components and the classification of each source can be found in  Table~\ref{tab:TableA1} and Table~\ref{tab:TableA2}.
As an example, Figure \ref{fig6} shows two representative sources of the \textit{group A} and \textit{group B} cases. Both sources are associated with 3mm continuum emission and along the filamentary structures. In the case of I16071-5142, representative of \textit{group A}, the broad component exhibits a more extended distribution than the narrow one, while toward I12326-6245 (representing \textit{group B}) the narrow component is distributed across a wider extension than the broad component.

\begin{table}
\caption{The number of sources from \textit{group A} and \textit{group B} that fall into the four categories defined in Section \ref{general}}
\renewcommand\tabcolsep{24pt}
\label{tab:Table1}
\begin{threeparttable}
\begin{tabular}{ccc}
\hline
\hline
Category  & \multicolumn{1}{c}{ \textit{group A}}   
&  \multicolumn{1}{c}{ \textit{group B}}     
\\
\hline
(1) &5 & 1 \\
(2) &42 & 12 \\
(3)&1 & 2\\
(4)&46 & 9 \\
\hline
\end{tabular}
 \begin{tablenotes}
        \footnotesize
        \item[]Note. 
    (1) Sources without UC H\textsc{ii} regions and not associated with filament skeletons.
    (2) Sources without UC H\textsc{ii} regions but associated with filament skeletons.
    (3) Sources hosting UC H\textsc{ii} regions but not associated with filament skeletons.
    (4) Sources hosting UC H\textsc{ii} regions and associated with filament skeletons. 
    \textit{group A}: The broad SiO is distributed more extensively than the narrow SiO. \textit{group B}: The narrow SiO is distributed more extensively than the broad SiO.
 \end{tablenotes}
\end{threeparttable}
\end{table}
We have performed the statistics about the number of sources from \textit{group A} and \textit{group B} that fall into the categories presented in Section \ref{general}, i.e. within Categories (1), (2), (3), and (4). This can be seen from Table~\ref{tab:Table1}. 
Fewer sources fall into categories (1) and (3), where sources with broad and narrow SiO emission tend not to be associated with filament skeletons (regardless they present UC H\textsc{ii} regions or not). Within these two categories, we find that five sources without UC H\textsc{ii} regions and one source with UC H\textsc{ii} regions present a more extended distribution of the broad SiO emission than for the narrow components (sources from {\it group A}; see Table~\ref{tab:Table1}). In contrast, only one source without an UC H\textsc{ii} region and two sources containing UC H\textsc{ii} regions display a more extended distribution of the narrow SiO component than for the broad SiO emission. 
For categories (2) and (4), for which sources are associated with filament skeletons, 42 objects without UC H\textsc{ii} regions and 46 sources with UC H\textsc{ii} regions present more extended broad SiO emission, while 12 sources without UC H\textsc{ii} regions and 9 sources with UC H\textsc{ii} regions sources exhibit more extended narrow SiO emission. Therefore, SiO emission (both narrow and broad) tend to appear for sources associated with filaments (categories (2) and (4)) with similar percentages of {\it group A} sources (75-85\%, i.e. 42 sources out of 54 in category (2) and 46 objects out of 55) and of {\it group B} sources (15-25\%, i.e. 12 sources out of 54 in category (2) and 9 objects out of 55).

\subsubsection{Properties of the narrow and broad SiO emission in sources hosting UC H\textsc{ii} regions}
\label{sec4.3.3}

\citet{Cosentino2020} found narrow SiO emission in molecular clouds affected by the expansion of nearby H\textsc{ii} regions. However, in that study, the lack of high-angular resolution observations prevents us from establishing the origin of the narrow SiO emission. Therefore, an interesting item to analyse using the high-angular resolution data of the ATOMS program is the emission of narrow and broad SiO toward sources hosting UC H\textsc{ii} regions. In this way, we can understand the properties of the SiO emission such as incidence of the broad and narrow components, spatial distribution and kinematics of the gas, in regions affected by expanding H\textsc{ii} regions with much higher resolution as done by \citet{Cosentino2020}. 
In Figure \ref{fig7}, we depict three representative sources of the subsample of 58 sources with UC H\textsc{ii} regions, which present different spatial distributions between SiO and 3mm continuum emission. Due to the contribution of free-free emission, the 3mm continuum emission is spatially associated with H40$\alpha$ emission, as reported by \citet{2022MNRASRong}.
In Figure \ref{fig7}, we use the half-peak value of the 3 mm continuum emission to validate the spatial relationship between SiO and the material associated with the UC H\textsc{ii} region as seen at these wavelengths.
We classify these sources with UC H\textsc{ii} regions into three groups: (i) the SiO emission peak falls within the half-intensity peak contour of the 3mm continuum emission; we name these sources as ‘\textit{coincident}’ (9 sources); (ii) the SiO emission peaks offset with respect to the half-intensity peak contour of the 3mm continuum emission; we name these sources as ‘\textit{offset}’ (40 sources); 
and (iii) the SiO emission surrounds the 3mm continuum emission and it is clearly associated with extended emission either at 1.3 GHz \citep[obtained from][]{2023arXiv231207275G} and/or 8 $\upmu$m emission, as if material had been compressed by the expansion of the H\textsc{ii} region; we label these sources as ‘\textit{surrounded}’ (9 sources). In Appendix~\ref{Appendix C}, we present the spatial distribution of the 1.3 GHz and 8 $\upmu$m emission. The classification of UC H\textsc{ii} sources is provided in Table~\ref{tab:TableA2}. 

In Figure~\ref{fig7}, the case of I11298-6155 (\textit{coincident})
demonstrates that both broad and narrow components coincide with 3mm continuum emission, and the narrow components are more extended than the broad ones in the coexisting position. 
Toward I15411-5352 (\textit{offset}), it is clear that both components are separated from the peak position of 3mm continuum emission, and the narrow ones present more compact distribution than the broad ones, similar to the source I16071-5142. 
In source I09002-4732 (\textit{surrounded}), the broad SiO components are distributed around but far away from the 3mm continuum emission, as if it were associated with compressed material produced by the expansion of the H\textsc{ii} region seen at 1.3 GHz and/or 8 $\upmu$m emission in Figure \ref{figC1}. The sources I09002-4732 and I15411-5352 exhibit associations with filamentary structures.

We utilize the distribution of sources with UC H\textsc{ii} regions as a function of velocity offset in comparison with the systemic velocity of the cloud, and as a function of line width to study the kinetic structure for three categories statistically.
In Figure~\ref{fig8}, the distributions of both SiO components are similar to the SiO distributions in Figure~\ref{fig5} (see \ref{histograms}). 
We employed the KS test to compare the velocity offset and line width across these three categories considering both broad and narrow components together. 
For the velocity offset comparisons between \textit{coincident} and \textit{offset}, \textit{offset} and \textit{surrounded}, and \textit{coincident} and \textit{surrounded}, the P-values returned by the KS test were 2.44$\times{10^{-9}}$, 2.49$\times{10^{-3}}$, and 5.89$\times{10^{-6}}$, respectively. For the line width comparisons among these three categories, the P-values returned by the KS test from these three categories were 0.09, 6.98$\times{10^{-18}}$, and 8.41$\times{10^{-11}}$, respectively.
These results suggest that line widths from \textit{surrounded} sources are likely drawn from different distributions with line widths from \textit{coincident} and \textit{offset} sources. Notably, the line widths of \textit{surrounded} sources tend to be narrower than those of \textit{coincident} and \textit{offset} sources, with median values of 3.17, 3.84, and 3.80 km s$^{-1}$ for narrow components, and 6.48, 9.82, and 9.07 km s$^{-1}$ for broad components, respectively. However, the velocity offset distributions across the three categories do not exhibit significant differences.


\subsection{SiO luminosity}
\label{sec4.4}

We calculated the luminosity of SiO emission ($L_\textup{SiO}$) for the broad and narrow components across each velocity decomposed source. $L_\textup{SiO}$ is estimated using the following formula \citep{nguy2013low}:
\begin{equation}
\begin{split}
L_\textup{SiO}& =4\pi{d^2}\times{\int{T_\textup{MB}\mathrm{d}{\upsilon}}}\\
& \simeq{1.9\times{10^{-5}}L_{\odot}\times\left ({\frac{d}{\textup{6\,kpc}}}\right)^{2}{\frac{\int{T_\textup{MB}\mathrm{d}{\upsilon}}}{1 \mathrm{K}\,\mathrm{km}\, \mathrm{s^{-1}}}}}
\label{eq:eq1}
\end{split}
\end{equation}

\begin{equation}
\begin{split}
&\int{T_\textup{MB}\mathrm{d}{\upsilon}} [\mathrm{K\, km\, s^{-1}}] = Q({\upsilon_{0}}) \int{F\mathrm{d}{\upsilon}} [\mathrm{Jy\, km\, s^{-1}}]\\ 
&Q({\upsilon_{0}}) =
\frac{c^2}{2k{\upsilon_{0}}^2}\times\frac{4 \ln2}{\uppi{\theta_{b}}^2}\\
&\simeq{1.22\times{10^{6}}
{\left(\frac{\theta_{b}}{\textup{arcsec}}\right)^{-2}}
\left(\frac{\upsilon_0}{\mathrm{GHz}}\right)^{-2}}
\label{eq:eq2}
\end{split}
\end{equation}
where $d$ is the distance to the source and $\int{F\mathrm{d}{\upsilon}}$ and $\int{T_\textup{MB}\mathrm{d}{\upsilon}}$ is the integrated intensity of SiO emission in units of [Jy km s$^{-1}$] and [K km s$^{-1}$], respectively. $k$ is the Boltzmann constant and c is the speed of light. The SiO line frequency $\nu_0$ is 86.847 GHz. ${\theta_{b}}$ is the beam FWHM. We calculated $\int{T_\textup{MB}\mathrm{d}{\upsilon}}$ using the results obtained from \texttt{scousepy}. First, we multiplied the intensity peak value by the linewidth (FWHM)  obtained for the SiO line profile toward each pixel and inferred from the Gaussian fitting results of \texttt{scousepy}, and by a constant factor of 1.064. Then, we added all these values for all pixels in each source. After this, we considered the number of beams contained within our source area, so that we converted the integrated spectrum from units of Jy beam$^{-1}$ km s$^{-1}$ to Jy km s$^{-1}$. Subsequently, we applied Equation ~\ref{eq:eq2} to convert these values from Jy km s$^{-1}$ to K km s$^{-1}$. 

$L_\textup{SiO}$ for both components, $d$, and $\int{F\mathrm{d}{\upsilon}}$ values can be found in Table~\ref{tab:TableA1}.
The derived range of $L_\textup{SiO}$ (as derived from the SiO 2-1 transition) for the broad SiO emission in the sources across the massive sample is 2.2$\times{10^{-6}}$ -- 1.3$\times{10^{-3}}$ $L_{\odot}$, with a median value of 7.6$\times{10^{-5}}$ $L_{\odot}$, and the range of $L_\textup{SiO}$ for the narrow SiO emission is 1.5$\times{10^{-6}}$ -- 9.7$\times{10^{-4}}$ $L_{\odot}$, with a median value of 3.9$\times{10^{-5}}$ $L_{\odot}$. For the two low-mass sources, I08076-3556 and I11590-6452, the $L_\textup{SiO}$ values for the broad SiO emission are 2.2$\times 10^{-6}$ and 2.7$\times 10^{-7}$ $L_{\odot}$, respectively. For the narrow SiO emission, these values are 2.2$\times 10^{-6}$ and 2.2$\times 10^{-7}$ $L_{\odot}$, respectively.

Toward source I17136-3617, no narrow SiO emission is observed. Except for this source,
the mean contribution of the narrow SiO component to the total $L_\textup{SiO}$ is 33.78$\%$. Note, however, that the $L_\textup{SiO}$ estimated for the narrow SiO component toward sources I08076-3556 and I16297-4757 represent, respectively, 50.8$\%$ and 52.2$\%$ of the total, i.e. more than half of the total emission of SiO observed toward these two sources. This result may imply that I08076-3556 and I16297-4757 are at an early stage of evolution characterized by young shocks and by the interaction of the magnetic precursor \citep{2004Jimnez,2005Jimnez}. The fact that narrow SiO emission represents more than half of the total SiO emission toward two sources in our large sample may indicate that this early stage in the evolution of massive stars is short and, as a result, the statistics are small. Our analysis of the luminosity of SiO (2-1) emission therefore indicates that the majority of $L_\textup{SiO}$ (above 66$\%$) can be attributed to broad SiO emission, consistent with strong outflows being its most probable origin.


\subsection{SiO column densities and abundances derived toward the ATOMS sample}
In this Section, we estimate the SiO total column densities and molecular abundances with respect to H$_2$ for the broad and narrow SiO components detected toward all decomposed sources.
Assuming local thermodynamic equilibrium (LTE), the total column density of SiO can be calculated as follows \citep{csengeri2016atlasgal}:  

\begin{equation}
\begin{split}
N_\textup{tot}& = \frac{3 k^2}{4 \uppi^3 h \nu^2} \frac{1}{S{\mu}^2}{T_\textup{ex}}
{{\rm e}^{\frac{E_{u}}{kT_\textup{ex}}}}{\int{T_\textup{MB}\mathrm{d}{\upsilon}} \frac{\tau}{1-e^{-\tau}}} \\
\end{split}
\label{eq:eq3}
\end{equation}

where $h$ represents the Planck constant and $k$ is the Boltzmann constant. Dipole moment term ($S\mu^{2}$) is 19.2 for the SiO (2-1) line, and the SiO line frequency $\nu$ is 86.847 GHz. The upper-level energy $E_u/k$ is 9 K, and $\tau$ is the optical depth. We use the excitation temperature ($T_\textup{ex}$) of 15 K for the narrow component, which is slightly higher than the 9 K value estimated for IRDCs \citep{jimenez2010parsec}, and which is likely closer to the actual value in the UC H\textsc{ii} candidate sample; and 50 K for the broad component, as estimated for the shocked gas in molecular outflows \citep{2005Jimnez}. Note that the derived column densities for narrow SiO change by less than a factor 1.2 when assuming 10 K instead of 15 K, and that the SiO column densities change by more than a factor of 1.4 when assuming 75 K instead of 50 K for broad SiO emission.

Assuming optically thin emission, the Equation ~\ref{eq:eq3} can thus be written as:

\begin{equation}
\begin{split}
{N_\textup{SiO}}\ 
& {\simeq{2 \times {10^{11}}}\frac{2 k T_\textup{ex}}{ h \nu}
{{\rm e}^{\frac{E_{u}}{kT_\textup{ex}}}}{\int{T_\textup{MB}\mathrm{d}{\upsilon}}}}[\mathrm{cm^{-2}}]\\
\end{split}
\label{eq:eq4}
\end{equation}

The estimated SiO column densities ($N_\textup{SiO}$) calculated using Equation~\ref{eq:eq4}, range from 1.4$\times{10^{13}}$ -- 1.9$\times{10^{14}}$ cm$^{-2}$ for the broad component, and from 1.6$\times{10^{12}}$ -- 5.9$\times{10^{13}}$ cm$^{-2}$ for the narrow component.  
$N_\textup{SiO}$ for both components in sources with UC H\textsc{ii} regions range from 1.6$\times{10^{13}}$ -- 2.2$\times{10^{14}}$ cm$^{-2}$, and $N_\textup{SiO}$ for both components in sources without UC H\textsc{ii} regions range from 1.9$\times{10^{13}}$ -- 2.5$\times{10^{14}}$ cm$^{-2}$. This implies that the derived $N_\textup{SiO}$ values are similar in sources with and without UC H\textsc{ii} regions.
We derived a range of $N_\textup{SiO}$ from $1.6 \times 10^{13}$ to $2.5 \times 10^{14}$ cm$^{-2}$, with a median value of $6.4 \times 10^{13}$ cm$^{-2}$. \citet{csengeri2016atlasgal} and \citet{2023A&A...679A.123K}, using the same $T_\textup{ex}$ of 10 K, derived $N_\textup{SiO}$ values ranging from $1.6 \times 10^{12}$ to $7.9 \times 10^{13}$ cm$^{-2}$ from SiO (2-1) and $2.2 \times 10^{12}$ to $7.9 \times 10^{13}$ cm$^{-2}$ from SiO (1-0), similar to our results for the narrow component for an approximate excitation temperature.

We calculated the H$^{13}$CO$^+$ column density at the positions with the broadest and narrowest SiO line width as mentioned in Sect.~\ref{sec4.1}. 
Under the optically thin emission and LTE assumptions, considering an excitation temperature of 50 K for the position with the broadest SiO line width and 15 K for the position with the narrowest SiO line width, the H$^{13}$CO$^+$ column density can also be calculated using Equation~\ref{eq:eq3}. This time, however, S$\mu^{2}$ is 15.2, as obtained from the Cologne Database for Molecular Spectroscopy \citep{2001A&A...370L..49M}; the frequency $\upsilon$, is 86.75 GHz for H$^{13}$CO$^+$ ($J$=1-0), and the upper-level energy $E_u/k$ is 4.169 K. The column densities of H$^{13}$CO$^+$ derived across our sample range from 6.1$\times{10^{11}}$ to 2.8$\times{10^{14}}$ cm$^{-2}$.

In order to derive the SiO abundance relative to H$_2$, we calculate the ratio of $N_\textup{SiO}$/$N_{H^{13}CO^{+}}$ and consider a H$^{13}$CO$^+$ abundance of 4.2$\times$10$^{-11}$. This is consistent with reported H$^{13}$CO$^+$ abundances relative to H$_2$ in the Aquila, Ophiuchus, and Orion B clouds, which have mean values ranging from 1.5 to 5.8$\times$10$^{-11}$ \citep{2017A&A...604A..74S}, and in OMC 2-FIR 4 and Sagittarius A, the H$^{13}$CO$^+$ abundance also estimates 1.1 ± 0.1$\times$10$^{-11}$ \citep{2015ApJS..217....7S} and 1.8 ± 0.4$\times$10$^{-11}$ \citep{2011PASJ...63..763T}. Additionally, \citet{2013A&A...555A.112P} derived H$^{13}$CO$^+$ abundance of 5 $\times$10$^{-11}$ from Mopra observations towards SDC335.
By doing this, the SiO abundance is estimated to range from 1.5$\times{10^{-11}}$${\sim}$1.3$\times{10^{-9}}$ toward the positions with the broadest SiO line widths. 
Toward the positions with the narrowest SiO line widths, the SiO abundance varies from 6.4$\times{10^{-11}}$${\sim}$4.5$\times{10^{-9}}$, factors of 3-4 higher than those found for the broad components of SiO. This is consistent with what has been found in both low-mass and high-mass star-forming regions \citep[see e.g.][]{2004Jimnez,2005Jimnez,jimenez2010parsec,2018Cosentino,Cosentino2020}.
We report the derived values of the SiO abundance in Table~\ref{tab:TableA2}. The SiO abundance for each decomposed source falls within the range of the values calculated toward the positions with the broadest and narrowest SiO line widths found for each source.

\section{Discussion}
\label{sec5}
\subsection{The origin of the SiO emission}
To gain a deeper understanding on the origin of SiO emission (especially on the origin of narrow SiO), in this work we have used higher-angular resolution data toward a larger sample of sources (and in different physical environments) than used in previous works. From previous observations, it has been established that the broad components of SiO emission in high-mass star-forming regions are likely generated by high-velocity shocks ($v_\textup{s}$ ${\ge}$ 20 km s$^{^{-1}}$) associated with powerful outflows from massive protostars \citep{qiu2007high,duarte2014sio}. The narrow component of SiO, however, can be attributed to different mechanisms: 
i) less powerful outflows driven by low-mass protostars \citep{lefloch1998widespread}; ii) young shocks characterized by the interaction of the magnetic precursor of MHD shocks \citep{2004Jimnez,2005Jimnez}; iii) cloud-cloud collisions \citep{jimenez2010parsec,2013ApJ...773..123S,louvet2016tracing}; iv) gas inflows \citep{2013ApJ...773..123S}; and v) the expansion of UC H\textsc{ii} regions or supernova remnants \citep{Cosentino2020, Cosentino2022}. In this Section, we evaluate the possible origins for the broad and narrow SiO components detected across the source sample of the ATOMS survey.

\begin{figure}
\begin{center}
\begin{minipage}[t]{0.45\linewidth}
\vspace{2pt}
\centerline{\includegraphics[width=2\linewidth]{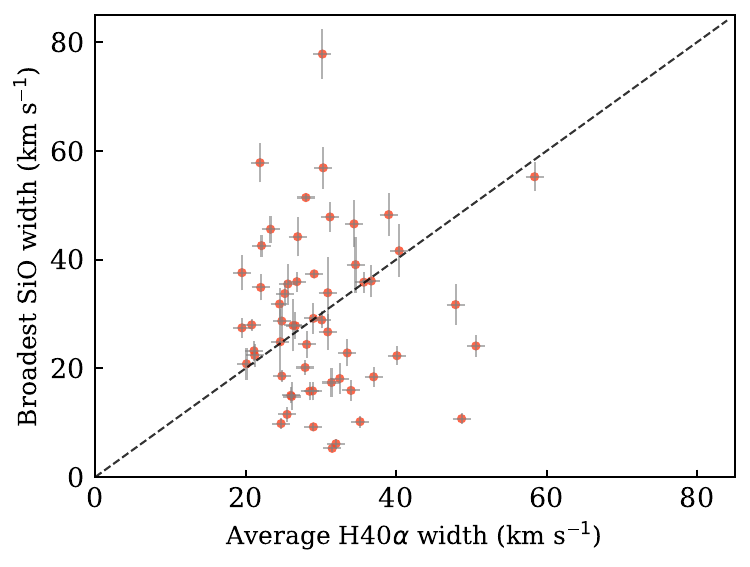}}
 \end{minipage}
\caption{The broadest SiO line width compared to the average H40$\alpha$ line width. The error bars in both plots represent the uncertainties associated with the line width.}
\label{fig9}
\end{center}
\end{figure}

\begin{figure*}
\begin{minipage}[t]{0.9\linewidth}
\vspace{2pt}
\centerline{\includegraphics[width=1.15\linewidth]{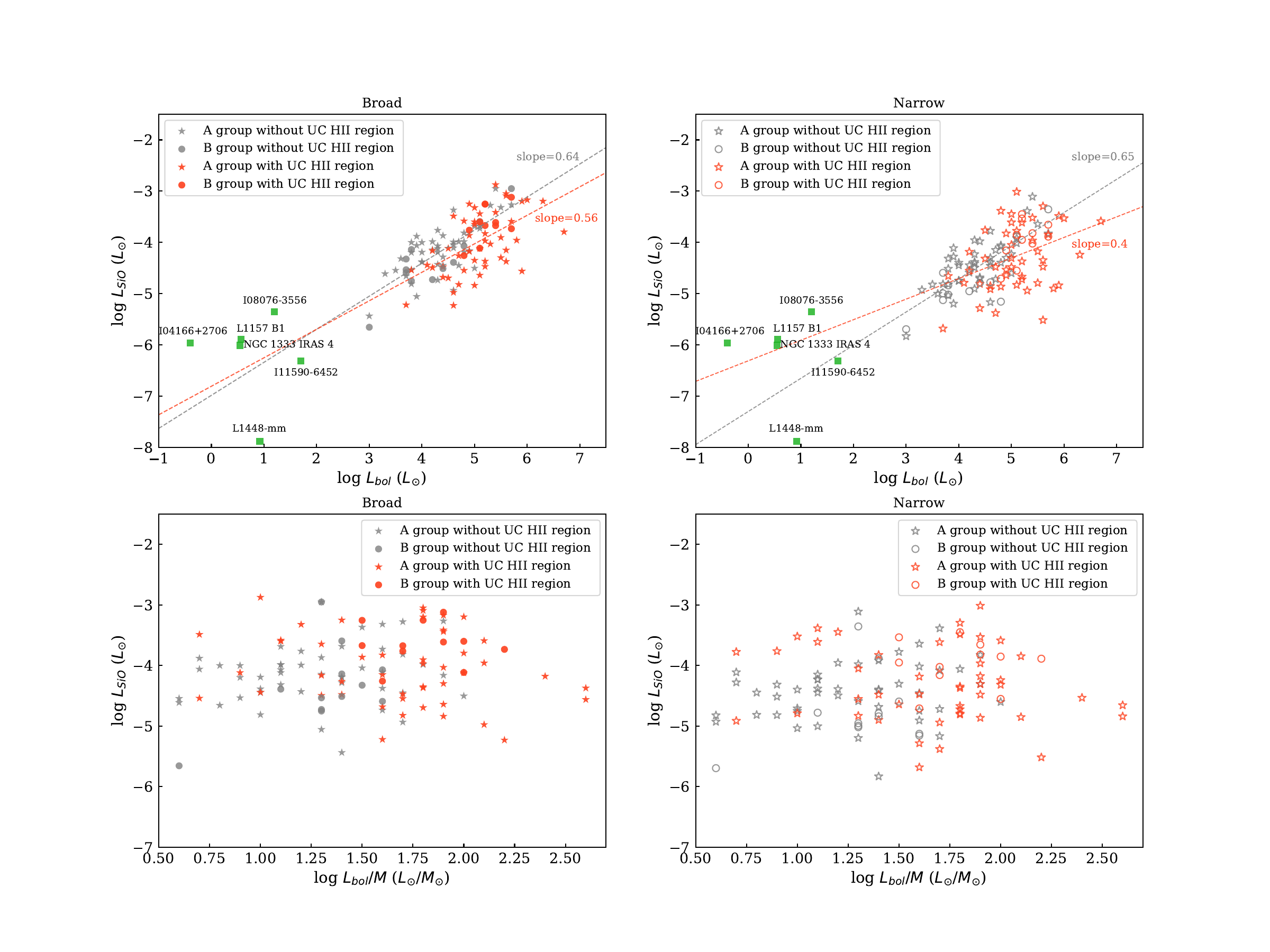}}
 \end{minipage}
\caption{{\it Upper panels}: The SiO luminosity ($L_\textup{sio}$) versus  bolometric luminosity ($L_\textup{bol}$). 
The filled and empty gray stars present broad and narrow SiO components in \textit{A groups} without UC H\textsc{ii} regions, and the filled and empty gray circles depict broad and narrow SiO
components in \textit{B groups} without UC H\textsc{ii} regions.
The filled and empty red stars show broad and narrow SiO components in \textit{A groups} hosting UC H\textsc{ii} regions, while the filled and empty red circles display broad and narrow SiO components in \textit{B groups} with UC H\textsc{ii} regions.
The filled green rectangles show the low-mass stars \citep{2022MNRAS.512.5214D,2011ApJ...739...80J,2020A&A...640A..74S,2009A&A...495..169S,2020A&ARv..28....1L}.
In the upper left panel, the gray line shows a linear fit of log $(L_\textrm{sio} / L_{\sun})$ $= (0.64\pm0.07)$ log $(L_\textrm{bol} / L_{\sun})$ - $6.98\pm0.32$ obtained for sources without UC H\textsc{ii} regions, while the red line shows the linear fit log $(L_\textrm{sio} / L_{\sun})$ $ = (0.56\pm0.11)$ log $(L_\textrm{bol} / L_{\sun})$ - $6.81\pm0.58$ obtained for sources hosting UC H\textsc{ii} sources. 
In the upper right panel, the gray line displays the linear fit 
log $(L_\textrm{sio} / L_{\sun})$ $ = (0.65\pm0.08)$ log $(L_\textrm{bol} / L_{\sun})$ - $7.30\pm0.33$ derived for sources without UC H\textsc{ii} regions, while the red line is used to show the linear fit
log $(L_\textrm{sio} / L_{\sun})$ $ = (0.40\pm0.13)$ log $(L_\textrm{bol} / L_{\sun})$ - $6.31\pm0.68$ inferred for sources hosting UC H\textsc{ii} regions. 
{\it Lower panels}: $L_\textup{sio}$ vs. $L_\textup{bol}/M$. No apparent correlation is seen either for the broad or for the narrow components. The symbols are the same as in the upper panels.}
\label{fig10}
\end{figure*}

\subsubsection{SiO emission as a probe of outflows}

For our sample, we have decomposed the emission of SiO detected across the ATOMS sources to investigate the morphology, spatial extent and kinematics of the broad and narrow SiO components (see Section \ref{sec4}). To distinguish the bright point sources in the mid-IR, we created the SiO moment 0 maps overlaid with multiple wavelengths (4.5, 8, and 24 $\upmu$m) in Figure~\ref{figC1}. 
Toward sources without UC H\textsc{ii} regions, 
by comparing the SiO moment 0 maps and the H$^{13}$CO$^{+}$ filament skeletons with the mid-IR images (4.5, 8, and 24 $\upmu$m), we can assess the fraction of SiO arising from molecular outflows. 
Toward sources with UC H\textsc{ii} regions, we use mid-IR images (4.5 and 8 $\upmu$m) and the maximum SiO line width to measure the SiO from UC H\textsc{ii} regions or molecular outflows.

The mid-IR 4.5 $\upmu$m emission is believed to be generated by H$_2$ vibrationally-excited emission typically associated with molecular outflows \citep{noriega04}, and the 8 $\upmu$m emission probes dust heated by protostars \citep{wynn82} or PAH emission from PDRs. 
The 24 $\upmu$m emission is usually used as the SFRs tracer \citep{2012ARA&A..50..531K}.
Therefore, we consider that the SiO emission from the sources without UC H\textsc{ii} regions but spatially coincident in the plane of the sky with one of the mid-IR emission (4.5, 8, and 24$\upmu$m) is likely generated by shocks in molecular outflows. From our 58 sources without UC H\textsc{ii} regions, we have found that 51 sources show SiO emission (especially the broad component) both again spatially coincident in the plane of the sky with H$^{13}$CO$^{+}$ filament skeletons (indicating the presence of high-density molecular gas, essential for forming new stars) and point objects in 4.5/8/24 $\upmu$m emission (potential outflow driving sources). Five sources I08448-4343, I08470-4243, I09094-4803, I17269-3312, and I18290-0924 only exhibit 4.5/8/24 $\upmu$m emission, while one source I16026-5035 only displays SiO emission associated with H$^{13}$CO$^{+}$ filament skeletons but no mid-IR emission. 
We note that only one source I18134-1942 lacks associations with dense filament skeletons and 4.5/8/24 $\upmu$m emission.

The average line width of the hydrogen recombination line (RRL) H40$\alpha$ is 28.06 km $s^{-1}$ from the \citet{zhang2023atoms}. We consider that SiO emission with a maximum SiO line width larger than that of RRL H40$\alpha$ arise from molecular outflows, while those with a maximum SiO line width smaller than that of RRL H40$\alpha$ may originate from outflows or UC H\textsc{ii} regions. In Figure \ref{fig9}, we plot the variation between the SiO line width at the positions ‘\textit{B}’ and the average line width of H40$\alpha$. The average line widths for the H40$\alpha$ lines have been calculated from the Gaussian fits of the source-averaged spectra \citep{zhang2023atoms}. Among 58 sources with UC H\textsc{ii} regions, approximately half of the sources (27 sources) exhibit a maximum SiO line width larger than that of RRL H40$\alpha$, while the remaining 31 sources have a maximum SiO line width narrower than that of RRL H40$\alpha$. These results suggest that the SiO emission in half of the sources may be associated with outflows, while for the other half of sources the SiO emission may be produced either by outflows or by the expansion of UC H\textsc{ii} regions. This will be investigated further below.

As shown in Section \ref{sec4.3.3}, broad components dominate the emission of SiO across the ATOMS sample with linewidths as large as 60 km s$^{-1}$ and offset velocities going beyond $\pm$40 km s$^{-1}$ with respect to the rest ambient velocity. All this suggests that the majority of the SiO emission in the ATOMS sample (above 66\% in SiO luminosity; see Section \ref{sec4.4}) is related to outflow activity from embedded protostars.

\begin{figure*}
\begin{minipage}[t]{0.9\linewidth}
\vspace{2pt}
\centerline{\includegraphics[width=1.15\linewidth]{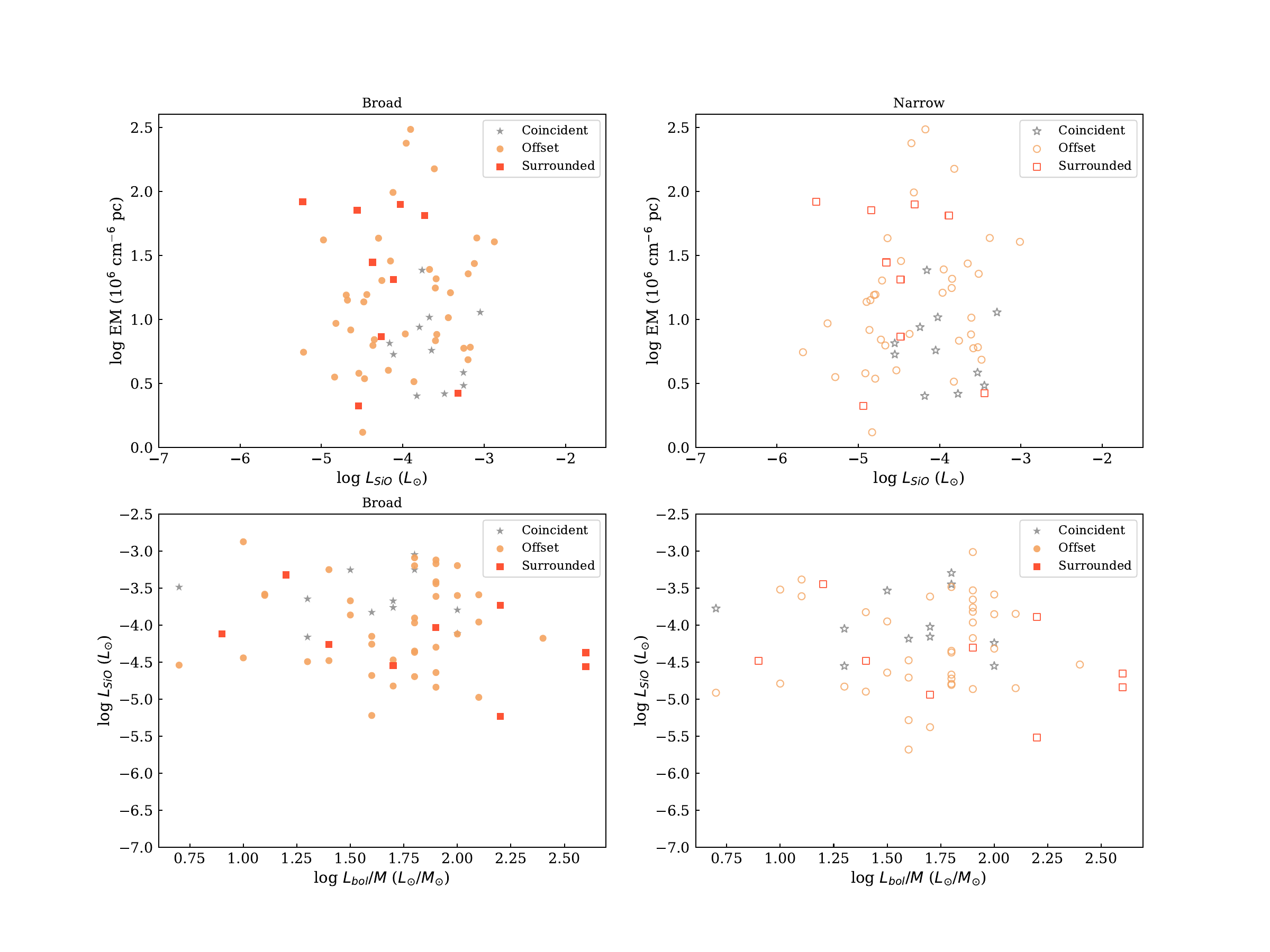}}
 \end{minipage}
\caption{{\it Upper}: The emission measure (EM) value vs. SiO luminosity ($L_\textup{SiO}$). The filled and empty gray stars indicate the broad and narrow SiO components associated with \textit{coincident} sources. The filled and empty orange circles designate the broad and narrow SiO components for \textit{offset} sources, and the filled and empty red rectangles display the broad and narrow SiO components for \textit{surrounded} sources. {\it Lower}: $L_\textup{SiO}$ versus $L_\textup{bol}/M$. Symbols are the same as in the upper panel.}
\label{fig11}
\end{figure*}
\subsection{Dependence of SiO on the evolutionary stage of the source}
\label{sec5.1.2}

In this Section, we investigate whether there is any dependence of the measured SiO luminosity with the evolutionary stage of the ATOMS sources. To do this, we have used the luminosity to mass ratio ($L_\textup{bol}/M$) parameter proposed by \citep{Molinari2008,molinari2016calibration} as a proxy of the evolutionary stage for high-mass protostars. 
Broad and narrow SiO luminosity against bolometric luminosity ($L_\textup{bol}$) and $L_\textup{bol}/M$ are presented in Figure~\ref{fig10}.
The values for $L_\textup{bol}/M$ and $L_\textup{bol}$ are given in \citet{liu2020atoms}. 

From Figure \ref{fig10} (upper panel), there is no significant difference between sources from the {\it A} (in stars) and {\it B groups}  (in circles), while {\it B group} sources may present a stronger correlation than sources from the {\it A group}, with the Spearman rank correlation coefficients of 0.55 and 0.89, respectively. 

Considering the evolution of SiO line profiles in molecular outflows, where narrow SiO is dominant at early stages and broad SiO becomes more prevalent as the evolution progresses \citep{jimenez2009evolution}, we propose that sources in {\it B group} may be younger than those in {\it A group}. As shown in Figure~\ref{fig10}, {\it B group} exhibits a strong correlation between $L_\textup{SiO}$ and $L_\textup{bol}$. These findings indicate a positive correlation between the outflow activity or the ability to detect SiO emission and the evolutionary stages at the early times, and during the evolutionary process, this correlation weakens. In Figure~\ref{fig10}, there may be a slight trend for sources with H\textsc{ii} regions (in red) to present higher SiO luminosities than sources without H\textsc{ii} regions (in gray), although the two groups seem to be well mixed. The P-value returned by the KS test for $L_\textup{SiO}$ between sources with H\textsc{ii} regions and those without is 0.02, indicating that the $L_\textup{SiO}$ from these two groups likely come from different distributions, and the median $L_\textup{SiO}$ for sources with H\textsc{ii} regions is 1.5 $\times 10^{-4}$ $L_{\odot}$, and for sources without H\textsc{ii} regions, these values is 1 $\times 10^{-4}$ $L_{\odot}$.

If all sources are taken together, Figure \ref{fig10} (upper panel) reveals a moderate positive trend between $L_\textup{SiO}$ and $L_\textup{bol}$ for both the broad and narrow SiO components, with Spearman rank correlation coefficients of 0.63 and 0.61, respectively. The positive correlation indicates that the brighter luminosity sources exhibit stronger SiO emission, implying more intense outflow activity. \citet{2023A&A...679A.123K} investigated SiO (1-0) emission toward 104 regions that consist of 57 IRDCs, 21 high-mass protostellar objects, and 26 UC H\textsc{ii} regions, using the Korean VLBI Network. In that paper, they found that sources with higher $L_\textup{SiO}$ tend to be associated with higher $L_\textup{bol}$, which is consistent with our results. However, note that for a given $L_\textup{bol}$, warmer sources in their sample tend to have lower $L_\textup{SiO}$. This trend can also be hinted at in our data by the lower slope obtained in the $L_\textup{SiO}$-$L_\textup{bol}$ relation for the sources with UC HII regions shown in Figure \ref{fig10}. 
The SiO broad and narrow components in sources lacking UC H\textsc{ii} regions exhibit a stronger correlation between $L_\textup{SiO}$ and $L_\textup{bol}$ with a Spearman rank correlation coefficient of 0.67. In contrast, this correlation appears slightly weaker in sources with UC H\textsc{ii} regions with a Spearman rank correlation coefficient of 0.55. These results suggest that the ability to detect SiO emission in sources at later stages diminishes. 
Furthermore, the weak trend in these correlations and higher $L_\textup{SiO}$ values, implies that the SiO emission from sources with UC H\textsc{ii} regions might be affected by the photo-chemistry (i.e. photo-dissociation) induced by the UV field produced by the UC H\textsc{ii} regions.

\begin{figure*}
\begin{minipage}[t]{0.32\linewidth}
\vspace{2pt}
\centerline{\includegraphics[width=1.2\linewidth]{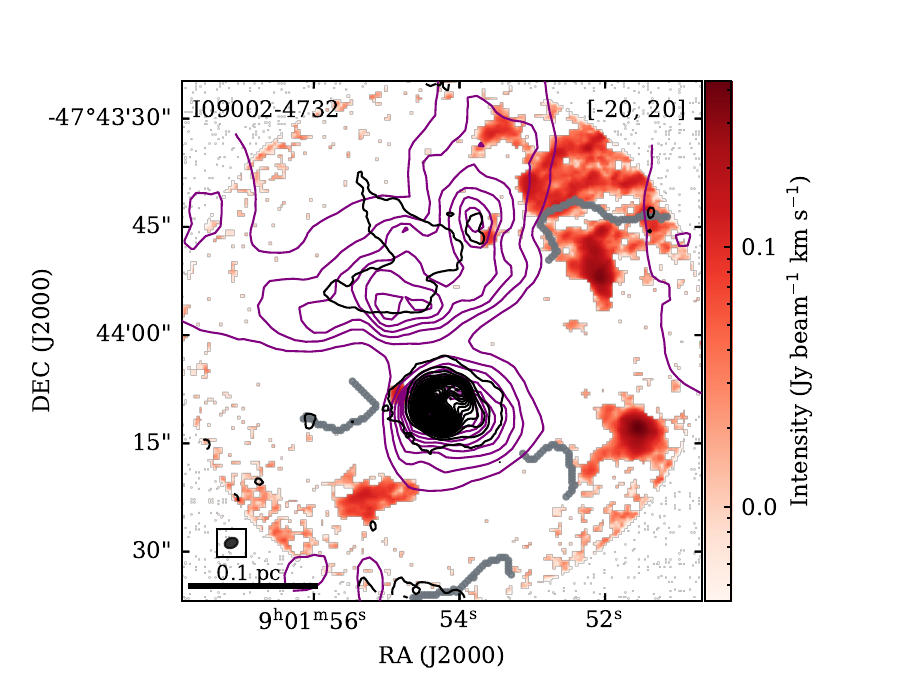}}
 \end{minipage}
\begin{minipage}[t]{0.32\linewidth}
\vspace{2pt}
\centerline{\includegraphics[width=1.2\linewidth]{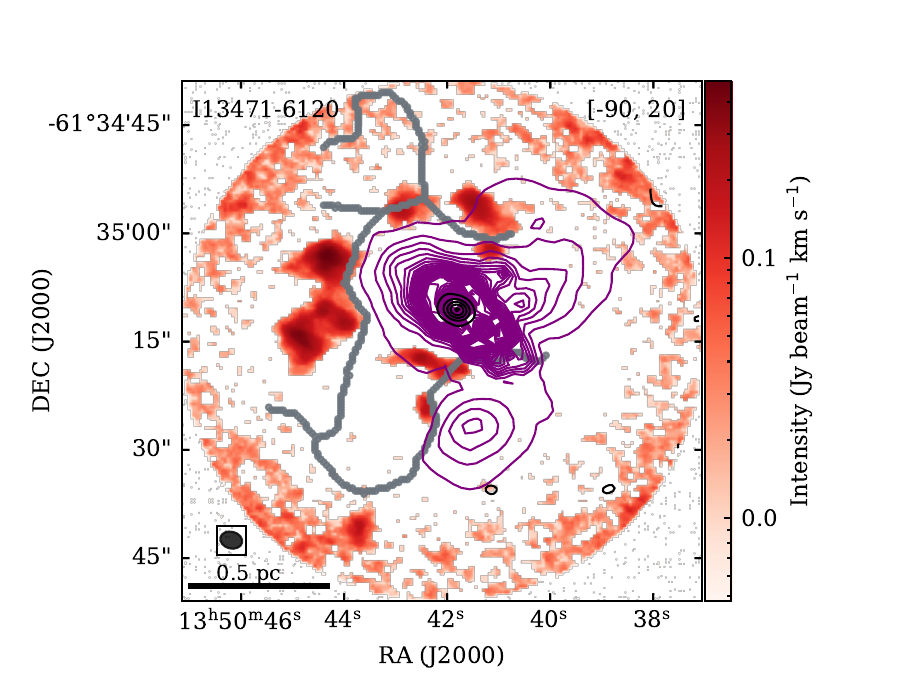}}
 \end{minipage}
 \begin{minipage}[t]{0.32\linewidth}
\vspace{2pt}
\centerline{\includegraphics[width=1.2\linewidth]{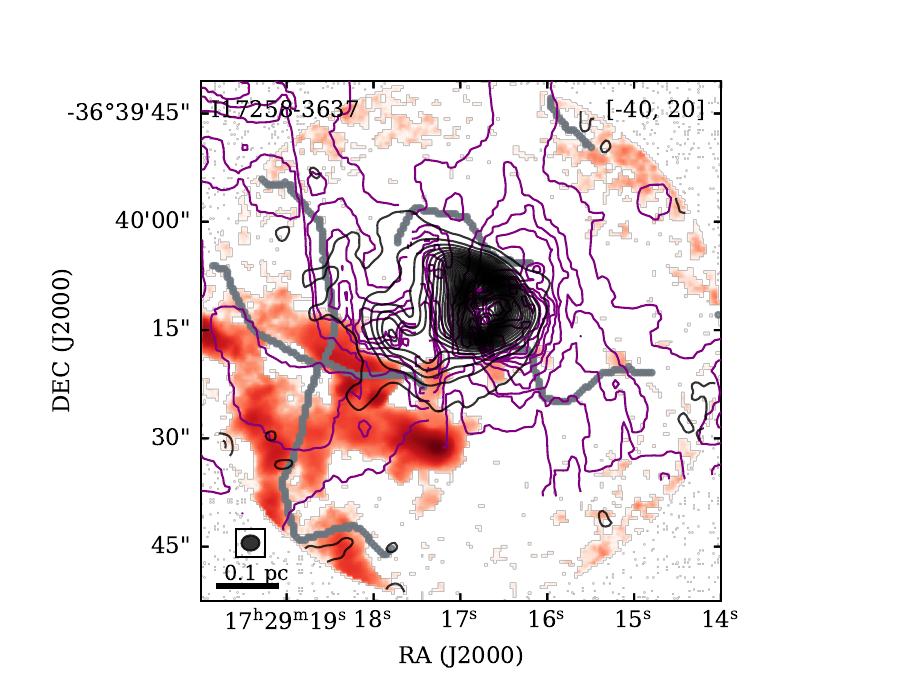}}
 \end{minipage}
\caption{Three sources with expanding UC H\textsc{ii} regions.
The background in red is SiO (2-1) integrated intensity maps. The black contours are 3 mm continuum emission, and contours are from 5$\sigma$ to the peak values in the step of 10$\sigma$. 
The purple contours are 8 $\upmu$m emission, and contours are from 5$\sigma$ to the peak values in the step of 50$\sigma$.
The bold gray lines represent the filament skeleton identified by H$^{13}$CO$^+$ emission. 
The shown field of view is 72$^{\prime}$$^{\prime}$ corresponding to the FOV of the ALMA observations. The source name and integrated velocity ranges (in km s$^{-1}$) are shown on the upper left and right corners of each panel, respectively. The beam size and the scale bar are presented in the lower left corner.}
\label{fig12}
\end{figure*}
In the upper panel of Figure \ref{fig10}, we also plot six low-mass stars so that their $L_\textup{SiO}$ and $L_\textup{bol}$ can be compared to the relationship between $L_\textup{SiO}$ and $L_\textup{bol}$ found for our sample. This sub-group of six low-mass sources include objects I08076-3556 and I11590-6452 also contained within the ATOMS sample, and four other sources NGC 1333 IRAS 4, L1448-mm, L1157 B1, and I04166+2706  \citep{2022MNRAS.512.5214D,2011ApJ...739...80J,2020A&A...640A..74S,2009A&A...495..169S,2020A&ARv..28....1L}.
Low-mass stars are represented by filled green squares in Figure \ref{fig10}. From this Figure, it is clear that they exhibit lower values of $L_\textup{SiO}$ and $L_\textup{bol}$ when compared to the high-mass sources of the ATOMS sample. Note that only one source, I11590-6452 fall into our trend.  
For the rest of low-mass sources, there is no obvious correlation between these two parameters. 
The number of low-mass sources to compare with is, however, very limited, and therefore, we cannot draw any robust conclusions.

\citet{duarte2014sio} suggested an evolutionary trend for the younger sources in the Cygnus X star-forming cluster to present a higher fraction of narrow SiO emission (up to $\sim$60\%) with respect to more evolved sources. To test this idea using the ATOMS source sample, in Figure \ref{fig10} (lower panels), we plot the derived SiO luminosity, $L_\textup{SiO}$, as a function of the $L_\textup{bol}/M$ ratio, considered a proxy of the evolutionary stage of the source \citep[see][]{Molinari2008}. 
If we just only consider the information about the SiO luminosity, from the lower panels of Figure \ref{fig10}, it is clear that there is no trend between $L_\textup{SiO}$ and $L_\textup{bol}/M$ either within all groups or for both the narrow and broad components of SiO.
This is consistent with the results
reported by \citet{csengeri2016atlasgal}, \citet{2019ApJ...878...29L}, and \citet{2023A&A...679A.123K}. In these papers, SiO abundance does not vary significantly at different evolutionary stages.
However, as discussed in Sections \ref{sec5.3} and \ref{sec5.4} below, if the spatial distribution of the broad and narrow SiO emission is taken into account for individual sources, the location and fraction of narrow SiO detected toward high-mass SFRs (and in particular toward sources hosting UC H\textsc{ii} regions) may depend on the evolutionary stage of the source.

\subsection{The effect of UC H\textsc{ii} regions on the observed SiO emission}
\label{sec5.3}

The UC H\textsc{ii} regions inject large amounts of ultraviolet photons into their surrounding medium, profoundly impacting the hosting molecular cloud. This interaction can cause both positive feedback (by triggering new star formation) and negative feedback (by disrupting and dispersing the parental molecular cloud where the UC H\textsc{ii} region is embedded, inhibiting future star formation). In our velocity decomposed source sample, \citet{zhang2023atoms} identified 58 UC H\textsc{ii} regions and calculated the emission measure (EM) using the H40$\alpha$ emission. 
In Sect.~\ref{sec4.3.2}, we classified these sources into three groups: "\textit{coincident} (9 sources)", "\textit{offset} (40 sources)", and "\textit{surrounded} (9 sources)". 
Based on the morphology of the three example sources presented in Figure~\ref{fig7}, we could imagine an evolutionary scenario in which: i) Massive stars are formed first and drive powerful outflows. This would be consistent with what we see toward I13291-6249, where the SiO emission peaks toward the 3mm continuum emission peak. ii) A subsequent stage in which low-mass stars are formed in the surroundings of massive stars. This would explain why the SiO emission peak appears offset with respect to the peak of the 3mm continuum emission (as seen in source I15411-5352). Note that this is a similar scenario to what has been observed in IRDCs, with massive protostars forming before their low-mass counterparts \citep[see][]{Zhang2015}.
The shocks at the evolutionary stages i) and ii) would be young so that the narrow SiO emission could be attributed to the interaction of the magnetic precursor of MHD shock waves, as suggested by \citet{2004Jimnez,2009Jimnez}. iii) The most evolved stage in which the UC H\textsc{ii} region has expanded disrupting the molecular cloud. The expansion of the H\textsc{ii} region generates SiO emission away from the 3mm continuum emission, and that surrounds the UC H\textsc{ii} region being seen at the edge of the photon-dominated regions (PDRs) probed by 8 $\upmu$m emission (this would be the case of source I09002-4732). The mostly broad SiO emission, although some narrow emission is also detected, can be seen toward certain regions at the PDR interface (see Figure~\ref {fig7}).

In Figure~\ref{fig11}, we plot the EM against $L_\textup{SiO}$ for the broad and narrow components (upper panels), and $L_\textup{SiO}$ versus $L_\textup{bol}/M$ also for the broad and narrow components of SiO (lower panels). 
From the upper panels, we find that there are no apparent differences in EM values between the \textit{coincident} and \textit{offset} groups for both SiO velocity components. 
By comparing each pair of groups, we obtained P-values from the KS test for $L_\textup{SiO}$ across these three categories. Notably, only $L_\textup{SiO}$ from \textit{offset} and \textit{surrounded} sources may originate from the same distributions with the P-value of 0.512. All other comparisons presented P-values of 0.046 and 0.036, suggesting that $L_\textup{SiO}$ from those sources likely come from different distributions. 
Furthermore, for both the broad and narrow components, their derived $L_\textup{SiO}$ values for \textit{coincident} sources are larger than those inferred for \textit{offset} and \textit{surrounded} sources, with median values of 3.1 $\times 10^{-4}$ $L_{\odot}$, 1.4 $\times 10^{-4}$ $L_{\odot}$, and 8.8 $\times 10^{-5}$ $L_{\odot}$, respectively. As suggested in Section \ref{sec4.3.3}, this may be due to the \textit{coincident} sources being at an earlier stage of evolution, driving more powerful outflows. 

From the lower panels of Figure \ref{fig11}, one can find a trend for the \textit{surrounded} sources 
to present lower SiO luminosities than \textit{coincident} sources. Since the SiO luminosity is lower for evolved UC H\textsc{ii} regions in the \textit{surrounded} group, this implies that the expansion of H\textsc{ii} regions has a negative effect on subsequent star formation activity as probed by SiO. The \textit{coincident} and \textit{offset} groups appear relatively well mixed although there is a small trend for the \textit{offset} sources to be older than the \textit{coincident} ones as postulated above. The median $L_\textup{bol}/M$ values are 50 and 63, respectively, with orange circles reaching higher $L_\textup{bol}/M$ values than the gray stars (see Figure \ref{fig11}), which is consistent with our discussion before.

\subsection{SiO emission produced by the expansion of UC H\textsc{ii} regions}
\label{sec5.4}
We identified nine sources hosting UC H\textsc{ii} regions that present SiO emission not associated with 3mm continuum emission and that appear surrounding the ionized gas (these sources are classified as \textit{surrounded} as described in Section \ref{sec4.3.3}). The ID of these sources in Table \ref{tab:TableA2} is 5, 12, 18, 24, 47 ,77, 98, 143, and 146.
We hypothesize that the SiO emission detected toward these sources is generated by the expansion of UC H\textsc{ii} regions. In Figure~\ref{fig12}, we show the SiO moment 0 maps overlaid on the 8 $\upmu$m emission for these sources showing the full extension of the material affected by the UV radiation coming from the H\textsc{ii} region. Black contours and bold gray lines mean the same as in Figure~\ref {fig3}. 

From Figure \ref{fig12}, it is clear that the 8 $\upmu$m emission engulfs the 3 mm continuum emission and delineates the interface between the ionized gas and the SiO emission in these sources. This suggests that the expansion of the UC H\textsc{ii} region pushes molecular gas against the surrounding molecular cloud inducing low-velocity shocks and generating SiO emission by the sputtering of dust grains. 
In our \textit{surrounded} sources, only in source I15567-5236, the narrow SiO dominates and presents more extended emission than the broad SiO. For the rest of the objects, broad SiO is more dominant than narrow SiO. 

\section{Conclusions}
In this work, we used ALMA data to investigate the origin of SiO (2-1) emission toward 146 massive star-forming regions. We decomposed the SiO line profiles into broad and narrow Gaussian velocity components, and studied their relative spatial distribution and kinematics to study their different origin. The main results are summarized as follows:

(1) Among the entire sample, we detected SiO emission in 136 sources and decomposed 118 sources. We calculated the $L_\textup{SiO}$ in broad and narrow components across each decomposed source and found the majority of $L_\textup{SiO}$ (above 66$\%$) is attributed to the broad components arising from strong outflows.
By comparing the filamentary skeletons in SiO moment 0 maps with 4.5, 8, and 24 $\upmu$m emission, and considering the broad components dominating the emission of SiO across the sample, we conclude that most SiO emission from our sources originates from the outflow activity. This conclusion supports the idea that SiO emission is a good outflow tracer. We found that SiO emission in nine sources may arise from the expansion of UC H\textsc{ii} regions.

(2) According to the different spatial distribution of the broad and narrow SiO emission, we classified 118 sources into two groups: (\textit{A}) The broad SiO is distributed more extensively than the narrow SiO in 94 sources; (\textit{B}) The narrow SiO is distributed more extensively than the broad SiO in 24 sources. 
Across all groups, We observed a moderate positive correlation between $L_\textup{SiO}$ and $L_\textup{bol}$ in both broad and narrow components,
suggesting that higher luminosity sources show more intense outflow activity.
Sources from \textit{B} groups present a
stronger correlation than sources from \textit{A} groups. These findings indicate a positive correlation between the outflow activity and the evolutionary stages at the early times, and during the evolutionary process, this correlation weakens. Furthermore, the \textit{A} group and \textit{B} group did not display clear evolutionary stages.
The sources without UC H\textsc{ii} regions seem to show a stronger correlation between $L_\textup{SiO}$ and $L_\textup{bol}$ and the lower $L_\textup{SiO}$ values compared with the UC H\textsc{ii} sources. This suggests that SiO emission might be influenced by UV-photochemistry induced by UC H\textsc{ii} regions. 

(3) Among the 58 sources hosting UC H\textsc{ii} regions, we divided them into three groups depending on the separation between the 3mm continuum emission and SiO emission: 
(1) The SiO emission coincides with the 3mm continuum emission in 9 sources, named ‘\textit{coincident}’; 
(2) The SiO emission is separated with the 3mm continuum emission in 40 sources, labeled ‘\textit{offset}’; 
(3) The SiO emission surrounds the 3mm continuum emission in 9 sources, named ‘\textit{surrounded}’.  
The \textit{coincident} group exhibited higher SiO luminosity values than the other groups in both broad and narrow components. This implies that UC H\textsc{ii} regions could have a negative effect on the shock activity within the \textit{offset} and \textit{surrounded} groups.
\label{sec6}

\section*{Acknowledgements}
R. L. gratefully acknowledges financial support from the China Scholarship Council (No.202204910347).
This work has been supported by the National Key R\&D Program
of China (No. 2022YFA1603100). 
Tie Liu acknowledges the supports by the National Key R\&D Program of China (No. 2022YFA1603100), National Natural Science Foundation of China (NSFC) through grants No.12073061 and No.12122307, and the Tianchi Talent Program of Xinjiang Uygur Autonomous Region.
I.J-.S, J.M.-P., V.M.R., L.C., A.M., A.M.H, A.L.G., M.S.-N., and D.S.A, acknowledge funding from grants No. PID2019-105552RB-C41 and PID2022-136814NB-I00 from the MICIU/AEI/10.13039/501100011033 and by “ERDF/EU". A.M. has received support from grant PRE2019-091471, funded by MCIN/AEI/10.13039/501100011033 and by “ESF, Investing in your future”.
PS was partially supported by a Grant-in-Aid for Scientific Research (KAKENHI Number JP22H01271 and JP23H01221) of JSPS. 
V. M. R. has also received support from project RYC2020-029387-I funded by MICIU/AEI/10.13039/501100011033 and by "ESF, Investing in your future", and from the Consejo Superior de Investigaciones Cient{\'i}ficas (CSIC) and the Centro de Astrobiolog{\'i}a (CAB) through project number 20225AT015 (Proyectos intramurales especiales del CSIC).
M.S.N. acknowledges a Juan de la Cierva Postdoctoral Fellow proyect JDC2022-048934-I, funded by MCIN/AEI/10.13039/501100011033 and by the European Union “NextGenerationEU”/PRTR”.
G.G. and L.B. gratefully acknowledge support by the ANID BASAL project FB210003.
DSA acknowledges the funds provided by the Comunidad de Madrid through the Grant PIPF-2022/TEC-25475, and also extends his gratitude for the financial support provided by the Consejo Superior de Investigaciones Cient{\'i}ficas (CSIC) and the Centro de Astrobiolog{\'i}a (CAB) through the project 20225AT015 (Proyectos intramurales especiales del CSIC).
C.W.L. is supported by the Basic Science Research Program through the National Research Foundation of Korea (NRF) funded by the Ministry of Education, Science and Technology (NRF-2019R1A2C1010851) and by the Korea Astronomy and Space Science Institute grant funded by the Korea government (MSIT; Project No. 2023-1-84000).
MJ acknowledges the support of the Research Council of Finland Grant No.348342. This work is supported by the Ministry of Science and Technology of China through grant 2010DFA02710, the Key Project of International 
Cooperation, and by the National Natural Science Foundation of China through grants 11503035, 11573036, 11373009, 11433008, 11403040 and 11403041.

This paper makes use of the following ALMA data: ADS/
JAO.ALMA 2019.1.00685.S. ALMA is a partnership of ESO
(representing its member states), NSF (USA), and NINS
(Japan), together with NRC (Canada), MOST and ASIAA
(Taiwan), and KASI (Republic of Korea), in cooperation
with the Republic of Chile. The Joint ALMA Observatory is
operated by ESO, AUI/NRAO, and NAOJ.
The Spitzer GLIMPSE legacy survey data can be downloaded at NASA/IPAC Infrared Science Archive (https://irsa.
ipac.caltech.edu/irsaviewer/).

\section*{Data Availability}
The data underlying this article are available in the ALMA archive. 


\bibliographystyle{mnras}
\bibliography{paper} 


\appendix{}

\section{}
\label{Appendix A}
\begin{table*}
\renewcommand\tabcolsep{2pt}
\caption{Basic parameters of the detected sources in the ATOMS survey.}
\label{tab:TableA1}
\scalebox{0.86}{
\begin{threeparttable}
\begin{tabular}{p{0.5cm}cccccp{1.6cm}p{1.6cm}lp{1.3cm}lcp{0.5cm}lccc}
\hline
\hline
ID  & \multicolumn{1}{c}{IRAS}   
&  Distance  & $v_{\rm LSR}$   
& \multicolumn{1}{c}{log $L_\textup{bol}$} 
& \multicolumn{1}{c}{$L_\textup{bol}/M$} &  
\multicolumn{2}{c}{SiO(2-1)}  & \multicolumn{3}{c}{H$^{13}$CO$^{+}$(1-0)} & Contribution$_{\rm N}$ & \multicolumn{3}{c}{Number}  &\multicolumn{1}{c}{$\sigma$ rms} 
& Outflow$^{\rm a}$ \\ 
\cline{7-11} \cline{13-15} 
 &  &   &   &  & &   
\multicolumn{1}{c}{$\int{T_\textup{B}\mathrm{d}{\upsilon}}$}   &  \multicolumn{1}{c}{$\int{T_\textup{N}\mathrm{d}{\upsilon}}$}
& 
\multicolumn{1}{c}{$\int{T_\textup{B}\mathrm{d}{\upsilon}}$}   
&  FWHM$_{\rm B}$  &FWHM$_{\rm N}$   & & ${\rm B}$ 
& ${\rm N}$
& Area \\
&
&\multicolumn{1}{c}{(kpc)}         
&\multicolumn{1}{c}{(km s$^{-1}$)}  
&\multicolumn{1}{c}{($L_{\odot}$)} 
& \multicolumn{1}{c}{($L_{\odot}$ $M_{\odot}$$^{-1}$)} 
&\multicolumn{1}{c}{(Jy km s$^{-1}$)} 
&\multicolumn{1}{c}{(Jy km s$^{-1}$)}
&\multicolumn{1}{c}{(Jy km s$^{-1}$)}
&\multicolumn{1}{c}{(km s$^{-1}$)}
&\multicolumn{1}{c}{(km s$^{-1}$)}
&\multicolumn{1}{c}{($\%$)}
& & & 
&\multicolumn{1}{c}{(Jy km s$^{-1}$)}
\\
\hline
1   & I08076-3556 & 0.4   & 5.9    & 1.2  & 3.16   & 0.359±0.002 & 0.370±0.002 & 0.059±0.004 & 0.89±0.07   & 0.72±0.07  & 50.8 & 136           & 58            & 2           & 0.008 & $\checkmark$       \\
2   & I08303-4303 & 2.3   & 14.3   & 3.8  & 25.12  & 0.376±0.001 & 0.075±0.001 & 0.120±0.009 & 3.7±0.3     & 2.29±0.17  & 16.6 & 47            & 59            & 2           & 0.01  & $\checkmark$       \\
3   & I08448-4343 & 0.7   & 3.7    & 3    & 25.12  & 0.209±0.001 & 0.084±0.001 & 0.067±0.006 & 2.38±0.2    & 2.3±0.3    & 28.7 & 165           & 86            & 4           & 0.01  & $\checkmark$       \\
4   & I08470-4243 & 2.1   & 12     & 4    & 39.81  & 0.265±0.002 & 0.113±0.003 & 0.062±0.008 & 2.1±0.3     & 1.20±0.11  & 29.9 & 244           & 140           & 2           & 0.009 & $\checkmark$       \\
5   & I09002-4732 & 1.2   & 3.1    & 4.6  & 158.49 & 0.119±0.001 & 0.062±0.001 & 0.039±0.007 & 1.7±0.3     & 1.59±0.24  & 34.1 & 153           & 30            & 4           & 0.01  &         \\
6   & I09018-4816 & 2.6   & 10.3   & 4.7  & 50.12  & 0.154±0.001 & 0.083±0.001 & 0.162±0.009 & 2.93±0.15   & 1.81±0.13  & 35.0 & 103           & 66            & 4           & 0.01  & $\checkmark$       \\
7   & I09094-4803 & 9.6   & 74.6   & 4.6  & 31.62  & 0.137±0.002 & 0.054±0.001 & 0.056±0.012 & 2.6±0.6     & 4.2±0.3    & 28.1 & 68            & 34            & 2           & 0.01  & $\checkmark$       \\
8   & I10365-5803 & 2.4   & -19    & 4.3  & 39.81  & 0.229±0.004 & 0.155±0.004 & 0.274±0.013 & 1.44±0.07   & 1.77±0.09  & 40.3 & 94            & 54            & 4           & 0.009 & $\checkmark$       \\
9   & I11298-6155 & 10    & 32.9   & 5.2  & 63.1   & 0.430±0.005 & 0.273±0.001 & 0.428±0.013 & 5.39±0.16   & 2.1±0.5    & 38.8 & 86            & 123           & 2           & 0.012 & $\checkmark$       \\
10  & I11332-6258 & 1.9   & -15.4  & 3.7  & 39.81  & 0.55±0.01   & 0.161±0.004 & 0.451±0.023 & 3.76±0.19   & 2.01±0.08  & 22.6 & 47            & 86            & 6           & 0.012 & $\checkmark$       \\
11  & I11590-6452 & 0.4   & -4.3   & 1.7  & 3.98   & 0.136±0.001 & 0.112±0.001 & 0.049±0.009 & 2.1±0.4     & 0.5±0.2    & 45.1 & 134           & 148           & 2           & 0.012 & $\checkmark$       \\
12  & I12320-6122 & 3.43  & -42.5  & 5.6  & 398.11 & 0.362±0.006 & 0.188±0.006 & 0.217±0.011 & 2.7±0.13    & 3.75±0.23  & 34.2 & 161           & 87            & 4           & 0.012 & $\checkmark$       \\
13  & I12326-6245 & 4.61  & -39.6  & 5.4  & 79.43  & 1.167±0.007 & 0.724±0.001 & 0.46±0.03   & 7.4±0.5     & 2.0±0.4    & 38.3 & 80            & 155           & 6           & 0.012 & $\checkmark$       \\
14  & I12383-6128 & 3.27  & -39.1  & 3.8  & 5.01   & 0.270±0.007 & 0.11±0.02   & 0.087±0.013 & 2.0±0.3     & 0.57±0.09  & 29.7 & 88            & 23            & 2           & 0.013 &         \\
15  & I12572-6316 & 11.57 & 30.9   & 4.6  & 5.01   & 0.243±0.003 & 0.125±0.003 & 0.224±0.012 & 1.75±0.09   & 1.75±0.19  & 34.0 & 165           & 65            & 2           & 0.012 & $\checkmark$       \\
16  & I13079-6218 & 3.8   & -42.6  & 5.1  & 39.81  & 1.269±0.005 & 0.654±0.003 & 1.02±0.03   & 6.26±0.19   & 1.08±0.11  & 34.0 & 132           & 127           & 6           & 0.012 & $\checkmark$       \\
17  & I13080-6229 & 3.8   & -35.6  & 5.1  & 79.43  & 0.155±0.003 & 0.093±0.003 & 0.668±0.014 & 3.27±0.07   & 3.48±0.16  & 37.5 & 59            & 45            & 2           & 0.013 &         \\
18  & I13111-6228 & 3.8   & -38.8  & 4.8  & 50.12  & 0.191±0.002 & 0.077±0.003 & 0.69±0.03   & 2.89±0.12   & 3.18±0.15  & 28.8 & 153           & 51            & 2           & 0.012 &         \\
19  & I13134-6242 & 3.8   & -31.5  & 4.6  & 31.62  & 0.614±0.006 & 0.336±0.006 & 0.099±0.011 & 2.9±0.3     & 1.62±0.08  & 35.3 & 120           & 77            & 6           & 0.012 &         \\
20  & I13140-6226 & 3.8   & -33.9  & 3.8  & 7.94   & 0.426±0.003 & 0.205±0.002 & 0.409±0.013 & 3.0±0.1     & 1.01±0.08  & 32.5 & 155           & 116           & 4           & 0.012 & $\checkmark$       \\
22  & I13291-6249 & 7.61  & -34.7  & 5.2  & 31.62  & 0.357±0.006 & 0.187±0.002 & 0.199±0.013 & 4.4±0.3     & 1.43±0.07  & 34.5 & 89            & 104           & 6           & 0.013 & $\checkmark$       \\
23  & I13295-6152 & 3.89  & -44.4  & 3.3  & 3.98   & 0.154±0.004 & 0.075±0.003 & 0.600±0.013 & 2.85±0.06   & 0.76±0.08  & 32.6 & 71            & 64            & 4           & 0.012 &         \\
24  & I13471-6120 & 5.46  & -56.7  & 5.3  & 79.43  & 0.260±0.003 & 0.138±0.012 & 0.244±0.008 & 1.64±0.06   & 1.70±0.09  & 34.6 & 111           & 15            & 4           & 0.013 & $\checkmark$       \\
25  & I13484-6100 & 5.4   & -55    & 4.8  & 50.12  & 0.432±0.014 & 0.235±0.006 & 1.11±0.04   & 4.32±0.15   & 1.25±0.22  & 35.2 & 109           & 107           & 6           & 0.013 & $\checkmark$       \\
26  & I14013-6105 & 4.12  & -48.1  & 5    & 63.1   & 0.213±0.002 & 0.091±0.002 & 0.409±0.012 & 2.24±0.07   & 4.4±0.3    & 29.8 & 61            & 32            & 4           & 0.013 & $\checkmark$       \\
28  & I14164-6028 & 3.19  & -46.5  & 3.7  & 31.62  & 0.372±0.006 & 0.200±0.004 & 0.710±0.021 & 4.49±0.13   & 1.8±0.1    & 35.0 & 65            & 69            & 4           & 0.013 & $\checkmark$       \\
29  & I14206-6151 & 3.29  & -50    & 3.7  & 19.95  & 0.22±0.01   & 0.075±0.007 & 0.147±0.012 & 1.69±0.14   & 1.18±0.12  & 25.7 & 23            & 24            & 2           & 0.012 &         \\
30  & I14212-6131 & 3.44  & -50.5  & 4    & 10     & 0.435±0.004 & 0.271±0.004 & 0.346±0.015 & 4.6±0.2     & 1.46±0.09  & 38.4 & 126           & 98            & 6           & 0.012 & $\checkmark$       \\
31  & I14382-6017 & 7.69  & -60.7  & 5.2  & 39.81  & 0.194±0.003 & 0.085±0.003 & 0.268±0.009 & 1.92±0.06   & 1.7±0.4    & 30.6 & 115           & 77            & 2           & 0.011 &         \\
32  & I14453-5912 & 2.82  & -40.2  & 4.2  & 19.95  & 0.309±0.001 & 0.142±0.001 & 0.23±0.01   & 2.10±0.09   & 2.1±0.5    & 31.5 & 103           & 46            & 6           & 0.011 & $\checkmark$       \\
33  & I14498-5856 & 3.16  & -49.3  & 4.4  & 25.12  & 0.234±0.003 & 0.124±0.001 & 0.65±0.03   & 3.03±0.12   & 2.4±0.6    & 34.5 & 115           & 125           & 4           & 0.011 & $\checkmark$       \\
34  & I15122-5801 & 9.26  & -60.9  & 5    & 12.59  & 0.179±0.004 & 0.051±0.005 & 0.160±0.011 & 2.83±0.19   & 4.1±0.5    & 22.2 & 41            & 19            & 2           & 0.011 &         \\
35  & I15254-5621 & 4     & -67.3  & 5.1  & 100    & 0.345±0.002 & 0.219±0.002 & 0.098±0.009 & 3.0±0.3     & 1.77±0.17  & 38.9 & 131           & 55            & 2           & 0.011 & $\checkmark$       \\
36  & I15290-5546 & 6.76  & -87.5  & 5.7  & 79.43  & 1.187±0.009 & 0.346±0.003 & 1.88±0.04   & 10.87±0.22  & 2.05±0.22  & 22.6 & 53            & 102           & 4           & 0.011 &         \\
38  & I15394-5358 & 1.82  & -41.6  & 3.7  & 6.31   & 0.487±0.003 & 0.338±0.002 & 0.936±0.024 & 3.8±0.1     & 2.8±0.5    & 41.0 & 128           & 73            & 8           & 0.011 & $\checkmark$       \\
39  & I15408-5356 & 1.82  & -39.7  & 4.9  & 100    & -           & -           & -           & -           & -          & -    & -             & -             & -           & 0.012 &         \\
40  & I15411-5352 & 1.82  & -41.5  & 4.5  & 63.1   & 0.447±0.002 & 0.345±0.002 & 0.328±0.013 & 2.5±0.1     & 1.54±0.06  & 43.6 & 171           & 117           & 2           & 0.012 &         \\
41  & I15437-5343 & 4.98  & -83    & 4.6  & 39.81  & 0.230±0.002 & 0.103±0.002 & 0.39±0.03   & 2.74±0.21   & 1.46±0.16  & 31.0 & 161           & 83            & 2           & 0.012 & $\checkmark$       \\
42  & I15439-5449 & 3.29  & -54.6  & 4.4  & 25.12  & 0.224±0.002 & 0.085±0.002 & 0.312±0.013 & 1.96±0.08   & 2.5±0.1    & 27.6 & 73            & 35            & 4           & 0.012 &         \\
43  & I15502-5302 & 5.8   & -91.4  & 5.8  & 125.89 & 0.237±0.003 & 0.097±0.007 & 0.321±0.012 & 2.8±0.1     & 3.6±0.3    & 29.0 & 46            & 14            & 4           & 0.012 &         \\
44  & I15520-5234 & 2.65  & -41.3  & 5.1  & 79.43  & -           & -           & -           & -           & -          & -    & -             & -             & -           & 0.012 & $\checkmark$       \\
45  & I15522-5411 & 2.73  & -46.7  & 3.8  & 7.94   & 0.285±0.002 & 0.148±0.001 & 0.23±0.01   & 2.18±0.09   & 0.83±0.06  & 34.1 & 132           & 114           & 2           & 0.012 &         \\
46  & I15557-5215 & 4.03  & -67.6  & 3.9  & 5.01   & 0.583±0.003 & 0.344±0.004 & 0.040±0.006 & 1.01±0.15   & 1.12±0.14  & 37.1 & 124           & 53            & 6           & 0.012 & $\checkmark$       \\
47  & I15567-5236 & 5.99  & -107.1 & 5.7  & 158.49 & 0.368±0.002 & 0.258±0.001 & 0.84±0.03   & 7.7±0.3     & 1.53±0.07  & 41.2 & 51            & 126           & 4           & 0.012 &         \\
49  & I15584-5247 & 4.41  & -76.8  & 4.2  & 12.59  & 0.379±0.002 & 0.131±0.002 & 0.163±0.008 & 1.67±0.08   & 1.74±0.08  & 25.7 & 113           & 50            & 4           & 0.012 & $\checkmark$       \\
50  & I15596-5301 & 10.11 & -72.1  & 5.5  & 39.81  & 0.336±0.001 & 0.160±0.001 & 0.356±0.011 & 3.4±0.1     & 1.88±0.22  & 32.3 & 184           & 105           & 4           & 0.012 & $\checkmark$       \\
51  & I16026-5035 & 4.53  & -78.3  & 4.9  & 79.43  & 0.237±0.002 & 0.171±0.002 & 0.170±0.011 & 2.98±0.19   & 1.33±0.17  & 41.9 & 108           & 66            & 2           & 0.012 &         \\
52  & I16037-5223 & 9.84  & -80    & 5.6  & 63.1   & 0.603±0.002 & 0.307±0.001 & 0.867±0.021 & 4.51±0.11   & 2.40±0.21  & 33.7 & 143           & 108           & 2           & 0.012 &         \\
53  & I16060-5146 & 5.3   & -91.6  & 5.8  & 79.43  & -           & -           & -           & -           & -          & -    & -             & -             & -           & 0.013 & $\checkmark$       \\
54  & I16065-5158 & 3.98  & -63.3  & 5.4  & 50.12  & 0.940±0.004 & 0.420±0.003 & 1.63±0.04   & 7.6±0.2     & 2.6±0.3    & 30.9 & 161           & 175           & 4           & 0.012 & $\checkmark$       \\
55  & I16071-5142 & 5.3   & -87    & 4.8  & 12.59  & 0.652±0.007 & 0.605±0.007 & 0.16±0.01   & 2.07±0.13   & 1.72±0.23  & 48.1 & 250           & 131           & 4           & 0.013 & $\checkmark$       \\
56  & I16076-5134 & 5.3   & -87.7  & 5.3  & 50.12  & 1.312±0.003 & 1.022±0.002 & 0.555±0.017 & 3.65±0.11   & 2.5±0.1    & 43.8 & 122           & 85            & 4           & 0.012 & $\checkmark$       \\
57  & I16119-5048 & 3.1   & -48.2  & 4.3  & 12.59  & 0.549±0.002 & 0.424±0.002 & 0.590±0.013 & 2.61±0.06   & 1.23±0.06  & 43.6 & 90            & 50            & 4           & 0.012 &         \\
59  & I16158-5055 & 3.57  & -49.2  & 5.2  & 50.12  & 0.184±0.002 & 0.088±0.002 & 0.11±0.01   & 2.46±0.24   & 1.98±0.13  & 32.3 & 54            & 24            & 4           & 0.012 & $\checkmark$       \\
60  & I16164-5046 & 3.57  & -57.3  & 5.5  & 63.1   & 0.676±0.001 & 0.363±0.002 & 0.677±0.021 & 3.72±0.12   & 2.81±0.13  & 34.9 & 208           & 90            & 4           & 0.014 & $\checkmark$       \\
61  & I16172-5028 & 3.57  & -51.9  & 4.3  & 63.1   & -           & -           & -           & -           & -          & -    & -             & -             & -           & 0.023 & $\checkmark$       \\
62  & I16177-5018 & 3.57  & -50.2  & 5.5  & 79.43  & 0.272±0.002 & 0.123±0.009 & 0.164±0.015 & 2.43±0.22   & 2.51±0.14  & 31.2 & 55            & 4             & 1           & 0.013 &         \\
63  & I16272-4837 & 2.92  & -46.6  & 4.3  & 12.59  & 0.650±0.002 & 0.316±0.002 & 0.120±0.009 & 2.91±0.21   & 2.1±0.4    & 32.7 & 223           & 93            & 6           & 0.012 & $\checkmark$       \\
64  & I16297-4757 & 5.03  & -79.6  & 4.9  & 31.62  & 0.349±0.005 & 0.381±0.003 & 0.816±0.016 & 3.74±0.07   & 2.50±0.08  & 52.2 & 146           & 114           & 2           & 0.012 & $\checkmark$       \\
65  & I16304-4710 & 11.32 & -62.8  & 4.9  & 25.12  & 0.282±0.003 & 0.130±0.001 & 0.295±0.011 & 2.9±0.1     & 1.84±0.09  & 31.5 & 44            & 37            & 2           & 0.012 &         \\
66  & I16313-4729 & 4.71  & -73.7  & 6.7  & 100    & 0.464±0.001 & 0.166±0.001 & 0.888±0.017 & 5.3±0.1     & 2.15±0.18  & 26.4 & 113           & 63            & 2           & 0.012 &         \\
67  & I16318-4724 & 7.68  & -119.8 & 5.2  & 31.62  & 0.606±0.005 & 0.318±0.002 & 1.141±0.023 & 5.34±0.11   & 2.68±0.12  & 34.4 & 85            & 100           & 4           & 0.012 &         \\
68  & I16330-4725 & 10.99 & -75.1  & 6.3  & 100    & 0.337±0.002 & 0.160±0.002 & 1.78±0.05   & 3.7±0.1     & 4.37±0.17  & 32.2 & 59            & 26            & 2           & 0.012 &         \\
69  & I16344-4658 & 12.09 & -49.5  & 5.4  & 19.95  & 0.490±0.005 & 0.340±0.003 & 1.33±0.03   & 5.23±0.11   & 3.57±0.16  & 41.0 & 154           & 138           & 2           & 0.012 & $\checkmark$       \\
70  & I16348-4654 & 12.09 & -46.5  & 5.4  & 10     & 0.583±0.003 & 0.424±0.002 & 0.660±0.018 & 4.40±0.12   & 2.0±0.3    & 42.1 & 183           & 104           & 4           & 0.012 & $\checkmark$       \\
71  & I16351-4722 & 3.02  & -41.4  & 4.9  & 50.12  & 1.213±0.002 & 0.489±0.002 & 0.83±0.09   & 7.5±0.8     & 1.55±0.09  & 28.7 & 76            & 137           & 6           & 0.012 & $\checkmark$       \\
72  & I16362-4639 & 3.01  & -38.8  & 3.6  & 12.59  & 0.338±0.003 & 0.070±0.002 & 0.102±0.006 & 1.78±0.11   & 2.26±0.11  & 17.1 & 70            & 25            & 1           & 0.012 &         \\
73  & I16372-4545 & 4.16  & -57.3  & 4.2  & 19.95  & 0.262±0.002 & 0.094±0.002 & 0.49±0.03   & 3.13±0.17   & 2.46±0.12  & 26.4 & 70            & 26            & 2           & 0.012 & $\checkmark$       \\
74  & I16385-4619 & 7.11  & -117   & 5.1  & 79.43  & 0.456±0.005 & 0.309±0.003 & 0.337±0.016 & 3.76±0.18   & 2.2±0.1    & 40.4 & 70            & 63            & 4           & 0.012 &         \\
75  & I16424-4531 & 2.63  & -34.2  & 3.9  & 15.85  & -           & -           & -           & -           & -          & -    & -             & -             & -           & 0.012 & $\checkmark$       \\
76  & I16445-4459 & 7.95  & -121.3 & 5    & 12.59  & 0.252±0.002 & 0.172±0.001 & 0.272±0.012 & 2.55±0.11   & 1.01±0.11  & 40.7 & 40            & 18            & 4           & 0.012 &         \\        
\end{tabular}
\end{threeparttable}
}
\end{table*}
\begin{table*}
\renewcommand\tabcolsep{2pt}
\scalebox{0.86}{
\begin{threeparttable}
\begin{tabular}{p{0.5cm}cccccp{1.6cm}p{1.6cm}lp{1.3cm}lcp{0.5cm}lccc}
\hline
\hline
ID  & \multicolumn{1}{c}{IRAS}   
&  Distance  & $v_{\rm LSR}$   
& \multicolumn{1}{c}{log $L_\textup{bol}$} 
& \multicolumn{1}{c}{$L_\textup{bol}/M$} &  
\multicolumn{2}{c}{SiO(2-1)}  & \multicolumn{3}{c}{H$^{13}$CO$^{+}$(1-0)} & Contribution$_{\rm N}$ & \multicolumn{3}{c}{Number}  &\multicolumn{1}{c}{$\sigma$ rms} 
& Outflow$^{\rm a}$ \\ 
\cline{7-11} \cline{13-15} 
 &  &   &   &  & &   
\multicolumn{1}{c}{$\int{T_\textup{B}\mathrm{d}{\upsilon}}$}   &  \multicolumn{1}{c}{$\int{T_\textup{N}\mathrm{d}{\upsilon}}$}
& 
\multicolumn{1}{c}{$\int{T_\textup{B}\mathrm{d}{\upsilon}}$}   
&  FWHM$_{\rm B}$  &FWHM$_{\rm N}$   & & ${\rm B}$ 
& ${\rm N}$
& Area \\
&
&\multicolumn{1}{c}{(kpc)}         
&\multicolumn{1}{c}{(km s$^{-1}$)}  
&\multicolumn{1}{c}{($L_{\odot}$)} 
& \multicolumn{1}{c}{($L_{\odot}$ $M_{\odot}$$^{-1}$)} 
&\multicolumn{1}{c}{(Jy km s$^{-1}$)} 
&\multicolumn{1}{c}{(Jy km s$^{-1}$)}
&\multicolumn{1}{c}{(Jy km s$^{-1}$)}
&\multicolumn{1}{c}{(km s$^{-1}$)}
&\multicolumn{1}{c}{(km s$^{-1}$)}
&\multicolumn{1}{c}{($\%$)}
& & & 
&\multicolumn{1}{c}{(Jy km s$^{-1}$)}
\\
\hline
77  & I16458-4512 & 3.56  & -50.4  & 4.5  & 7.94   & 0.380±0.002 & 0.165±0.002 & 0.15±0.01   & 2.86±0.19   & 1.57±0.17  & 30.2 & 122           & 65            & 6           & 0.012 & $\checkmark$       \\
78  & I16484-4603 & 2.1   & -32    & 5    & 100    & 0.451±0.002 & 0.360±0.002 & 0.43±0.01   & 3.01±0.07   & 1.15±0.09  & 44.4 & 83            & 31            & 6           & 0.012 & $\checkmark$       \\
79  & I16487-4423 & 3.26  & -43.4  & 4.4  & 25.12  & 0.310±0.001 & 0.123±0.001 & 0.266±0.012 & 2.92±0.13   & 0.59±0.12  & 28.5 & 116           & 54            & 4           & 0.012 & $\checkmark$       \\
80  & I16489-4431 & 3.26  & -41.3  & 3.8  & 7.94   & 0.588±0.002 & 0.285±0.002 & 0.31±0.05   & 4.0±0.6     & 2.30±0.17  & 32.6 & 134           & 71            & 4           & 0.012 & $\checkmark$       \\
81  & I16506-4512 & 2.42  & -26.2  & 5    & 79.43  & 0.155±0.002 & 0.056±0.001 & 0.075±0.008 & 1.89±0.21   & 1.23±0.11  & 26.4 & 35            & 19            & 4           & 0.012 &         \\
82  & I16524-4300 & 3.43  & -40.8  & 4.4  & 10     & 0.244±0.001 & 0.134±0.001 & 0.480±0.019 & 1.92±0.08   & 1.6±0.3    & 35.5 & 190           & 45            & 6           & 0.012 & $\checkmark$       \\
83  & I16547-4247 & 2.74  & -30.4  & 4.8  & 39.81  & -           & -           & -           & -           & -          & -    & -             & -             & -           & 0.012 & $\checkmark$       \\
84  & I16562-3959 & 2.38  & -12.6  & 5.7  & 316.23 & -           & -           & -           & -           & -          & -    & -             & -             & -           & 0.012 & $\checkmark$       \\
85  & I16571-4029 & 2.38  & -15    & 4.3  & 25.12  & 0.903±0.001 & 0.547±0.001 & 0.315±0.019 & 3.98±0.23   & 2.9±0.2    & 37.7 & 128           & 85            & 4           & 0.012 & $\checkmark$       \\
86  & I17006-4215 & 2.21  & -23.2  & 4.4  & 39.81  & 0.346±0.002 & 0.234±0.004 & 0.071±0.006 & 1.02±0.09   & 2.6±0.3    & 40.3 & 87            & 25            & 4           & 0.012 & $\checkmark$       \\
87  & I17008-4040 & 2.38  & -17    & 4.6  & 39.81  & -           & -           & -           & -           & -          & -    & -             & -             & -           & 0.012 & $\checkmark$       \\
88  & I17016-4124 & 1.37  & -27.1  & 5.3  & 31.62  & -           & -           & -           & -           & -          & -    & -             & -             & -           & 0.012 & $\checkmark$       \\
89  & I17136-3617 & 1.34  & -10.6  & 4.6  & 125.89 & 0.353±0.001 & 0           & 0.278±0.013 & 2.25±0.11   & 1.5±0.1    & 0.0  & 81            & 81            & 2           & 0.011 & $\checkmark$       \\
90  & I17143-3700 & 12.67 & -31.1  & 5.6  & 63.1   & 0.333±0.003 & 0.189±0.002 & 1.51±0.03   & 4.85±0.09   & 2.69±0.08  & 36.2 & 114           & 107           & 2           & 0.011 & $\checkmark$       \\
91  & I17158-3901 & 3.38  & -15.2  & 4.8  & 25.12  & 0.346±0.001 & 0.214±0.002 & 0.032±0.005 & 1.18±0.19   & 3.1±0.3    & 38.3 & 172           & 34            & 6           & 0.011 & $\checkmark$       \\
92  & I17160-3707 & 10.53 & -69.5  & 6    & 79.43  & 0.372±0.003 & 0.162±0.005 & 0.447±0.013 & 3.88±0.11   & 1.83±0.19  & 30.4 & 97            & 45            & 4           & 0.011 & $\checkmark$       \\
93  & I17175-3544 & 1.34  & -5.7   & 4.8  & 50.12  & -           & -           & -           & -           & -          & -    & -             & -             & -           & 0.013 & $\checkmark$       \\
94  & I17204-3636 & 3.32  & -18.2  & 4.2  & 19.95  & 0.375±0.001 & 0.153±0.001 & 0.447±0.014 & 2.65±0.08   & 1.42±0.18  & 28.9 & 84            & 44            & 4           & 0.011 & $\checkmark$       \\
95  & I17220-3609 & 8.01  & -93.7  & 5.7  & 25.12  & -           & -           & -           & -           & -          & -    & -             & -             & -           & 0.013 & $\checkmark$       \\
96  & I17233-3606 & 1.34  & -2.7   & 4.6  & 39.81  & -           & -           & -           & -           & -          & -    & -             & -             & -           & 0.011 & $\checkmark$       \\
97  & I17244-3536 & 1.36  & -10.2  & 3.7  & 39.81  & 0.196±0.001 & 0.068±0.002 & 0.119±0.011 & 1.2±0.1     & 1.01±0.24  & 25.7 & 68            & 7             & 4           & 0.011 &         \\
98  & I17258-3637 & 2.59  & -11.9  & 5.9  & 398.11 & 0.246±0.006 & 0.130±0.002 & 0.686±0.022 & 2.32±0.07   & 1.5±0.1    & 34.5 & 105           & 55            & 6           & 0.012 & $\checkmark$       \\
99  & I17269-3312 & 4.38  & -21    & 4.7  & 12.59  & 0.320±0.002 & 0.221±0.003 & 0.036±0.006 & 1.5±0.3     & 1.7±0.3    & 40.8 & 158           & 46            & 4           & 0.011 & $\checkmark$       \\
100 & I17271-3439 & 3.1   & -18.2  & 5.6  & 39.81  & 0.439±0.001 & 0.208±0.001 & 0.177±0.018 & 1.76±0.18   & 3.3±0.3    & 32.2 & 148           & 66            & 8           & 0.011 & $\checkmark$       \\
101 & I17278-3541 & 1.33  & -0.4   & 3.8  & 19.95  & 0.605±0.006 & 0.327±0.003 & 0.112±0.012 & 8.1±0.9     & 1.22±0.07  & 35.1 & 85            & 110           & 6           & 0.012 & $\checkmark$       \\
102 & I17439-2845 & 8     & 18.7   & 5.7  & 79.43  & 0.392±0.001 & 0.108±0.001 & 0.10±0.01   & 4.5±0.5     & 2.04±0.05  & 21.6 & 51            & 22            & 4           & 0.009 & $\checkmark$       \\
104 & I17455-2800 & 10    & -15.6  & 5.9  & 63.1   & 0.293±0.001 & 0.151±0.001 & 0.140±0.007 & 2.87±0.15   & 1.76±0.14  & 34.0 & 116           & 61            & 4           & 0.009 &         \\
105 & I17545-2357 & 2.93  & 7.9    & 4.1  & 10     & 0.279±0.003 & 0.126±0.002 & 0.098±0.011 & 4.6±0.5     & 1.1±0.3    & 31.0 & 86            & 78            & 4           & 0.011 & $\checkmark$       \\
106 & I17589-2312 & 2.97  & 21.3   & 4    & 10     & 0.308±0.001 & 0.133±0.001 & 0.205±0.012 & 3.66±0.21   & 1.93±0.05  & 30.2 & 84            & 75            & 6           & 0.011 & $\checkmark$       \\
107 & I17599-2148 & 2.99  & 18.6   & 5.2  & 63.1   & 0.320±0.003 & 0.160±0.002 & 0.641±0.015 & 2.37±0.05   & 1.98±0.06  & 33.3 & 39            & 18            & 2           & 0.011 & $\checkmark$       \\
108 & I18032-2032 & 5.15  & 4.3    & 5.4  & 79.43  & 0.950±0.001 & 0.268±0.001 & 1.42±0.04   & 4.59±0.12   & 3.3±0.3    & 22.0 & 97            & 81            & 6           & 0.01  & $\checkmark$       \\
109 & I18056-1952 & 8.55  & 66.7   & 5.7  & 19.95  & 1.002±0.003 & 0.396±0.001 & 3.04±0.08   & 12.2±0.3    & 4.5±0.3    & 28.3 & 67            & 189           & 4           & 0.011 & $\checkmark$       \\
111 & I18079-1756 & 1.83  & 18     & 3.9  & 19.95  & 0.148±0.001 & 0.107±0.003 & 0.054±0.009 & 0.79±0.13   & 2.61±0.14  & 42.0 & 177           & 26            & 4           & 0.011 & $\checkmark$       \\
112 & I18089-1732 & 2.5   & 33.5   & 4.3  & 15.85  & 0.334±0.002 & 0.288±0.002 & 0.347±0.018 & 2.31±0.12   & 2.52±0.21  & 46.3 & 73            & 45            & 4           & 0.011 & $\checkmark$       \\
113 & I18110-1854 & 3.37  & 37     & 4.8  & 39.81  & 0.273±0.004 & 0.097±0.001 & 0.148±0.012 & 3.4±0.3     & 6.2±0.4    & 26.1 & 14            & 57            & 4           & 0.011 & $\checkmark$       \\
114 & I18116-1646 & 3.94  & 48.5   & 5.1  & 100    & 0.279±0.003 & 0.102±0.003 & 0.498±0.015 & 2.98±0.09   & 2.6±0.2    & 26.8 & 17            & 18            & 2           & 0.011 & $\checkmark$       \\
115 & I18117-1753 & 2.57  & 36.7   & 4.6  & 12.59  & 0.347±0.004 & 0.142±0.002 & 0.36±0.02   & 5.1±0.3     & 2.6±0.3    & 29.1 & 162           & 192           & 6           & 0.011 & $\checkmark$       \\
116 & I18134-1942 & 1.25  & 10.6   & 3    & 3.98   & 0.079±0.001 & 0.072±0.001 & 0.028±0.011 & 1.9±0.7     & 1.6±0.3    & 47.8 & 5             & 25            & 2           & 0.011 &         \\
117 & I18139-1842 & 3.02  & 39.8   & 4.9  & 251.19 & 0.409±0.003 & 0.181±0.002 & 0.500±0.018 & 3.52±0.13   & 0.87±0.08  & 30.7 & 136           & 116           & 2           & 0.011 & $\checkmark$       \\
118 & I18159-1648 & 1.48  & 22     & 3.8  & 10     & 0.397±0.001 & 0.237±0.002 & 0.082±0.009 & 1.58±0.18   & 1.52±0.07  & 37.4 & 173           & 59            & 6           & 0.011 & $\checkmark$       \\
119 & I18182-1433 & 4.71  & 59.1   & 4.3  & 15.85  & 0.441±0.002 & 0.282±0.001 & 0.82±0.05   & 3.74±0.21   & 2.3±0.2    & 39.1 & 82            & 81            & 4           & 0.011 & $\checkmark$       \\
120 & I18223-1243 & 3.37  & 44.8   & 4.2  & 19.95  & 0.094±0.001 & 0.056±0.001 & 0.333±0.024 & 4.0±0.3     & 1.28±0.07  & 37.2 & 1             & 13            & 2           & 0.011 & $\checkmark$       \\
121 & I18228-1312 & 3.21  & 32.3   & 4.7  & 50.12  & 0.083±0.001 & 0.023±0.001 & 0.05±0.01   & 0.74±0.16   & 3.5±0.3    & 21.8 & 50            & 7             & 2           & 0.011 &         \\
122 & I18236-1205 & 2.17  & 25.9   & 3.5  & 3.98   & 0.344±0.004 & 0.181±0.005 & 0.216±0.015 & 3.9±0.3     & 1.08±0.13  & 34.5 & 124           & 60            & 4           & 0.011 & $\checkmark$       \\
123 & I18264-1152 & 3.33  & 43.2   & 3.9  & 5.01   & 0.446±0.003 & 0.270±0.005 & 0.18±0.01   & 2.36±0.13   & 1.56±0.14  & 37.7 & 112           & 54            & 6           & 0.011 & $\checkmark$       \\
124 & I18290-0924 & 5.34  & 84.2   & 4    & 6.31   & 0.194±0.001 & 0.069±0.001 & 0.137±0.015 & 3.0±0.3     & 3.39±0.23  & 26.4 & 72            & 16            & 4           & 0.01  & $\checkmark$       \\
126 & I18311-0809 & 6.06  & 113    & 5    & 19.95  & 0.341±0.001 & 0.135±0.001 & 0.414±0.013 & 2.74±0.09   & 4.7±0.3    & 28.4 & 54            & 28            & 4           & 0.01  &         \\
128 & I18316-0602 & 2.09  & 42.8   & 4    & 12.59  & -           & -           & -           & -           & -          & -    & -             & -             & -           & 0.01  & $\checkmark$       \\
129 & I18317-0513 & 2.18  & 42.1   & 4.8  & 39.81  & 1.000±0.003 & 0.082±0.001 & 0.103±0.009 & 3.5±0.3     & 1.70±0.15  & 7.6  & 24            & 64            & 2           & 0.01  & $\checkmark$       \\
130 & I18317-0757 & 4.79  & 80.7   & 5.2  & 63.1   & 0.260±0.003 & 0.103±0.003 & 0.236±0.012 & 3.44±0.17   & 2.03±0.08  & 28.4 & 32            & 22            & 2           & 0.01  &         \\
131 & I18341-0727 & 6.04  & 112.7  & 5.1  & 25.12  & 0.315±0.007 & 0.184±0.006 & 0.231±0.008 & 3.43±0.11   & 1.7±0.3    & 36.9 & 103           & 63            & 4           & 0.01  & $\checkmark$       \\
132 & I18411-0338 & 7.41  & 102.8  & 5.1  & 25.12  & 0.282±0.002 & 0.146±0.001 & 0.490±0.015 & 4.81±0.15   & 2.3±0.2    & 34.1 & 60            & 70            & 4           & 0.01  & $\checkmark$       \\
133 & I18434-0242 & 5.16  & 97.2   & 5.7  & 125.89 & 0.590±0.001 & 0.327±0.001 & 0.423±0.015 & 2.13±0.08   & 1.74±0.05  & 35.7 & 96            & 63            & 4           & 0.01  & $\checkmark$       \\
135 & I18445-0222 & 5.16  & 86.9   & 4.6  & 15.85  & 0.233±0.002 & 0.094±0.002 & 0.509±0.027 & 3.50±0.19   & 0.81±0.09  & 28.6 & 17            & 14            & 2           & 0.01  & $\checkmark$       \\
136 & I18461-0113 & 5.16  & 96.1   & 4.4  & 19.95  & 0.314±0.002 & 0.243±0.001 & 0.27±0.01   & 3.22±0.12   & 2.10±0.12  & 43.7 & 101           & 82            & 4           & 0.01  & $\checkmark$       \\
137 & I18469-0132 & 5.16  & 87     & 4.8  & 63.1   & 0.239±0.001 & 0.202±0.001 & 0.168±0.007 & 2.01±0.08   & 2.53±0.09  & 45.7 & 92            & 42            & 4           & 0.01  & $\checkmark$       \\
138 & I18479-0005 & 12.96 & 14.6   & 6.1  & 79.43  & -           & -           & -           & -           & -          & -    & -             & -             & -           & 0.01  &         \\
140 & I18507+0110 & 1.56  & 57.2   & 4.8  & 39.81  & -           & -           & -           & -           & -          & -    & -             & -             & -           & 0.011 & $\checkmark$       \\
141 & I18507+0121 & 1.56  & 57.9   & 3.5  & 7.94   & -           & -           & -           & -           & -          & -    & -             & -             & -           & 0.01  & $\checkmark$       \\
142 & I18517+0437 & 2.36  & 43.9   & 4.6  & 50.12  & 0.135±0.001 & 0.079±0.001 & 0.34±0.01   & 1.61±0.05   & 1.49±0.23  & 37.0 & 92            & 41            & 6           & 0.01  & $\checkmark$       \\
143 & I18530+0215 & 4.67  & 74.1   & 4.7  & 25.12  & 0.157±0.002 & 0.095±0.001 & 0.359±0.012 & 2.15±0.07   & 2.6±0.3    & 37.7 & 65            & 46            & 4           & 0.01  & $\checkmark$       \\
144 & I19078+0901 & 11.11 & 2.9    & 6.9  & 79.43  & -           & -           & -           & -           & -          & -    & -             & -             & -           & 0.011 &         \\
145 & I19095+0930 & 6.02  & 43.7   & 5.1  & 100    & 0.443±0.002 & 0.248±0.001 & 0.599±0.019 & 7.36±0.24   & 4.5±0.5    & 35.9 & 82            & 137           & 4           & 0.011 & $\checkmark$       \\
146 & I19097+0847 & 8.47  & 58     & 5    & 15.85  & 0.425±0.002 & 0.319±0.001 & 0.303±0.014 & 3.90±0.17   & 1.2±0.3    & 42.9 & 70            & 38            & 4           & 0.011 &        
\end{tabular}
\end{threeparttable}
}
 \begin{tablenotes}
        \footnotesize
        \item[] Symbol ‘-’ indicates that this source exhibits absorption features.
        \item[a] The identified outflow source from Baug et al. (in prep).
  \end{tablenotes}
\end{table*}

\begin{table*}
\renewcommand\tabcolsep{2.2pt}
\caption{Basic parameters of the velocity decomposed sources in the ATOMS survey.}
\label{tab:TableA2}
\begin{threeparttable}
\scalebox{0.84}{
\begin{tabular}{ccllllllccccccccc}
\hline
\hline
ID  & IRAS       
& \multicolumn{2}{c}{$L_\textup{SiO}$$\times{10^{-5}}$}   
& \multicolumn{2}{c}{$N$(SiO)$\times{10^{12}}$}  & \multicolumn{2}{c}{$X$(SiO)$\times{10^{-10}}$}   
& Filaments 
& \multicolumn{3}{c}{Infrared emission }  & \textit{A} group 
& \textit{B} group 
& \multicolumn{3}{c}{UC H\textsc{ii}}\\
 &   & Broad   & Narrow   & Broad   & Narrow  & $B$ position    & $N$ position  &  & 4.5$\upmu$m & 8$\upmu$m & 24$\upmu$m & &  
& \textit{concident} 
& \textit{offset}
& \textit{surrounded} \\
&   
&\multicolumn{2}{c}{($L_{\odot}$)}   
&\multicolumn{2}{c}{(cm$^{-2}$)}
& &    &   &  &  & & &  \\
\hline\\
1   & I08076-3556 & 0.216±0.001   & 0.224±0.001   & 13.56±0.06 & 5.62±0.02   & 5.9±0.4   & 16.8±0.8  &           & $\checkmark$     & $\checkmark$   &      & $\checkmark$ &   &            &        &            \\
2   & I08303-4303 & 7.25±0.02     & 1.45±0.02     & 13.75±0.05 & 1.10±0.01   & 2.3±0.1   & 7.6±0.4   &           & $\checkmark$     & -   & -    &   & $\checkmark$ &            &        &            \\
3   & I08448-4343 & 0.367±0.002   & 0.148±0.002   & 7.50±0.04  & 1.21±0.01   & 2.5±0.2   & 7.2±0.3   & N         &       &     &      & $\checkmark$ &   &            &        &            \\
4   & I08470-4243 & 4.21±0.04     & 1.80±0.05     & 9.57±0.09  & 1.64±0.04   & 3.4±0.3   & 11.7±1.5  & N         & $\checkmark$     & $\checkmark$   & $\checkmark$    & $\checkmark$ &   &            &        &            \\
5   & I09002-4732 & 0.589±0.004   & 0.305±0.004   & 4.10±0.03  & 0.85±0.01   & 2.5±0.1   & 5.8±0.3   &           & $\checkmark$     & $\checkmark$   &      & $\checkmark$ &   &            &        & $\checkmark$          \\
6   & I09018-4816 & 3.56±0.03     & 1.91±0.02     & 5.27±0.04  & 1.14±0.01   & 0.8±0.1   & 4.7±0.5   &           & $\checkmark$     & $\checkmark$   & -    & $\checkmark$ &   &            &        &            \\
7   & I09094-4803 & 42.9±0.6      & 16.8±0.5      & 4.67±0.07  & 0.73±0.02   & 2.0±0.1   & 4.5±0.9   & N         & $\checkmark$     & $\checkmark$   & -    & $\checkmark$ &   &            &        &            \\
8   & I10365-5803 & 1.84±0.03     & 1.24±0.04     & 3.20±0.06  & 0.87±0.03   & 0.7±0.3   & 2.4±0.9   &           & $\checkmark$     & $\checkmark$   & $\checkmark$    & $\checkmark$ &   &            &        &            \\
9   & I11298-6155 & 56.0±0.6      & 35.50±0.17    & 5.61±0.06  & 1.43±0.01   & 0.9±0.3   & 20.0±0.8  & N         & $\checkmark$     & $\checkmark$   &      &   & $\checkmark$ & $\checkmark$          &        &            \\
10  & I11332-6258 & 2.57±0.05     & 0.75±0.02     & 7.12±0.13  & 0.83±0.02   & 0.9±0.3   & 5.1±2.5   &           & $\checkmark$     & $\checkmark$   & -    &   & $\checkmark$ &            &        &            \\
11  & I11590-6452 & 0.0270±0.0003 & 0.0220±0.0002 & 1.68±0.02  & 0.554±0.004 & 2.5±0.1   & 12.4±0.3  &           & $\checkmark$     & $\checkmark$   & $\checkmark$    &   & $\checkmark$ &            &        &            \\
12  & I12320-6122 & 4.25±0.07     & 2.21±0.07     & 3.62±0.06  & 0.76±0.03   & 1.4±0.5   & 2.7±0.8   &           & $\checkmark$     & $\checkmark$   &      & $\checkmark$ &   &            &        & $\checkmark$          \\
13  & I12326-6245 & 24.43±0.14    & 15.14±0.02    & 11.52±0.07 & 2.867±0.004 & 2.2±0.2   & 38.0±1.1  &           & $\checkmark$     & $\checkmark$   &      &   & $\checkmark$ &            & $\checkmark$      &            \\
14  & I12383-6128 & 2.88±0.08     & 1.22±0.22     & 2.70±0.07  & 0.46±0.08   & 2.5±0.8   & 22±7      &           &       &     &      & $\checkmark$ &   &            & $\checkmark$      &            \\
15  & I12572-6316 & 32.6±0.4      & 16.8±0.4      & 2.44±0.03  & 0.51±0.01   & 0.9±0.3   & 3.4±0.5   &           & $\checkmark$     & $\checkmark$   &      & $\checkmark$ &   & $\checkmark$          &        &            \\
16  & I13079-6218 & 18.68±0.07    & 9.63±0.04     & 12.97±0.05 & 2.68±0.01   & 1.0±0.1   & 44.6±1.7  &           & $\checkmark$     &     & $\checkmark$    & $\checkmark$ &   &            &        &            \\
17  & I13080-6229 & 2.29±0.05     & 1.37±0.04     & 1.59±0.04  & 0.38±0.01   & 0.2±0.2   & 1.4±0.7   &           & $\checkmark$     & $\checkmark$   &      & $\checkmark$ &   &            & $\checkmark$      &            \\
18  & I13111-6228 & 2.85±0.04     & 1.15±0.05     & 1.98±0.02  & 0.32±0.01   & 0.2±0.1   & 1.9±0.7   &           & $\checkmark$     & $\checkmark$   &      & $\checkmark$ &   &            &        & $\checkmark$          \\
19  & I13134-6242 & 9.14±0.09     & 4.99±0.09     & 6.34±0.07  & 1.39±0.03   & 5.2±0.5   & 9.8±2.5   &           & $\checkmark$     & $\checkmark$   & $\checkmark$    & $\checkmark$ &   &            &        &            \\
20  & I13140-6226 & 6.35±0.04     & 3.06±0.02     & 4.41±0.03  & 0.85±0.01   & 0.9±0.2   & 35.7±2.9  &           & $\checkmark$     & $\checkmark$   & $\checkmark$    & $\checkmark$ &   &            &        &            \\
22  & I13291-6249 & 21.3±0.3      & 11.21±0.14    & 3.69±0.06  & 0.78±0.01   & 1.5±0.3   & 6.4±1.9   &           & $\checkmark$     & $\checkmark$   &      &   & $\checkmark$ &            & $\checkmark$      &            \\
23  & I13295-6152 & 2.44±0.06     & 1.18±0.05     & 1.62±0.04  & 0.32±0.01   & 0.2±0.3   & 4.8±1.1   &           & $\checkmark$     & $\checkmark$   & $\checkmark$    & $\checkmark$ &   &            &        &            \\
24  & I13471-6120 & 9.3±0.1       & 4.9±0.4       & 3.13±0.03  & 0.67±0.06   & 0.9±0.6   & 2.4±1.1   &           & $\checkmark$     & $\checkmark$   &      & $\checkmark$ &   &            &        & $\checkmark$          \\
25  & I13484-6100 & 15.3±0.5      & 8.30±0.22     & 5.25±0.17  & 1.15±0.03   & 0.3±0.3   & 25.9±4.5  &           & $\checkmark$     & $\checkmark$   & $\checkmark$    & $\checkmark$ &   &            &        &            \\
26  & I14013-6105 & 4.46±0.04     & 1.90±0.04     & 2.64±0.02  & 0.45±0.01   & 0.4±0.2   & 1.9±0.3   &           & $\checkmark$     & $\checkmark$   &      & $\checkmark$ &   &            & $\checkmark$      &            \\
28  & I14164-6028 & 4.73±0.07     & 2.55±0.05     & 4.66±0.07  & 1.01±0.02   & 0.4±0.3   & 4.6±1.3   &           & $\checkmark$     & $\checkmark$   & $\checkmark$    &   & $\checkmark$ &            &        &            \\
29  & I14206-6151 & 2.93±0.14     & 1.01±0.09     & 2.72±0.13  & 0.38±0.04   & 1.2±0.7   & 4.4±2.1   &           & $\checkmark$     & -   & -    &   & $\checkmark$ &            &        &            \\
30  & I14212-6131 & 6.43±0.06     & 4.01±0.05     & 5.44±0.05  & 1.36±0.02   & 1.1±0.3   & 9.2±1.5   &           & $\checkmark$     & $\checkmark$   & $\checkmark$    & $\checkmark$ &   &            &        &            \\
31  & I14382-6017 & 14.91±0.23    & 6.57±0.21     & 2.53±0.04  & 0.45±0.01   & 0.6±0.3   & 8.5±0.6   &           & $\checkmark$     & $\checkmark$   &      & $\checkmark$ &   & $\checkmark$          &        &            \\
32  & I14453-5912 & 3.21±0.01     & 1.48±0.01     & 4.05±0.02  & 0.75±0.01   & 1.1±0.1   & 20.9±0.4  &           & $\checkmark$     & $\checkmark$   &      & $\checkmark$ &   &            & $\checkmark$      &            \\
33  & I14498-5856 & 3.08±0.03     & 1.63±0.02     & 3.10±0.04  & 0.66±0.01   & 0.3±0.1   & 8.4±0.4   &           & $\checkmark$     & $\checkmark$   & $\checkmark$    &   & $\checkmark$ &            &        &            \\
34  & I15122-5801 & 20.6±0.5      & 5.9±0.5       & 2.41±0.06  & 0.28±0.03   & 0.9±0.4   & 2.7±0.8   &           & $\checkmark$     & $\checkmark$   & $\checkmark$    & $\checkmark$ &   &            &        &            \\
35  & I15254-5621 & 7.58±0.05     & 4.82±0.05     & 4.75±0.03  & 1.21±0.01   & 3.0±0.2   & 4.1±0.3   &           & $\checkmark$     & $\checkmark$   &      & $\checkmark$ &   &            & $\checkmark$      &            \\
36  & I15290-5546 & 75.9±0.6      & 22.14±0.18    & 16.65±0.13 & 1.95±0.02   & 0.5±0.2   & 26.5±1.9  &           & $\checkmark$     & $\checkmark$   &      &   & $\checkmark$ &            & $\checkmark$      &            \\
38  & I15394-5358 & 2.20±0.01     & 1.53±0.01     & 6.67±0.04  & 1.86±0.01   & 0.5±0.1   & 26±1      &           & $\checkmark$     & $\checkmark$   & $\checkmark$    & $\checkmark$ &   &            &        &            \\
40  & I15411-5352 & 2.02±0.01     & 1.56±0.01     & 6.11±0.03  & 1.90±0.01   & 1.2±0.2   & 4.4±0.5   &           & $\checkmark$     & $\checkmark$   &      & $\checkmark$ &   &            & $\checkmark$      &            \\
41  & I15437-5343 & 7.80±0.08     & 3.50±0.08     & 3.15±0.03  & 0.57±0.01   & 0.5±0.1   & 4.7±0.5   &           & $\checkmark$     & $\checkmark$   & $\checkmark$    & $\checkmark$ &   &            &        &            \\
42  & I15439-5449 & 3.32±0.03     & 1.27±0.03     & 3.08±0.03  & 0.47±0.01   & 0.6±0.1   & 1.5±0.4   &           & $\checkmark$     & $\checkmark$   &      & $\checkmark$ &   &            & $\checkmark$      &            \\
43  & I15502-5302 & 11.02±0.15    & 4.5±0.3       & 3.28±0.05  & 0.54±0.04   & 0.6±0.4   & 3.6±1.1   &           & $\checkmark$     & $\checkmark$   &      & $\checkmark$ &   &            & $\checkmark$      &            \\
45  & I15522-5411 & 2.94±0.02     & 1.52±0.01     & 3.96±0.03  & 0.82±0.01   & 1.0±0.2   & 7.8±0.8   &           & $\checkmark$     & $\checkmark$   & $\checkmark$    & $\checkmark$ &   &            &        &            \\
46  & I15557-5215 & 13.14±0.07    & 7.75±0.08     & 8.11±0.04  & 1.92±0.02   & 12.5±0.5  & 30.9±1.5  &           & $\checkmark$     & $\checkmark$   & $\checkmark$    & $\checkmark$ &   &            &        &            \\
47  & I15567-5236 & 18.51±0.09    & 12.96±0.04    & 5.17±0.02  & 1.45±0.01   & 0.4±0.1   & 3.2±0.3   &           & $\checkmark$     & $\checkmark$   &      &   & $\checkmark$ &            &        & $\checkmark$          \\
49  & I15584-5247 & 10.35±0.05    & 3.58±0.04     & 5.33±0.02  & 0.74±0.01   & 1.8±0.2   & 4.4±0.6   &           & $\checkmark$     & $\checkmark$   & $\checkmark$    & $\checkmark$ &   &            &        &            \\
50  & I15596-5301 & 48.1±0.2      & 22.87±0.19    & 4.71±0.02  & 0.90±0.01   & 0.8±0.1   & 7.3±0.3   &           & $\checkmark$     & $\checkmark$   & $\checkmark$    & $\checkmark$ &   &            &        &            \\
51  & I16026-5035 & 6.90±0.07     & 4.98±0.06     & 3.37±0.03  & 0.98±0.01   & 1.2±0.2   & 10.7±0.8  &           &       & -   & -    & $\checkmark$ &   &            &        &            \\
52  & I16037-5223 & 81.21±0.23    & 41.32±0.12    & 8.41±0.02  & 1.72±0.01   & 0.6±0.1   & 12.1±0.4  &           & $\checkmark$     & $\checkmark$   &      & $\checkmark$ &   &            & $\checkmark$      &            \\
54  & I16065-5158 & 21.2±0.1      & 9.48±0.07     & 13.44±0.06 & 2.41±0.02   & 0.5±0.1   & 9.4±0.4   &           & $\checkmark$     & $\checkmark$   &      &   & $\checkmark$ & $\checkmark$          &        &            \\
55  & I16071-5142 & 26.2±0.3      & 24.3±0.3      & 9.4±0.1    & 3.48±0.04   & 3.7±0.7   & 18.0±1.5  &           & $\checkmark$     & $\checkmark$   &      & $\checkmark$ &   &            & $\checkmark$      &            \\
56  & I16076-5134 & 52.82±0.12    & 41.12±0.08    & 18.85±0.04 & 5.89±0.01   & 2.1±0.2   & 10.4±0.6  &           & $\checkmark$     & $\checkmark$   & $\checkmark$    & $\checkmark$ &   &            &        &            \\
57  & I16119-5048 & 7.61±0.03     & 5.88±0.02     & 7.94±0.04  & 2.46±0.01   & 0.8±0.2   & 6.3±0.6   &           & $\checkmark$     & $\checkmark$   & $\checkmark$    & $\checkmark$ &   &            &        &            \\
59  & I16158-5055 & 3.39±0.04     & 1.61±0.04     & 2.66±0.03  & 0.51±0.01   & 1.4±0.2   & 2.1±0.4   &           & $\checkmark$     &     &      & $\checkmark$ &   &            & $\checkmark$      &            \\
60  & I16164-5046 & 12.47±0.03    & 6.69±0.04     & 9.81±0.02  & 2.11±0.01   & 0.8±0.1   & 2.6±0.1   &           & $\checkmark$     & $\checkmark$   &      & $\checkmark$ &   &            & $\checkmark$      &            \\
62  & I16177-5018 & 5.04±0.03     & 2.29±0.18     & 3.96±0.02  & 0.72±0.06   & 1.4±0.3   & 3±1       &           & $\checkmark$     & $\checkmark$   &      & $\checkmark$ &   &            & $\checkmark$      &            \\
63  & I16272-4837 & 8.56±0.02     & 4.15±0.03     & 10.06±0.03 & 1.96±0.02   & 4.5±0.2   & 25.2±0.5  &           & $\checkmark$     & $\checkmark$   & $\checkmark$    & $\checkmark$ &   &            &        &            \\
64  & I16297-4757 & 13.7±0.2      & 14.96±0.12    & 5.44±0.08  & 2.38±0.02   & 0.4±0.3   & 3.9±1.6   &           & $\checkmark$     & $\checkmark$   &      & $\checkmark$ &   &            & $\checkmark$      &            \\
65  & I16304-4710 & 56.3±0.6      & 25.91±0.24    & 4.40±0.05  & 0.81±0.01   & 0.8±0.2   & 3.7±0.8   &           & $\checkmark$     & $\checkmark$   &      & $\checkmark$ &   &            & $\checkmark$      &            \\
66  & I16313-4729 & 16.04±0.05    & 5.74±0.05     & 7.25±0.02  & 1.04±0.01   & 0.4±0.1   & 6.7±0.3   &           & $\checkmark$     & $\checkmark$   &      & $\checkmark$ &   & $\checkmark$          &        &            \\
67  & I16318-4724 & 55.7±0.5      & 29.25±0.17    & 9.47±0.08  & 2.00±0.01   & 0.4±0.2   & 6.8±1.1   &           & $\checkmark$     & $\checkmark$   &      &   & $\checkmark$ & $\checkmark$          &        &            \\
68  & I16330-4725 & 63.7±0.4      & 30.2±0.4      & 5.28±0.03  & 1.01±0.01   & 0.16±0.04 & 0.6±0.1   &           & $\checkmark$     & $\checkmark$   &      & $\checkmark$ &   &            & $\checkmark$      &            \\
69  & I16344-4658 & 112.1±1.3     & 77.9±0.6      & 7.69±0.09  & 2.14±0.02   & 0.3±0.2   & 7.5±1.8   &           & $\checkmark$     & $\checkmark$   & $\checkmark$    & $\checkmark$ &   &            &        &            \\
70  & I16348-4654 & 133.6±0.6     & 97.2±0.5      & 9.16±0.04  & 2.68±0.01   & 0.8±0.1   & 13.2±0.4  &           & $\checkmark$     & $\checkmark$   &      & $\checkmark$ &   &            & $\checkmark$      &            \\
71  & I16351-4722 & 17.29±0.03    & 6.97±0.02     & 19.01±0.03 & 3.08±0.01   & 1.17±0.02 & 9.8±0.3   &           & $\checkmark$     & $\checkmark$   &      &   & $\checkmark$ & $\checkmark$          &        &            \\
72  & I16362-4639 & 4.82±0.04     & 1.00±0.03     & 5.33±0.04  & 0.44±0.01   & 2.5±0.4   & 5±1       &           & $\checkmark$     & $\checkmark$   & $\checkmark$    & $\checkmark$ &   &            &        &            \\
73  & I16372-4545 & 7.19±0.07     & 2.58±0.06     & 4.16±0.04  & 0.60±0.02   & 0.4±0.1   & 1.8±0.4   &           & $\checkmark$     & $\checkmark$   & $\checkmark$    & $\checkmark$ &   &            &        &            \\
74  & I16385-4619 & 36.3±0.4      & 24.53±0.22    & 7.19±0.09  & 1.95±0.02   & 1.2±0.3   & 4.7±1.1   &           & $\checkmark$     & $\checkmark$   &      & $\checkmark$ &   &            & $\checkmark$      &            \\
76  & I16445-4459 & 25.24±0.15    & 17.28±0.09    & 4.00±0.02  & 1.10±0.01   & 0.8±0.1   & 11.6±0.6  &           & $\checkmark$     & $\checkmark$   &      & $\checkmark$ &   &            & $\checkmark$      &            \\
77  & I16458-4512 & 7.64±0.03     & 3.31±0.04     & 6.04±0.03  & 1.05±0.01   & 2.0±0.2   & 7.5±0.4   &           & $\checkmark$     & $\checkmark$   &      & $\checkmark$ &   &            &        & $\checkmark$          \\
78  & I16484-4603 & 3.14±0.01     & 2.51±0.02     & 7.14±0.03  & 2.29±0.01   & 1.0±0.2   & 8.1±0.4   &           & $\checkmark$     & -   & -    & $\checkmark$ &   &            &        &            \\
79  & I16487-4423 & 5.20±0.02     & 2.07±0.02     & 4.91±0.02  & 0.79±0.01   & 0.9±0.1   & 30.9±0.7  &           & $\checkmark$     & $\checkmark$   & $\checkmark$    & $\checkmark$ &   &            &        &            \\
80  & I16489-4431 & 9.99±0.03     & 4.84±0.03     & 9.43±0.03  & 1.83±0.01   & 1.54±0.04 & 7.4±0.4   &           & $\checkmark$     & $\checkmark$   &      & $\checkmark$ &   &            &        &            \\       
\end{tabular}
}
\end{threeparttable}
\end{table*}
\begin{table*}
\renewcommand\tabcolsep{2.2pt}
\begin{threeparttable}
\scalebox{0.84}{
\begin{tabular}{ccllllllccccccccc}
\hline
\hline
ID  & IRAS       
& \multicolumn{2}{c}{$L_\textup{SiO}$$\times{10^{-5}}$}   
& \multicolumn{2}{c}{$N$(SiO)$\times{10^{12}}$}  & \multicolumn{2}{c}{$X$(SiO)$\times{10^{-10}}$}   
& Filaments 
& \multicolumn{3}{c}{Infrared emission }  & \textit{A} group 
& \textit{B} group 
& \multicolumn{3}{c}{UC H\textsc{ii}}\\
 &   & Broad   & Narrow   & Broad   & Narrow  & $B$ position    & $N$ position  &  & 4.5$\upmu$m & 8$\upmu$m & 24$\upmu$m & &  
& \textit{concident} 
& \textit{offset}
& \textit{surrounded} \\
&   
&\multicolumn{2}{c}{($L_{\odot}$)}   
&\multicolumn{2}{c}{(cm$^{-2}$)}
& &    &   &  &  & & &  \\
\hline\\
81  & I16506-4512 & 1.45±0.02     & 0.52±0.01     & 2.49±0.03  & 0.36±0.01   & 1.6±0.2   & 2.5±0.3   &           & $\checkmark$     & $\checkmark$   &      & $\checkmark$ &   &            & $\checkmark$      &            \\
82  & I16524-4300 & 3.57±0.02     & 1.96±0.01     & 3.04±0.02  & 0.67±0.01   & 0.4±0.1   & 8.2±0.2   &           & $\checkmark$     & $\checkmark$   & $\checkmark$    & $\checkmark$ &   &            &        &            \\
85  & I16571-4029 & 6.42±0.01     & 3.88±0.01     & 11.36±0.02 & 2.76±0.01   & 2.5±0.1   & 8.0±0.2   &           & $\checkmark$     & $\checkmark$   & $\checkmark$    & $\checkmark$ &   &            &        &            \\
86  & I17006-4215 & 2.09±0.01     & 1.41±0.03     & 4.28±0.02  & 1.16±0.02   & 4.3±0.4   & 6.6±0.5   &           & $\checkmark$     & $\checkmark$   &      & $\checkmark$ &   &            & $\checkmark$      &            \\
89  & I17136-3617 & 1.060±0.003   & 0             & 5.91±0.02  & 0           & 0.9±0.1   & 5.6±0.2   &           & $\checkmark$     & $\checkmark$   &      & $\checkmark$ &   &            & $\checkmark$      &            \\
90  & I17143-3700 & 89.5±0.7      & 50.8±0.4      & 5.59±0.05  & 1.27±0.01   & 0.2±0.1   & 2.3±0.6   &           & $\checkmark$     & $\checkmark$   &      & $\checkmark$ &   & $\checkmark$          &        &            \\
91  & I17158-3901 & 6.49±0.03     & 4.02±0.03     & 5.69±0.02  & 1.42±0.01   & 9.2±0.3   & 8.4±0.4   &           & $\checkmark$     & $\checkmark$   & $\checkmark$    & $\checkmark$ &   &            &        &            \\
92  & I17160-3707 & 67.7±0.6      & 29.5±0.9      & 6.12±0.05  & 1.07±0.03   & 0.7±0.3   & 6.5±0.7   &           & $\checkmark$     & $\checkmark$   &      & $\checkmark$ &   &            & $\checkmark$      &            \\
94  & I17204-3636 & 6.89±0.02     & 2.80±0.02     & 6.26±0.02  & 1.02±0.01   & 0.7±0.1   & 14.9±0.4  &           & $\checkmark$     & $\checkmark$   &      & $\checkmark$ &   & $\checkmark$          &        &            \\
97  & I17244-3536 & 0.603±0.007   & 0.21±0.01     & 3.27±0.02  & 0.45±0.02   & 1.3±0.1   & 7.6±0.3   &           & $\checkmark$     & $\checkmark$   &      & $\checkmark$ &   &            & $\checkmark$      &            \\
98  & I17258-3637 & 2.75±0.07     & 1.45±0.03     & 4.11±0.11  & 0.87±0.02   & 0.3±0.2   & 3±1       &           & $\checkmark$     & $\checkmark$   &      & $\checkmark$ &   &            &        & $\checkmark$          \\
99  & I17269-3312 & 10.31±0.05    & 7.11±0.08     & 5.39±0.03  & 1.49±0.02   & 7.9±0.3   & 22.4±0.8  & N         &       &     &      & $\checkmark$ &   &            &        &            \\
100 & I17271-3439 & 7.06±0.02     & 3.35±0.02     & 7.37±0.02  & 1.40±0.01   & 2.0±0.1   & 7.4±0.2   &           & $\checkmark$     & $\checkmark$   &      & $\checkmark$ &   &            & $\checkmark$      &            \\
101 & I17278-3541 & 1.77±0.02     & 0.96±0.01     & 10.05±0.09 & 2.18±0.02   & 4.5±0.4   & 10.0±1.6  &           & $\checkmark$     & $\checkmark$   & -    &   & $\checkmark$ &            &        &            \\
102 & I17439-2845 & 54.27±0.17    & 14.98±0.11    & 8.50±0.03  & 0.94±0.01   & 3.1±0.1   & 2.5±0.4   &           & $\checkmark$     & $\checkmark$   & $\checkmark$    & $\checkmark$ &   &            &        &            \\
104 & I17455-2800 & 63.37±0.21    & 32.7±0.3      & 6.35±0.02  & 1.32±0.01   & 1.7±0.1   & 7.0±0.4   &           & $\checkmark$     & $\checkmark$   &      & $\checkmark$ &   &            & $\checkmark$      &            \\
105 & I17545-2357 & 3.61±0.03     & 1.63±0.02     & 4.22±0.04  & 0.76±0.01   & 2.3±0.2   & 17.3±0.6  &           & $\checkmark$     & $\checkmark$   &      & $\checkmark$ &   &            & $\checkmark$      &            \\
106 & I17589-2312 & 4.11±0.02     & 1.78±0.01     & 4.68±0.02  & 0.81±0.01   & 1.2±0.1   & 1.7±0.3   &           & $\checkmark$     & $\checkmark$   & $\checkmark$    & $\checkmark$ &   &            &        &            \\
107 & I17599-2148 & 4.29±0.03     & 2.15±0.02     & 4.81±0.04  & 0.97±0.01   & 0.4±0.2   & 1.6±0.5   &           & $\checkmark$     & $\checkmark$   &      & $\checkmark$ &   &            & $\checkmark$      &            \\
108 & I18032-2032 & 38.52±0.03    & 10.86±0.03    & 14.56±0.01 & 1.65±0.01   & 0.51±0.02 & 10.2±0.1  &           & $\checkmark$     & $\checkmark$   &      & $\checkmark$ &   &            & $\checkmark$      &            \\
109 & I18056-1952 & 111.84±0.31   & 44.20±0.09    & 15.34±0.04 & 2.43±0.01   & 0.26±0.03 & 8.9±0.4   &           & $\checkmark$     & $\checkmark$   & $\checkmark$    &   & $\checkmark$ &            &        &            \\
111 & I18079-1756 & 0.88±0.01     & 0.64±0.02     & 2.64±0.02  & 0.77±0.02   & 2.4±0.2   & 1.3±0.3   &           & $\checkmark$     & $\checkmark$   & $\checkmark$    & $\checkmark$ &   &            &        &            \\
112 & I18089-1732 & 3.72±0.02     & 3.20±0.02     & 5.96±0.04  & 2.06±0.01   & 0.9±0.1   & 9.9±0.7   &           & $\checkmark$     & $\checkmark$   & $\checkmark$    & $\checkmark$ &   &            &        &            \\
113 & I18110-1854 & 5.53±0.07     & 1.96±0.03     & 4.89±0.06  & 0.69±0.01   & 1.5±0.2   & 1.9±0.3   &           & $\checkmark$     & $\checkmark$   &      &   & $\checkmark$ &            & $\checkmark$      &            \\
114 & I18116-1646 & 7.69±0.08     & 2.81±0.08     & 4.97±0.05  & 0.73±0.02   & 0.4±0.2   & 4.3±0.7   & N         & $\checkmark$     & $\checkmark$   &      &   & $\checkmark$ & $\checkmark$          &        &            \\
115 & I18117-1753 & 4.07±0.04     & 1.67±0.03     & 6.18±0.06  & 1.02±0.02   & 0.8±0.2   & 7.2±0.8   &           & $\checkmark$     & $\checkmark$   & $\checkmark$    &   & $\checkmark$ &            &        &            \\
116 & I18134-1942 & 0.220±0.004   & 0.202±0.002   & 1.42±0.03  & 0.52±0.01   & 2.6±0.1   & 4.1±0.3   & N         & -     & -   & -    &   & $\checkmark$ &            &        &            \\
117 & I18139-1842 & 6.65±0.04     & 2.94±0.03     & 7.31±0.05  & 1.30±0.01   & 0.7±0.1   & 13.0±0.9  &           & -     & $\checkmark$   &      & $\checkmark$ &   &            & $\checkmark$      &            \\
118 & I18159-1648 & 1.55±0.01     & 0.93±0.01     & 7.08±0.02  & 1.70±0.01   & 4.1±0.2   & 7.6±0.7   &           & $\checkmark$     & $\checkmark$   & $\checkmark$    & $\checkmark$ &   &            &        &            \\
119 & I18182-1433 & 17.27±0.07    & 11.07±0.03    & 7.80±0.03  & 2.01±0.01   & 0.47±0.03 & 15.7±0.7  &           & $\checkmark$     & $\checkmark$   & $\checkmark$    & $\checkmark$ &   &            &        &            \\
120 & I18223-1243 & 1.88±0.02     & 1.11±0.02     & 1.66±0.02  & 0.40±0.01   & 0.24±0.04 & 1.9±0.4   &           & $\checkmark$     & $\checkmark$   & $\checkmark$    &   & $\checkmark$ &            &        &            \\
121 & I18228-1312 & 1.51±0.03     & 0.42±0.02     & 1.47±0.02  & 0.16±0.01   & 1.4±0.1   & 1.4±0.3   &           &       &     &      & $\checkmark$ &   &            & $\checkmark$      &            \\
122 & I18236-1205 & 2.86±0.03     & 1.51±0.04     & 6.08±0.07  & 1.29±0.03   & 1.3±0.3   & 18.7±2.1  &           & $\checkmark$     &     & $\checkmark$    & $\checkmark$ &   &            &        &            \\
123 & I18264-1152 & 8.72±0.05     & 5.3±0.1       & 7.88±0.05  & 1.92±0.04   & 2.2±0.3   & 8.6±0.8   &           & $\checkmark$     & $\checkmark$   & $\checkmark$    & $\checkmark$ &   &            &        &            \\
124 & I18290-0924 & 10.00±0.04    & 3.58±0.04     & 3.52±0.01  & 0.51±0.01   & 1.11±0.05 & 2.3±0.2   & N         &       & $\checkmark$   &      & $\checkmark$ &   &            &        &            \\
126 & I18311-0809 & 22.59±0.08    & 8.94±0.06     & 6.17±0.02  & 0.98±0.01   & 0.7±0.1   & 5.3±0.4   &           & $\checkmark$     & $\checkmark$   &      & $\checkmark$ &   & $\checkmark$          &        &            \\
129 & I18317-0513 & 8.49±0.02     & 0.70±0.01     & 17.91±0.05 & 0.59±0.01   & 6.9±0.2   & 20.5±0.7  &           & $\checkmark$     & $\checkmark$   & -    &   & $\checkmark$ &            &        &            \\
130 & I18317-0757 & 10.74±0.11    & 4.26±0.12     & 4.69±0.05  & 0.75±0.02   & 0.9±0.2   & 2.4±0.8   & N         & -     & -   &      & $\checkmark$ &   &            & $\checkmark$      &            \\
131 & I18341-0727 & 20.7±0.5      & 12.1±0.4      & 5.68±0.13  & 1.34±0.04   & 1.2±0.9   & 17.3±2.6  &           & $\checkmark$     & $\checkmark$   & $\checkmark$    & $\checkmark$ &   &            &        &            \\
132 & I18411-0338 & 25.48±0.18    & 13.16±0.09    & 4.65±0.03  & 0.97±0.01   & 0.5±0.1   & 7.9±0.7   &           & $\checkmark$     & $\checkmark$   & $\checkmark$    &   & $\checkmark$ &            &        &            \\
133 & I18434-0242 & 25.69±0.05    & 14.23±0.02    & 9.67±0.02  & 2.151±0.004 & 1.2±0.1   & 3.6±0.2   &           & $\checkmark$     & $\checkmark$   &      & $\checkmark$ &   &            & $\checkmark$      &            \\
135 & I18445-0222 & 10.2±0.1      & 4.08±0.07     & 3.83±0.04  & 0.62±0.01   & 0.4±0.1   & 10±1      &           & $\checkmark$     & $\checkmark$   & $\checkmark$    & $\checkmark$ &   &            &        &            \\
136 & I18461-0113 & 13.57±0.07    & 10.51±0.04    & 5.11±0.03  & 1.59±0.01   & 1.0±0.1   & 6.8±0.6   &           & $\checkmark$     & $\checkmark$   & $\checkmark$    & $\checkmark$ &   &            &        &            \\
137 & I18469-0132 & 10.38±0.06    & 8.75±0.03     & 3.91±0.02  & 1.32±0.01   & 1.3±0.2   & 3.0±0.4   &           & $\checkmark$     & $\checkmark$   & $\checkmark$    & $\checkmark$ &   &            &        &            \\
142 & I18517+0437 & 1.17±0.01     & 0.68±0.01     & 2.10±0.01  & 0.49±0.01   & 0.3±0.1   & 4.3±0.2   &           & $\checkmark$     & $\checkmark$   & $\checkmark$    & $\checkmark$ &   &            &        &            \\
143 & I18530+0215 & 5.47±0.06     & 3.31±0.05     & 2.51±0.03  & 0.61±0.01   & 0.4±0.1   & 6.4±0.7   &           & $\checkmark$     & $\checkmark$   &      & $\checkmark$ &   &            &        & $\checkmark$          \\
145 & I19095+0930 & 25.15±0.09    & 14.06±0.07    & 6.96±0.02  & 1.56±0.01   & 0.6±0.1   & 12.9±0.4  &           & $\checkmark$     & $\checkmark$   &      &   & $\checkmark$ &            & $\checkmark$      &            \\
146 & I19097+0847 & 47.67±0.18    & 35.77±0.12    & 6.66±0.02  & 2.01±0.01   & 1.3±0.1   & 27.7±0.3  &           & $\checkmark$     & $\checkmark$   &      & $\checkmark$ &   &            &        & $\checkmark$         
\end{tabular}
}
\end{threeparttable}
 \begin{tablenotes}
        \footnotesize
        \item[] The symbol ‘-’ indicates that the corresponding emission is not detect in this source.
        \item[] The symbol ‘N’ in the Filaments column indicates that the SiO emission is not associated with the filament skeleton identified by \citet{zhou2022atoms} in this source.
        \item[] The 4.5, 8 and 24$\upmu$m emission were obtained from the GLIMPSE and MIPSGAL \citep{benjamin2003glimpse, churchwell2009spitzer,2009PASP..121...76C}.
  \end{tablenotes}
\end{table*}

Table \ref{tab:TableA1} presents the basic parameters of the observed sources, including the systemic velocity ($v_{\rm LSR}$ km s$^{-1}$), distance from the sun (kpc), bolometric luminosity($L_\textup{bol}$, $L_{\odot}$), luminosity to mass ratio ($L_\textup{bol}/M$ $L_{\odot}$ $M_{\odot}$$^{-1}$), velocity integrated intensity for SiO broad components ($\int{F_\textup{B}\mathrm{d}{\upsilon}}$ Jy km s$^{-1}$),velocity integrated intensity for SiO narrow components ($\int{F_\textup{N}\mathrm{d}{\upsilon}}$ Jy km s$^{-1}$), velocity integrated intensity for H$^{13}$CO$^{+}$ extracted from ‘\textit{B}’ ($\int{F_\textup{B}\mathrm{d}{\upsilon}}$ Jy km s$^{-1}$), H$^{13}$CO$^{+}$ line width extracted from ‘\textit{B}’ position (FWHM$_{\rm B}$ km s$^{-1}$), H$^{13}$CO$^{+}$ line width extracted from ‘\textit{N}’ position (FWHM$_{\rm N}$ km s$^{-1}$), the contribution of SiO narrow components in total velocity integrated intensity (Contribution$_N$ $\%$), the number of sub-areas for detected SiO broad components in each source (B), the number of sub-areas for detected SiO narrow components in each source (N), the size of sub-areas used to extract average spectra in each source (Area), 1$\sigma$ rms noise  (Jy km s$^{-1}$), identified outflow activity from Baug et al. (in prep).

Table \ref{tab:TableA2} shows the basic parameters of the velocity decomposed sources, 
including the SiO luminosity for broad and narrow components ($L_\textup{SiO}$, $L_{\odot}$), the SiO column density for broad and narrow components 
($N$(SiO) cm$^{-2}$), the abundance in the position with the broadest
SiO line width and in the position with narrowest SiO line width ($X$(SiO)), the SiO emission is associated with filaments skeleton, the infrared emission (4.5 $\upmu$m, 8 $\upmu$m, and 24 $\upmu$m), the two cases based on the morphology between the narrow and broad components (\textit{A} and \textit{B} groups), and the three conditions of decomposed sources that host UC H\textsc{ii} regions based on their spatial distribution of the SiO and 3mm continuum emission (\textit{coincident}, \textit{offset}, and \textit{surrounded}). 
\newpage

\section{}
\label{Appendix B}
\begin{figure*}
\begin{minipage}{0.29\linewidth}
    \vspace{3pt}
\centerline{\includegraphics[width=1\linewidth]{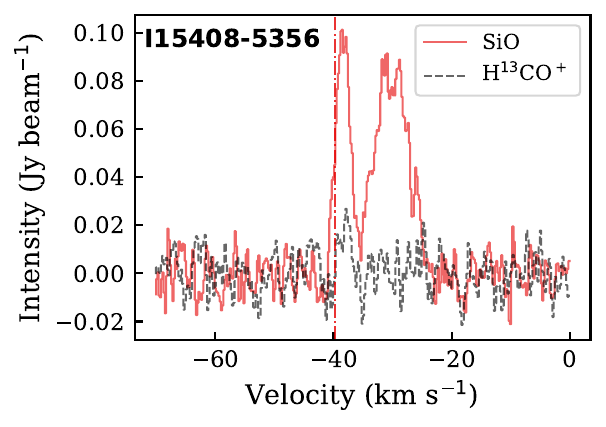}}
 \end{minipage}
\begin{minipage}{0.29\linewidth}
    \vspace{3pt}
\centerline{\includegraphics[width=1\linewidth]{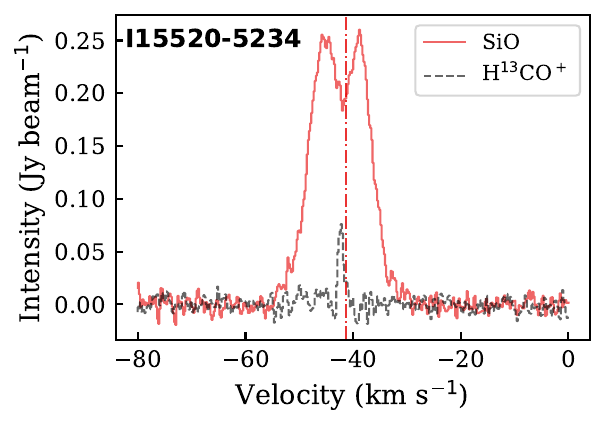}} 
 \end{minipage}
\begin{minipage}{0.29\linewidth}
    \vspace{3pt}
\centerline{\includegraphics[width=1\linewidth]{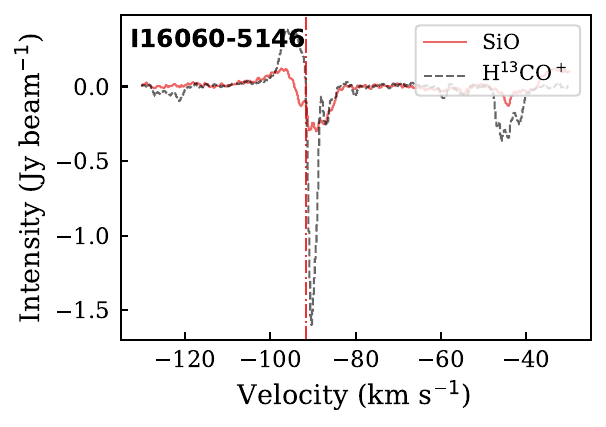}} 
 \end{minipage}
\begin{minipage}{0.29\linewidth}
    \vspace{3pt}
\centerline{\includegraphics[width=1\linewidth]{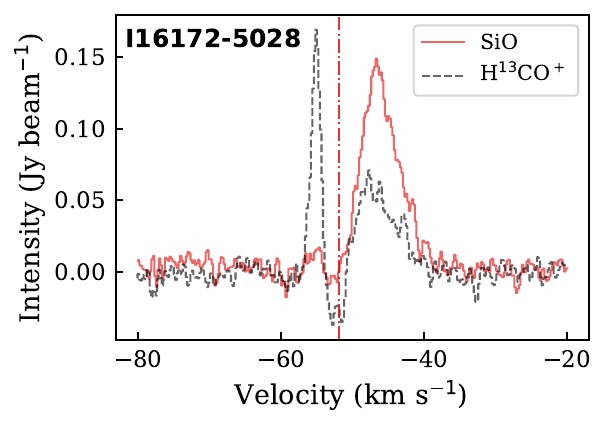}}
 \end{minipage}
\begin{minipage}{0.29\linewidth}
    \vspace{3pt}
\centerline{\includegraphics[width=1\linewidth]{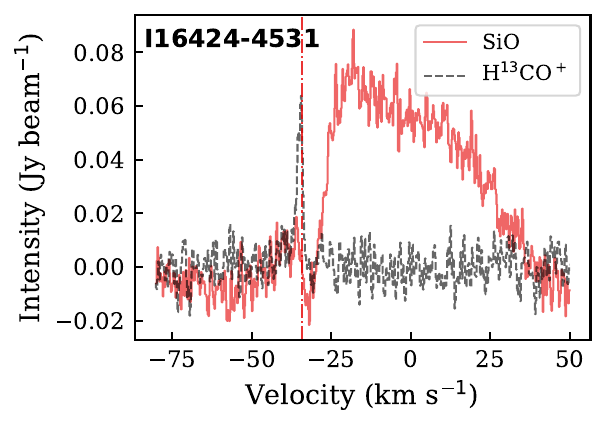}}
 \end{minipage}
\begin{minipage}{0.29\linewidth}
    \vspace{3pt}
\centerline{\includegraphics[width=1\linewidth]{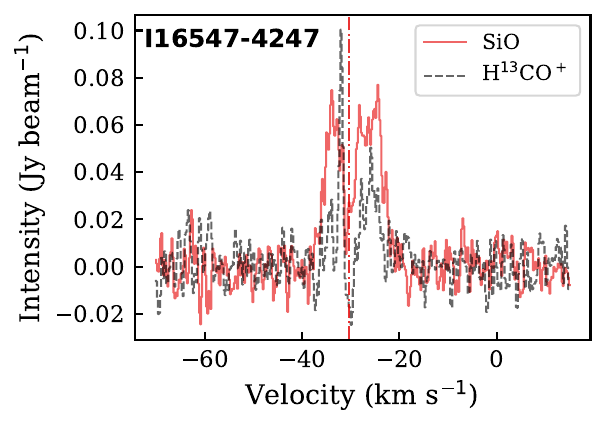}} 
 \end{minipage}
\begin{minipage}{0.29\linewidth}
    \vspace{3pt}
\centerline{\includegraphics[width=1\linewidth]{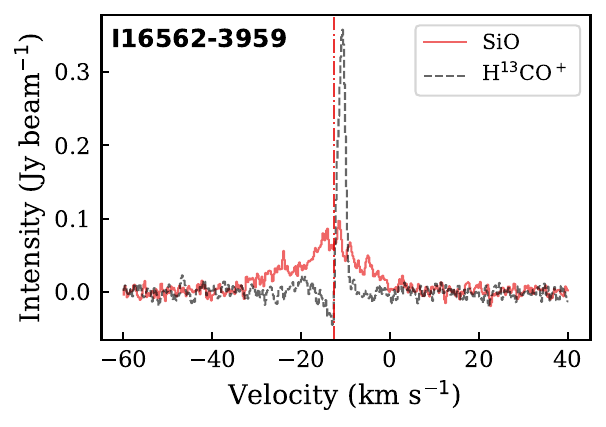}} 
 \end{minipage}
 \begin{minipage}{0.29\linewidth}
    \vspace{3pt}
\centerline{\includegraphics[width=1\linewidth]{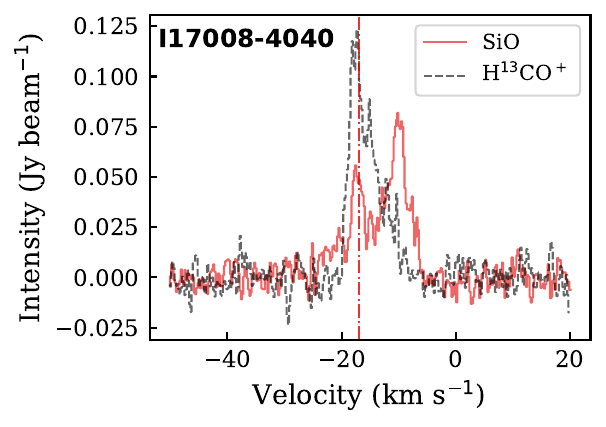}} 
 \end{minipage}
 \begin{minipage}{0.29\linewidth}
    \vspace{3pt}
\centerline{\includegraphics[width=1\linewidth]{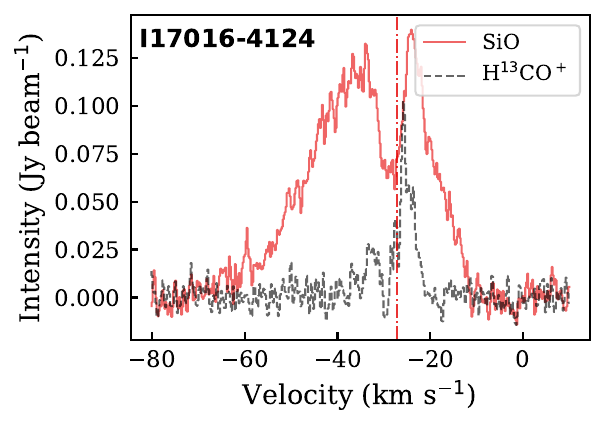}} 
 \end{minipage}
\begin{minipage}{0.29\linewidth}
    \vspace{3pt}
\centerline{\includegraphics[width=1\linewidth]{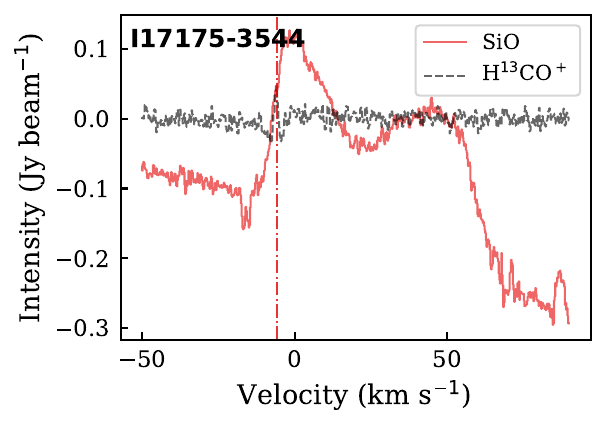}} 
 \end{minipage}
 \begin{minipage}{0.29\linewidth}
    \vspace{3pt}
\centerline{\includegraphics[width=1\linewidth]{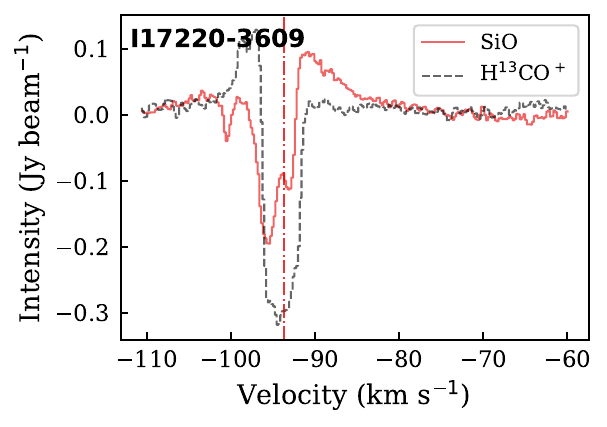}} 
 \end{minipage}
\begin{minipage}{0.29\linewidth}
    \vspace{3pt}
\centerline{\includegraphics[width=1\linewidth]{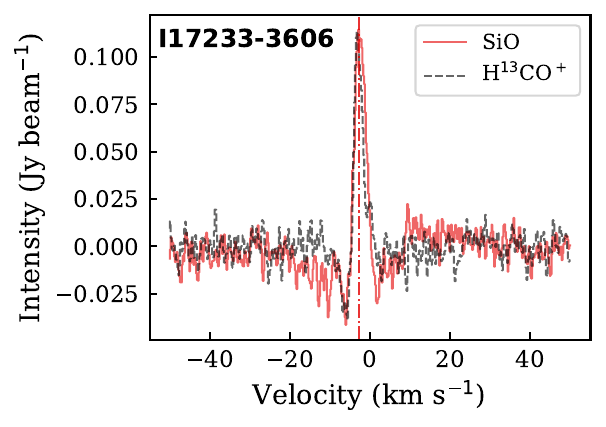}} 
 \end{minipage}
\begin{minipage}{0.29\linewidth}
    \vspace{3pt}
\centerline{\includegraphics[width=1\linewidth]{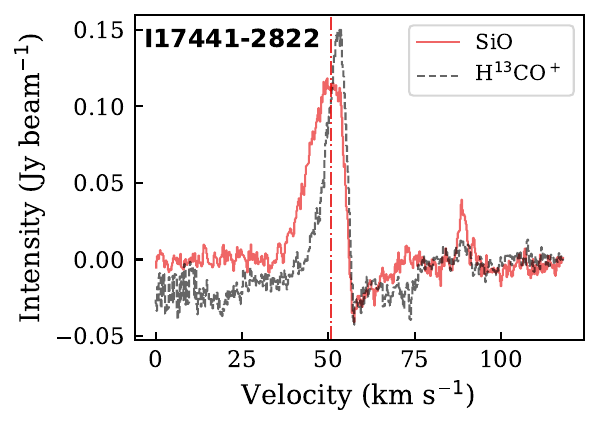}} 
 \end{minipage}
\begin{minipage}{0.29\linewidth}
    \vspace{3pt}
\centerline{\includegraphics[width=1\linewidth]{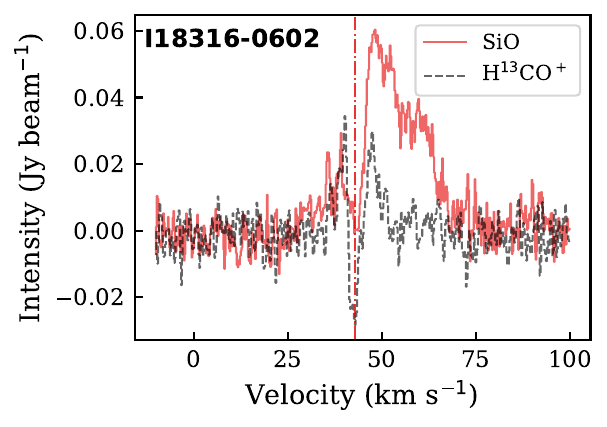}} 
 \end{minipage}
 \begin{minipage}{0.29\linewidth}
    \vspace{3pt}
\centerline{\includegraphics[width=1\linewidth]{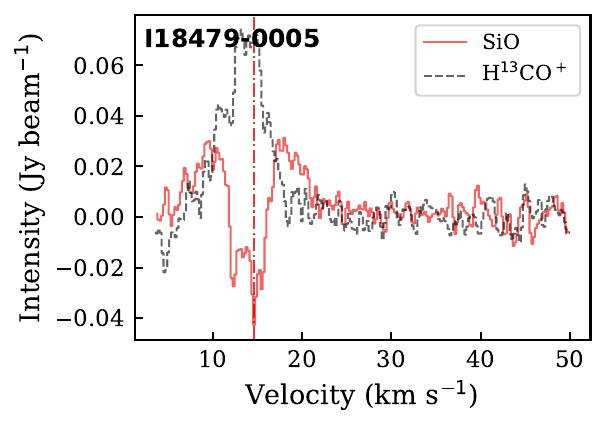}} 
 \end{minipage}
\begin{minipage}{0.29\linewidth}
    \vspace{3pt}
\centerline{\includegraphics[width=1\linewidth]{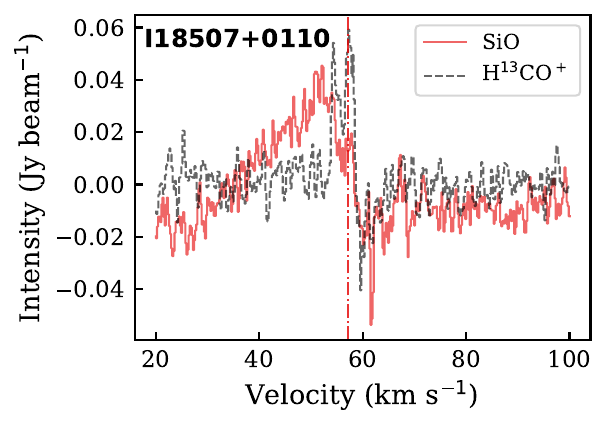}} 
 \end{minipage}
\begin{minipage}{0.29\linewidth}
    \vspace{3pt}
\centerline{\includegraphics[width=1\linewidth]{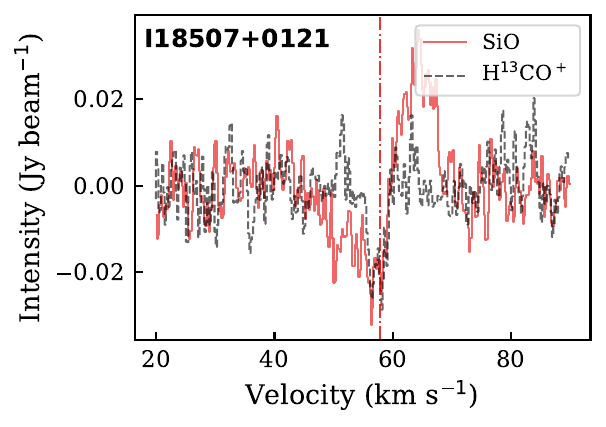}} 
 \end{minipage}
\begin{minipage}{0.29\linewidth}
    \vspace{3pt}
\centerline{\includegraphics[width=1\linewidth]{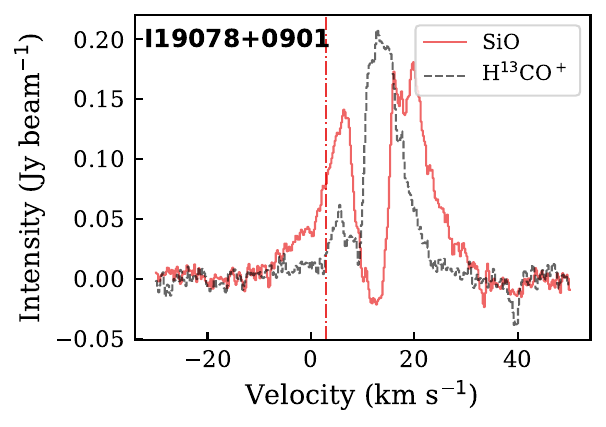}} 
 \end{minipage}
\caption{The observed SiO spectra with negative features. Red histograms and black dashed histograms represent the SiO and H$^{13}$CO$^+$ spectra, respectively.}
\label{figB1}
\end{figure*}
Figure~\ref{figB1} shows the SiO and H$^{13}$CO$^{+}$ spectra with absorption features. Sources with negative features around systemic velocity in both SiO and H$^{13}$ CO$^{+}$ line profiles are caused by the presence of strong continuum sources \citep{2022ApJ...927...54O,2024MNRAS.528.7383C}. These sources (I15520-5234, I16424-4531, and I18479-0005) only show SiO absorption, which may be due to the effects of large optical depth in the SiO (2-1) line emission.
\newpage

\section{}
\label{Appendix C}
\begin{figure*}
\begin{minipage}{0.29\linewidth}
    \vspace{3pt}
\centerline{\includegraphics[width=1.1\linewidth]{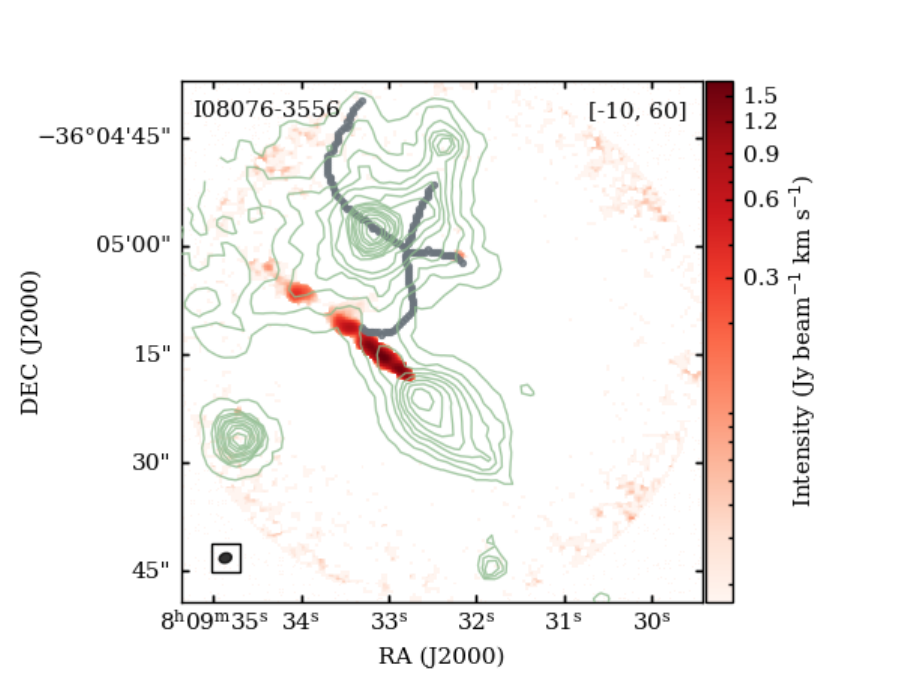}}
 \end{minipage}
\begin{minipage}{0.29\linewidth}
    \vspace{3pt}
\centerline{\includegraphics[width=1.1\linewidth]{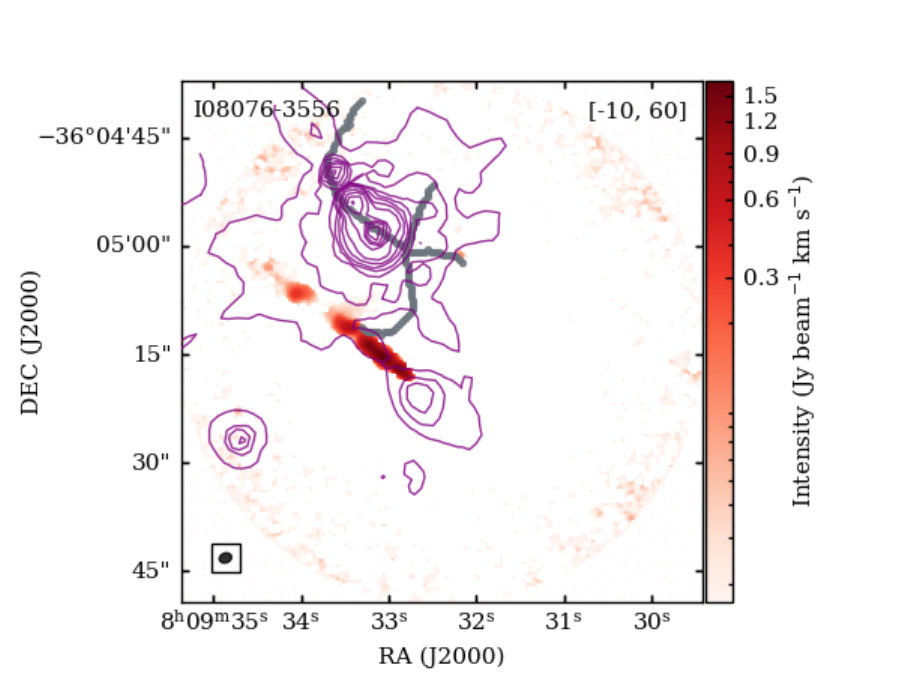}} 
 \end{minipage}
\begin{minipage}{0.29\linewidth}
    \vspace{3pt}
\centerline{\includegraphics[width=1.1\linewidth]{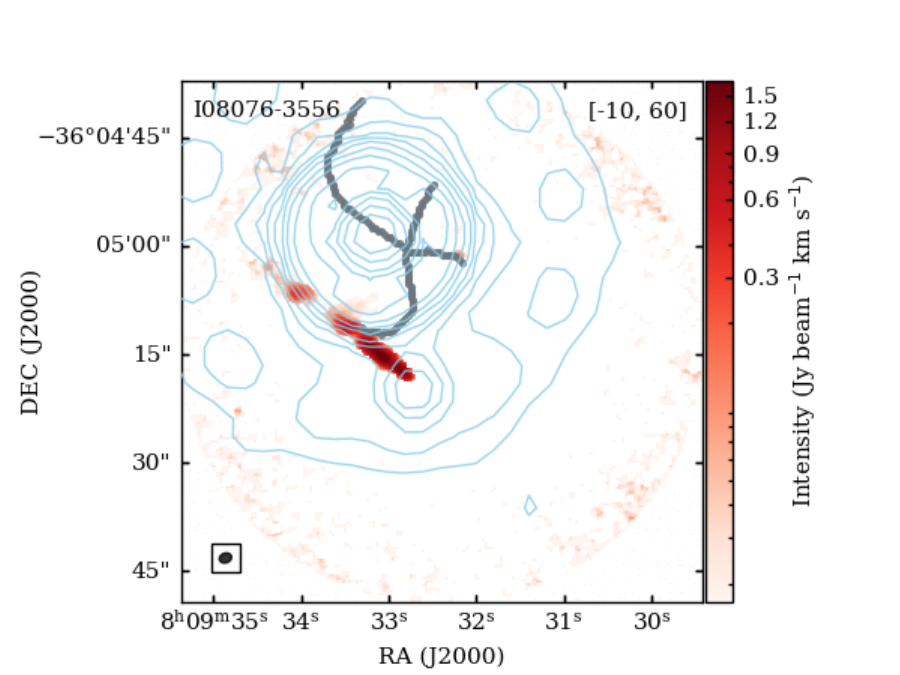}} 
 \end{minipage}
\begin{minipage}{0.29\linewidth}
    \vspace{3pt}
\centerline{\includegraphics[width=1.1\linewidth]{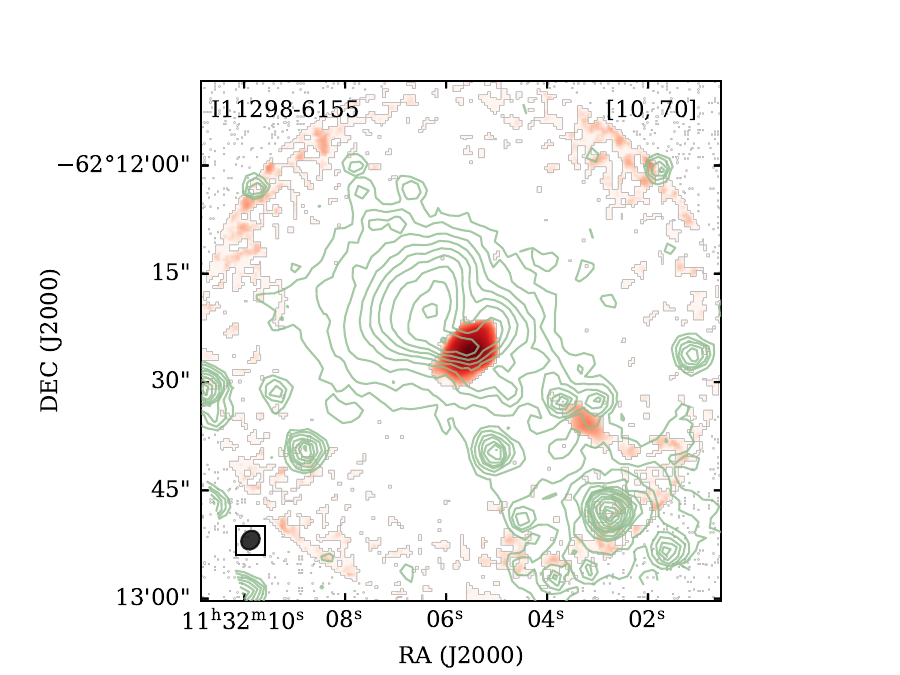}}
 \end{minipage}
\begin{minipage}{0.29\linewidth}
    \vspace{3pt}
\centerline{\includegraphics[width=1.1\linewidth]{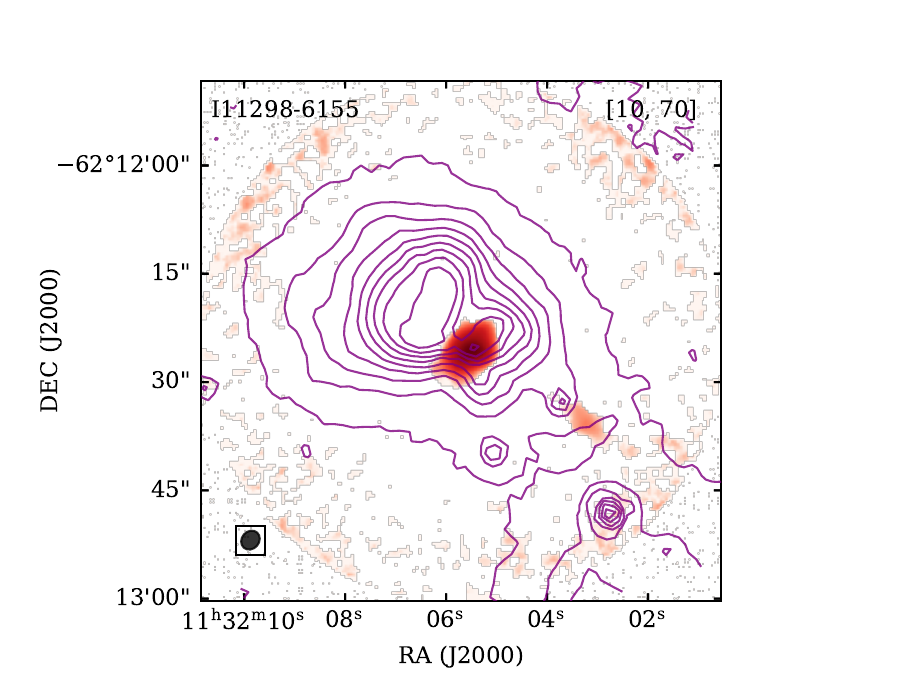}} 
 \end{minipage}
\begin{minipage}{0.29\linewidth}
    \vspace{3pt}
\centerline{\includegraphics[width=1.1\linewidth]{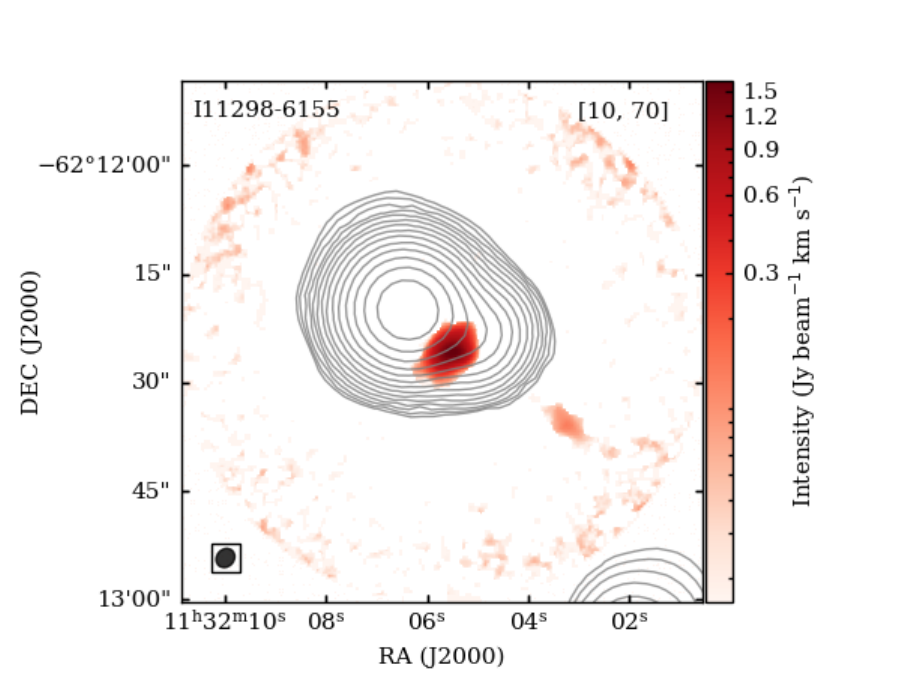}} 
 \end{minipage}
\caption{The background images correspond to the SiO (2-1) integrated intensity maps obtained within the ATOMS survey. {\it Upper panels}: One representative source without an UC H\textsc{ii} region. {\it Lower panels}: One representative source with an UC H\textsc{ii} region. The green, purple, blue, and gray contours are 4.5 $\upmu$m, 8 $\upmu$m, 24 $\upmu$m, and 1.3 GHz emission, respectively.
Contours range from 5$\sigma$ to the peak values, divided into 15 logarithmic steps (base 10). The 4.5, 8, and 24$\upmu$m emission were obtained from the GLIMPSE and MIPSGAL \citep{benjamin2003glimpse,churchwell2009spitzer,2009PASP..121...76C}, and the 1.3 GHz emission were utilized from \citet{2023arXiv231207275G}.  
The bold gray lines represent the filament skeleton identified in H$^{13}$CO$^+$ emission. 
The shown field of view is 72$^{\prime}$$^{\prime}$ corresponding to the FOV of the ALMA observations. The source name and integrated velocity ranges (in km s$^{-1}$) are shown on the upper left and right corners of each panel, respectively. The beam size is presented in the lower left corner. The same images are provided for all decomposed sources within the supplementary material.}
\label{figC1}
\end{figure*}
Figure~\ref{figC1} shows the SiO (2-1) integrated intensity maps overlaid with 4.5, 8, and 24 $\upmu$m emission toward sources lacking UC H\textsc{ii} regions and the SiO (2-1) integrated intensity maps overlaid with 4.5, 8, and 1.3 GHz emission toward sources with UC H\textsc{ii} regions, respectively. 
The images for decomposed sources, both with and without UC H\textsc{ii} regions, are available within the supplementary material. 

\bsp	
\label{lastpage}
\end{document}